\documentclass{aa}  
\usepackage{graphicx}
\usepackage{lscape}
\usepackage{txfonts}
\newcommand{\kms}{\mbox{km~s$^{-1}$}}
\newcommand{\vlsr}{\mbox{$\varv_{\rm lsr}$}}

\newcommand{\uchii}{UCHII}
\newcommand{\unidens}{cm$^{-2}$}
\newcommand{\lsol}{L$_\odot$}
\newcommand{\msol}{M$_\odot$}
\newcommand{\hii}{H\scriptsize{II}}
\begin{document} 

\title{Complex organic molecules uncover deeply embedded precursors of hot cores}
\subtitle{An APEX unbiased spectral survey of infrared quiet massive clumps}

   \author{L. Bouscasse
          \inst{1,2}
          \and
           T. Csengeri \inst{3}
           \and F. Wyrowski \inst{2}
           \and K. M. Menten \inst{2}
           \and S. Bontemps \inst{3}
           }

   \institute{ {IRAM, 300 rue de la Piscine, 38400 Saint-Martin-d'Hères, France} \label{inst1}
\and   {Max-Planck-Institut f\"ur Radiastronomie, Auf dem Hügel 69, 53121 Bonn, Germany} \label{inst2}
\and {Laboratoire d'astrophysique de Bordeaux, Univ. Bordeaux, CNRS, B18N, allée Geoffroy Saint-Hilaire, 33615 Pessac, France} \label{inst3} }

   \date{Received  ; accepted  }
  \abstract
  {During the process of star formation, the dense gas undergoes significant chemical evolution leading to the emergence of a rich variety of molecules associated with hot cores and hot corinos. However, the physical conditions and the chemical processes involved in this evolution are poorly constrained. In particular the early phases, corresponding to a stage prior to the emergence of any strong ionising emission from the protostar, are still poorly studied.}
   {We provide here a full inventory of the emission from complex organic molecules (COMs) to investigate the physical structure and chemical composition of six high-mass protostellar envelopes. We aim to investigate the conditions for the emergence of COMs in hot cores.}
   {We performed an unbiased spectral survey towards six infrared-quiet massive clumps between 159 GHz and 374 GHz with the APEX~12\,m telescope, covering the entire atmospheric windows at 2\,mm, 1.2\,mm, and 0.8\,mm. To identify the spectral lines, we used rotational diagrams and radiative transfer modelling assuming local thermodynamic equilibrium.}
   {We detect up to 11 COMs plus three isotopologues, of which at least five COMs (CH$_3$OH, CH$_3$CN, CH$_3$OCHO, CH$_3$OCH$_3$, CH$_3$CHO) are detected  
   towards all sources. Towards all the objects, most of the COM emission is found to be cold, with respect to the typical temperatures at which COMs are found, with a temperature of 30\,K and extended with a size of $\sim$0.3\,pc. Although for our sample of young massive clumps the bulk of the gas has a cold temperature, we also detect emission from COMs originating from the immediate vicinity of the protostar. This warm component of the envelope is best traced by methanol and methyl cyanide, in particular methyl cyanide traces a compact ($\sim$1$''$) and hottest ($T\sim$200\,K) component of the envelope. Only three out of the six sources exhibit a robustly detected  hot gas component (T$>$100\,K) traced by several COMs. We find a gradual emergence of the warm component in terms of size and temperature, together with an increasing molecular complexity, allowing us to establish an evolutionary sequence for our sample based on COMs. While they can already be well characterized by an emerging molecular richness, gas temperatures of COMs in the hot gas and molecular abundances suggest that COMs may become abundant in the gas phase at temperatures below the thermal desorption temperature.}
   {Our findings confirm that the sources of our sample of infrared-quiet massive clumps are in an early evolutionary stage during which the bulk of the gas is cold. The presence of COMs is found to be characteristic of these early evolutionary stages accompanying high-mass star and cluster formation. While the extent of the compact heated regions resembles that of hot cores, the molecular abundances, except for complex cyanides, resemble those of hot corinos and are lower than the peak COM abundances of hot cores. We suggest that the emergence of hot cores is preceded by a phase in which mostly O-bearing COMs appear first with similar abundances to hot corinos albeit with larger source sizes.} 


   \maketitle

%

\section{Introduction}

The formation of high-mass stars takes place in dense and compact cores within giant molecular clouds \citep{Beuther2007, Tan2014, Motte2018}. In hot molecular cores, associated with high-mass star and cluster formation, the high temperatures lead to the sublimation of icy dust mantles which reveals a large molecular richness including complex organic molecules (COMs) \citep{Kurtz2000, Cesaroni2005}. 
The molecular composition of hot cores has been observed to show significant variations \citep[e.g.][]{Widicus2017}, while the origin of this diversity remains poorly understood. In the early evolutionary phase, radiative feedback from the protostar is expected to be still limited to a compact region, while the bulk of the gas should consist of  chemically pristine star forming gas. The molecular content of this early warm-up phase should therefore reflect the chemical diversity 
for these early evolutionary phase.

Large area dust continuum surveys of the Galactic Plane, such as the Bolocam Galactic Plane Survey \citep{Aguirre2011}, the APEX Telescope Large Survey of the Galaxy \citep[ATLASGAL][]{Schuller2009,Csengeri2014}, Hi-GAL \citep{Molinari2010}, JCMT Plane Survey \citep{Eden2017}, delivered statistically significant samples of high-mass star forming regions. These large samples have been investigated in depth for their physical parameters leading to a  characterization into different evolutionary stages \citep{Csengeri2016, Koenig2017, Urquhart2018}. This allowed us to recognise samples of  
massive clumps  
in an early evolutionary phase. Massive clumps that are precursors to high-mass stars and clusters in an early evolutionary stage are believed to host protostars at the onset of radiative heating, which gives us the possibility to study the early warm-up phase chemistry.

Sites of high-mass star formation are particularly important for studies of the emergence of  molecular richness due to their high column densities,
and several studies underline the molecular diversity observed towards these sites \citep[e.g][]{Foster2011, Gerner2015, Taniguchi2019, Mininni2020, Peng2022}.
Consequently, a large number of molecular species in the interstellar medium have been first discovered towards active sites of high-mass star formation, for example Sgr~B2 \citep[e.g][]{Belloche2013, McGuire2016S, Belloche2017,Belloche2019}. While the precise formation processes of  COMs are still debated, it is  accepted that they can form efficiently  
at low temperatures when molecules stick to the grains, 
allowing the subsequent formation of COMs on the surface of grains and ices  \citep{Garrod2006, Garrod2008}. Recently, \citet{Garrod2022} showed that non-diffusive grain surface and bulk-ice reactions enhance the production of COMs at even lower temperatures compared to their previous models.
Towards hot cores and hot corinos, the emergence of the molecular richness in the gas phase is assumed to originate from the heated inner regions of the protostellar envelope where radiative heating liberates the grain surface products to the gas phase. At high temperatures, subsequent gas phase reactions contribute to the molecular evolution of the gas \citep[e.g.][]{Charnley1997}. 
However, observations of dark clouds and prestellar cores reveal the presence of COMs at low temperatures ($\sim$10--30\,K) even in the pre-collapse gas \citep{Oberg2010,Bacmann2012,Jaber2014, McGuire2020, Cernicharo2022}. The observed molecular abundances of COMs at such low temperatures suggest non-thermal desorption processes, such as  sputtering \citep{Dartois2020}, UV photons propagating through the outflow cavity \citep{Oberg2010}, shocks or supersonic turbulence \citep{Requenas-Torres2006}, cosmic ray radiation or secondary UV photons \citep{Bacmann2012} or chemical desorption \citep{Garrod2007}.
Complementary to that, efficient formation mechanisms of COMs at low temperatures in the gas phase have also been suggested \citep{Balucani2015, Vasyunin2017}.

Sensitive unbiased wide band spectral line surveys are extremely useful to characterise the chemical composition and the physical structure of star-forming regions \citep[e.g.][]{Herbst2009, Caux2011, Fontani2017}. Simultaneously observable large bandwidths grant access to several transitions of a given molecular species in particular of COMs, covering a large range of upper level energies. 
This allows us to estimate the rotational temperature, column density, and size of the emission of each molecule, even when this emission is spatially unresolved. 
The large instantaneous bandwidth of receivers has made this type of study feasible and considerably increased the number of line surveys such as ASAI carried out with the IRAM~30m telescope \citep{Lefloch2018}, and several others with the ALMA and NOEMA interferometers, such as PILS \citep{Jorgensen2016}, EMoCA and ReMoCA surveys \citep{Belloche2016,Belloche2019}, SOLIS \citep{Ceccarelli2017}, ATOMS \citep{Liu2020}, GUAPOS \citep{Mininni2020}.

The SPARKS project (Search for High-mass Protostars with ALMA revealed up to kilo-parsec scales, \citealt{Csengeri2017b, Csengeri2018}, Csengeri et al. in prep) targeted a homogenous and representative sample of infrared quiet\footnote{\citet{Csengeri2017} defines an infrared-quiet clump as a source with a flux density below 289~Jy at 22~$\mu$m at 1~kpc, possibly hosting  O-type stars or their precursors.} \citep[e.g.][]{Tackenberg2012, Csengeri2017}  massive clumps (with $L_{\rm bol}\lesssim10^4\,L_{\odot}$ and $M>650\,M_{\odot}$) based on the ATLASGAL survey, and revealed a sample of massive dense cores. From this sample, we selected the only six massive clumps hosting  high-mass protostars that are isolated, that is, appear as a single source on scales from 0.3\,pc down to 2000\,au with no other compact and bright sub-millimetre source in their vicinity. These early stage high-mass protostars, all associated with powerful outflows observed with ALMA, are excellent laboratories for studying the impact of the onset of radiative heating on the chemistry during the early warm-up phase.

Here we study the molecular composition, in particular that of COMs,  of these six massive protostellar envelopes using the Atacama Pathfinder EXperiment 12~m  submillimeter telescope (APEX, \citealt{Gusten2006}). We performed an unbiased spectral survey in the 2\,mm, 1.2\,mm, and 0.8\,mm atmospheric bands in order to study the physical conditions and the molecular composition of the dense gas in these sources. The paper is organised as follows: in Sect.\,\ref{sec:obs} we present the observations and the data reduction. In Sect.\,\ref{sec:results}, we show the resulting spectra and introduce our analysis methods. In Sect.\,\ref{sec:COMs_properties}, we describe the properties of COMs in our sample. In Sect.\,\ref{sec:discussion}, we discuss these properties. In Sect.\,\ref{sec:sec_comp} we compare the physical and chemical properties of our sample with samples of hot corinos and hot cores.
%

\section{Observations and data reduction}
\label{sec:obs}
We performed an unbiased spectral survey between 159~GHz and 374~GHz with the APEX telescope towards six massive protostellar objects from the SPARKS survey \citep{Csengeri2017b} that are associated with the ATLASGAL clumps G320.2325-0.2844 (G320.23), G328.2551-0.5321 (G328.25), G333.4659-0.1641 (G333.46), G335.5857-0.2906 (G335.58), G335.7896+0.1737 (G335.78), G343.7559-0.1640 (G343.75). We list their physical properties in Table \ref{tab:source_coord}.  Our approach was to cover the complete atmospheric transmission windows of the 2~mm band using the SEPIA receiver \citep{Belitsky2018}, the 1.2~mm band using the PI230 receiver and the 0.8~mm band using the FLASH$^+$ receiver \citep{Klein2014} at the APEX telescope. Observations of the G328.2551$-$0.5321 clump were discussed in detail in \citet{paperI}, here we present the same dataset for the entire sample of six sources.

In short, the observations were done in position switching mode with a reference position offset of 600$''$, $-$600$''$ in right ascension and declination with respect to the central position of the ATLASGAL clumps given in Table\,\ref{tab:source_coord}, a nearby position chosen to secure baseline stability.

\begin{table*}
\caption{Source properties.}
\label{tab:source_coord}
\footnotesize
\begin{tabular}{l c c c c c c}
\hline
 & G320.2325-0.2844 & G328.2551-0.5321\tablefootmark{c} & G333.4659-0.1641 & G335.5857-0.2906  &  G335.7896+0.1737 & G343.7559-0.1640\\
\hline
\hline
Short name & G320.23 & G328.25 & G333.46 & G335.58 & G335.78 & G343.75\\
R.A (J2000) & $15^{\rm h}09^{\rm m}51.948{\rm s}$           & $15^{\rm h}57^{\rm m}59.791{\rm s}$           & $16^{\rm h}21^{\rm m}20.190{\rm s}$           & $16^{\rm h}30^{\rm m}58.754{\rm s}$           & $16^{\rm h}29^{\rm m}47.335{\rm s}$           & $17^{\rm h}00^{\rm m}49.870{\rm s}$           \\ 
Dec. (J2000) & $-58^\circ25'38\rlap{.}''38$          & $-53^\circ58'00\rlap{.}{''}56$          & $-50^\circ09'46\rlap{.}{''}60$          & $-48^\circ43'53\rlap{.}{''}91$          & $-48^\circ15'52\rlap{.}{''}33$          & $-42^\circ26'09\rlap{.}{''}09$          \\ 
$v_{\rm lsr}$ [km/s] & --66.3 & --43.13 & --42.48  & --46.22 & --49.46  & --26.91 \\ 
distance [kpc]\tablefootmark{a} &  3.9 &   2.8 &   4.2 &   3.8 &   3.8 &   1.8 \\ 
$L_{\rm bol}$ [\lsol]\tablefootmark{b} & 5$\times$10$^3$ & 1.2$\times$10$^4$ & 5$\times$10$^3$ & 1.9$\times$10$^4$ & 8$\times$10$^3$  & 1.3$\times$10$^3$ \\
M$_{\rm clump}$ [\msol]\tablefootmark{c} & 1162 & 591 & 1997 & 1964 & 1617 & 330 \\ 
F$_{\rm \nu}$ [Jy/beam]\tablefootmark{d} &   6.3 &   8.3 &   8.1 &  11.2 &   8.2 &  10.4 \\ 
$FWHM$ size [\arcsec]\tablefootmark{d} &  26.6 &  25.7 &  28.5 &  26.6 &  28.1 &  23.9 \\ 
$T_{\rm dust}$ [K]\tablefootmark{e} & 21  &  22 & 25  & 23  & 23  &  21 \\
$N_{\rm H_2}$ [cm$^{-2}$]\tablefootmark{f} & 1.5$\times$10$^{23}$& 1.8$\times$10$^{23}$ & 1.5$\times$10$^{23}$& 2.3$\times$10$^{23}$& 1.7$\times$10$^{23}$& 2.4$\times$10$^{23}$\\
\hline
\end{tabular}
\tablefoot{
  \tablefoottext{a}{Kinematic distance estimates from \citet{Csengeri2017}}
  \tablefoottext{b}{Bolometric luminosity estimates from Csengeri et al. (in prep)}
  \tablefoottext{b}{Clump mass estimates from \citep{Csengeri2017}}
  \tablefoottext{d}{APEX/LABOCA measurements at 870~$\mu$m from \citet{Csengeri2014}}
  \tablefoottext{e}{Dust temperature estimates from \citet{Urquhart2018}}
  \tablefoottext{f}{Mean H$_2$ column density estimates, in the LABOCA beam of 19$''$, computed assuming the dust temperature indicated in the table and a dust absorption coefficient of 1.85~cm$^2$~g$^{-1}$ at 0.8~mm (see Sect.\,\ref{sec:H2} for more details).}
  }
\end{table*}

\begin{table*}[!ht]
\footnotesize
\centering
\caption{Parameters of the observations.}
\label{tab:parameters_data}
\begin{tabular}{c c c c c c c c c c c}     
\hline 
\hline
Band & Receiver & Frequency & Observation & HBPW\tablefootmark{a} & Spec. res. & Polars & Pwv & T$_{\rm sys}$ (<T$_{\rm sys}$>) & <$\sigma$>\tablefootmark{b} & $\eta _{\rm mb}$\tablefootmark{a} \\  
 &  & [GHz] & dates [MM.YY] & [\arcsec] & [\kms] & & [mm] & [K] & [mK] & [\%]  \\           
\hline 
2\,mm & SEPIA180 & 159-185 & 09.18 & 36 &  0.13 & 2 & 0.3--2.4 & 75--293 (101)\tablefootmark{c} & 11 & 79\\ 
 & PI230 & 186-212 & 09.18 & 31 & 0.09 &2&0.6--3.5 & 93--308 (144) & 18 & 71\\
1.2\,mm & PI230 &  202-278 & 03-04.17 & 27 & 0.08 & 2&0.1--1.7 & 99--377 (151) & 14 & 71\\
0.8\,mm & FLASH345 & 270-374 & 09.18 & 18 & 0.03 & 1& 0.1--1.6 & 142--594 (324)\tablefootmark{c} & 25 & 69\\ 
\hline 
\end{tabular}
\tablefoot{ 
\tablefoottext{a}{Given at the frequency of 180\,GHz, 196\,GHz, 230\,GHz, and 345\,GHz for SEPIA180, PI230 (4-8\,GHz IF), PI230 (4-12\,GHz IF), and FLASH345, respectively.}
\tablefoottext{b}{Given for a resolution of 0.7\,\kms on $T_{\rm MB}$ scale. All frequencies were covered twice, except for the one at the edges of the bands, corresponding to the frequencies 159--161\,GHz, 186--188\,GHz, 318--320\,GHz, 328--330\,GHz, and 372--374\,GHz, where the rms is higher by a factor $\sqrt2$.}
\tablefoottext{c}{excluding sidebands covering telluric water lines.}
}
\end{table*}

The pointing accuracy of $\sim$2$''$ was checked by regular pointings on IRAS15194$-$5115, NGC 6072 and RAFGL 2135. The telescope focus was checked with observations of Mars and Jupiter. 
To eliminate ghost lines, spikes or line contamination from the other sideband, we observed our spectral survey with a double frequency coverage (see Fig.~1 from \citealt{paperI}) and hence used 14, 24, and 12 setups for the 2\,mm, 1\,mm, and 0.8\,mm band, respectively. The half power beam width (HBPW) varies between 16$''$ at 374\,GHz and 39$''$ at 159\,GHz.
In total we covered a frequency range of 206\,GHz per source. 
The main observation parameters are summarized in Table\,\ref{tab:parameters_data}. The median system temperature is 250\,K. The deviation from this value depends on the receiver and on the weather during the observations. The data calibration was done with the APEX online calibrator. 

We performed the data reduction with the GILDAS\footnote{https://www.iram.fr/IRAMFR/GILDAS/} software. We used the same method for the reduction of the data as described in \citet{paperI}. In short, we averaged the two polarizations, subtracted a zero order baseline and averaged the spectra of each sideband of each setup. We compared the overlapping spectra obtained with the double frequency coverage of our spectral survey to remove the artefacts and the ghost lines.  
We converted the corrected antenna temperatures (T$_A$*) to main beam temperatures (T$_{MB}$) using calibration factors for the telescope main beam efficiency determined from planet observations at selected frequencies.  
These efficiencies were then interpolated for the
frequencies of all the spectra of the whole survey using the Ruze formula: $\eta _{mb}=\eta _{0}\,\exp{[-(\frac{4\pi\delta\nu}{c})^2]}$ where $\eta _{mb}$ is the main beam efficiency, $\eta _{0}$ depends on the receiver and is given in Table\,\ref{tab:parameters_data}, $\delta$ the surface accuracy, $c$ the speed of light, and $\nu$ the frequency. A forward efficiency of 95\% was used for all receivers. Finally, we smoothed the data to a common resolution of 0.7\,\kms.

\section{Results and analysis}
\label{sec:results}
\begin{sidewaysfigure*}
    \centering
    \includegraphics[width=1\linewidth,trim=0cm 3cm 3cm 0cm, clip]{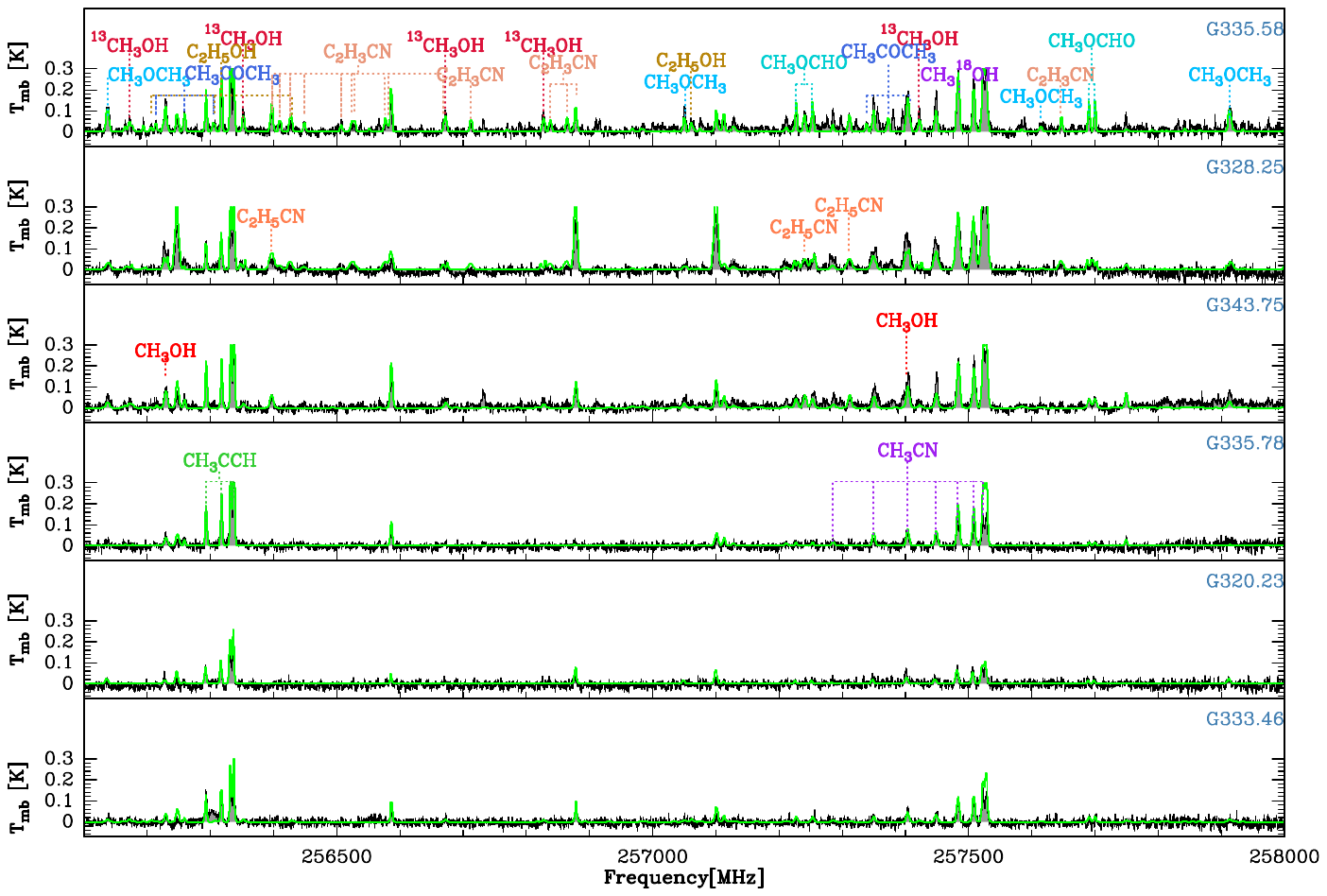}
    \caption{Spectrum towards all of our target sources between 256.1\,GHz and 258\,GHz. The filled grey histograms show the spectra, while the coloured labels show the detected transitions of all the light molecules and some transitions arising from COMs. The green line represents the LTE model 
    fitted with Weeds.}
    \label{fig:spectra}
\end{sidewaysfigure*}

Toward each source, we detect numerous spectral lines over the non-continuous 206\,GHz frequency range covered. Here we focus on identifying and analysing emission from COMs over this frequency range. As an example, Figure\,\ref{fig:spectra} shows a 2~GHz frequency range for all the six sources covering mostly rotational transitions from COMs.  
In total, we detect between 981 and 5433\,lines for each source. From these between 454 and 4584~lines originate from COMs over the total frequency range of 206\,GHz covered by the survey above the 3$\sigma$ level.

\subsection{Line identification}

We identified the observed spectral lines using the Cologne Database for Molecular Spectroscopy \citep[CDMS][]{Muller2005} and the JPL catalog \citep{Pickett1998} with the help of the Weeds package, which is a part of the 
GILDAS/CLASS program \citep{Maret2011}. These two databases provide the rest frequencies, the upper-level energies, $E_{\rm up}$, and the Einstein coefficients, $A_{\rm ul}$, for many molecular transitions. COMs have a plethora of lines in the covered bands due to the complexity of their energy level distributions. 
To identify the origin of the observed spectral lines, we followed the same procedure as described in \citet{paperI}.
In short, using the spectroscopic databases, we first identify the molecules with the brightest emission lines, which are typically simple. 
To identify and characterise emission from COMs, on which we focus in this work, we use an iterative procedure relying on synthetic local thermodynamic equilibrium models (LTE) and rotational diagrams.

As a first step we create a synthetic spectrum with a first guess of parameters, and extract the fitted properties of the molecular lines (see Sect.\,\ref{sec:full_line_prof}) coinciding with the  transition frequencies of the LTE model. The integrated intensities are used to build rotational diagrams (see  Sect.\,\ref{sec:rotational}) and, based on them, the model parameters for the synthetic spectra are improved. 

To obtain the final results, we visually checked whether all the expected lines were detected with intensities consistent with the synthetic spectrum. This procedure allowed us to firmly identify between 5 and 11 COMs (and three $^{13}$Cs substituted isotopologues for the most line rich sources) in  the sources of the sample, which are listed in Table\,\ref{tab:COM_detection}.

\begin{table*}
    \centering
    \caption{COMs detected towards the sample.}
    \label{tab:COM_detection}
    \begin{tabular}{l l c c c c c c}
    \hline
    \hline
    \multicolumn{2}{c}{Molecules} & G320.23 & G328.25 & G333.46 & G335.58 & G335.78 & G343.75 \\
    \hline
     CH$_3$OH, $v_t=0$  & methanol  & x & x & x & x & x & x \\
     $^{13}$CH$_3$OH & methanol & & x& x & x & x & x \\
     CH$_3\,^{18}$OH & methanol & &  &   & x &   &   \\
     CH$_3$OH, $v_t=1$  & methanol   &   & x &   & x &   & x \\
     CH$_3$OCHO  & methyl formate      & x & x & x & x & x & x \\
     CH$_3$OCH$_3$ & dimethyl ether  & x & x & x & x & x & x \\
     CH$_3$CHO & acetaldehyde & x & x & x & x & x & x \\
     C$_2$H$_5$OH & ethanol & & & x& x & &\\
     CH$_3$COCH$_3$ & acetone & & & & x & &\\
     HC(O)NH$_2$ & formamide & & x& x& x& & x\\
     CH$_3$CN & methyl cyanide & x & x & x & x & x & x \\
     $^{13}$CH$_3$CN & methyl cyanide & & & & x & & \\
     C$_2$H$_5$CN & ethyl cyanide & & x&  & x& & x\\
     C$_2$H$_3$CN & vinyl cyanide & & x&  & x& &  \\
     CH$_3$SH & methyl mercaptan & & x& x& x& & x\\
    \hline
    \end{tabular}
\end{table*}

\begin{table*}
\caption{Ranges of upper level energies of the COM lines detected in the survey.}\label{tab:COM_Eup}
{\vspace{0.1cm}}
\label{molecules}
\footnotesize
\centering
\begin{tabular}{l c c c c c c c c}  
\hline 
\hline
Molecule & $E_{\rm up}$\tablefootmark{a}  & Database & G320.23 & G328.25 & G333.46 & G335.58 & G335.78 & G343.75 \\
& min-max [K] &   & min-max [K] & min-max [K] & min-max [K] & min-max [K] & min-max [K] & min-max [K] \\
\hline 
CH$_3$OH, $\varv_{\rm t}=0$    &  17 -- 295     &   JPL & 17--295 & 17–-260 & 17 --295 & 17 -- 295 & 17 -- 295 & 17 -- 295\\
CH$_3$OH, $\varv_{\rm t}=1$    &  339 -- 1971   &   JPL & -- & 339–-611 & -- & 339 -- 546 & --&  348 -- 462 \\
$^{13}$CH$_3$OH                &  17 -- 844     &   JPL & -- & 17--50 & 35--69 &--& --& \\
CH$_3\,^{18}$OH                &  17 -- 844     &   JPL & -- & - & -- &16--115 &--& -- \\
CH$_3$OCH$_3$                  &  32 -- 3106    &  CDMS & 32 -- 167 & 15-475 & 33--143 & 32 -- 432 & 32 -- 167 & 32 -- 167 \\
CH$_3$OCHO                     &  79 -- 757     &   JPL & 91 -- 199 & 79--318 & 79--199& 79 -- 432 & 91 -- 199 & 79 -- 293\\
CH$_3$CHO                      &  26 -- 1063    &  CDMS & 43 -- 51 & 40--106 & 40--96 & 40 -- 129 & 41 -- 94 & 41 -- 106 \\
HC(O)NH$_2$                    &  36 -- 940     &  CDMS & -- & 36--380 & 85--91 & 36 -- 439 & -- & 36 -- 174\\
CH$_3$COCH$_3$                 &  34 -- 2327    & JPL & --  & -- & --  & 34 -- 288 & --& --\\  
C$_2$H$_5$OH                   &  29 -- 2889    & CDMS &  --  &-- & 33 -- 110 & 29 -- 296 & --& --\\
CH$_3$CN                       &  40 -- 2747    &  CDMS & 40 -- 297 & 40-745& 40 --392 & 40 -- 712 & 40 -- 626 & 40 -- 549 \\
C$_2$H$_5$CN                   &  34 -- 3255    &  CDMS &  -- & 43-425 &-- & 78 -- 519 & 86--103 & 76 -- 554\\
C$_2$H$_3$CN                   &  40 -- 2695    &  CDMS & --  & 169--350& -- & 74 -- 469 & -- & -- \\
CH$_3$SH                       &  55 -- 1317    &  CDMS & --  & 55--99 & 59--73 & 55 -- 96 & --& 56 -- 107 \\
\hline 
\end{tabular}
\tablefoot{ 
\tablefoottext{a}{The range of upper-level energies for the COMs is given in temperature units for lines with $A_{\rm ul}$ above 10$^{-4}$\,s$^{-1}$}.}
\end{table*}

\subsection{Line profiles}
\label{sec:full_line_prof}

\begin{figure*}[!ht]
    \centering
    \includegraphics[trim=10mm 15mm 149mm 197mm, clip, height=10.2cm,keepaspectratio]{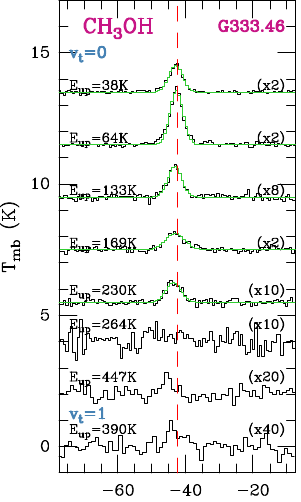}
    \includegraphics[trim=15mm 15mm 149mm 197mm, clip,height=10.2cm,keepaspectratio]{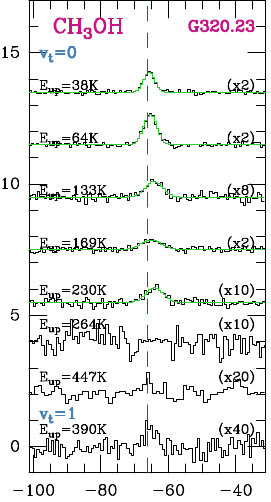}
    \includegraphics[trim=15mm 15mm 149mm 197mm, clip,height=10.2cm,keepaspectratio]{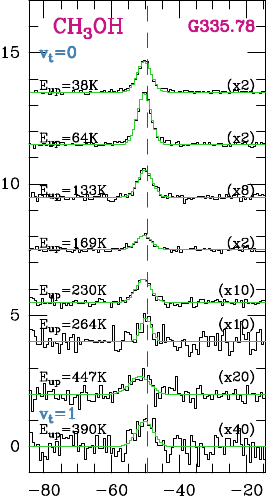}
    \includegraphics[trim=10mm 9mm 149mm 197mm, clip,height=10.9cm,keepaspectratio]{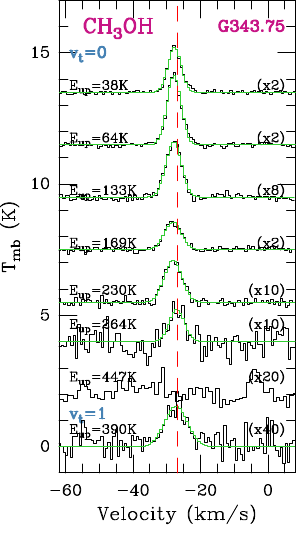}
    \includegraphics[trim=15mm 9mm 149mm 197mm, clip,height=10.9cm,keepaspectratio]{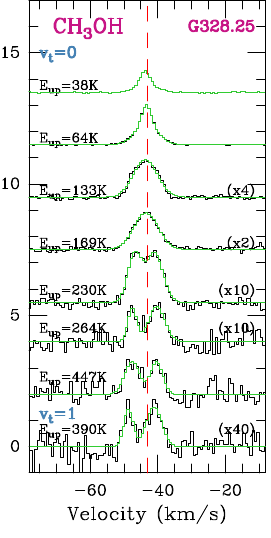}
    \includegraphics[trim=15mm 9mm 149mm 197mm, clip,height=10.9cm,keepaspectratio]{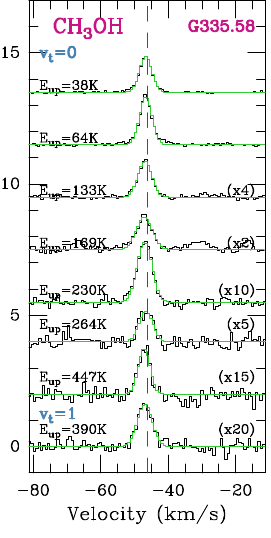}
    \caption{Line profile of methanol lines over a range of upper level energies towards the six objects. 
    The x-axis shows the LSR velocity range with respect to the rest frequency of each transition. The source \vlsr\ (Table\,\ref{tab:source_coord}) marked by a vertical red dashed line. The green line shows Gaussian fits to the individual spectral lines.}
    \label{fig:CH3OH_lineprofile}
\end{figure*}

Our spectral resolution of 0.7\,\kms\ is sufficient to well resolve the lines. The spectra exhibit a variety of line profiles showing broad and narrow velocity components,
and the line profiles may vary as a function of upper level energy, which is demonstrated in Fig.\,\ref{fig:CH3OH_lineprofile} for CH$_3$OH. 
We see that towards G333.46, G320.23, and G335.78 only transitions with low upper level energies are detected with sufficient signal-to-noise ratio, while towards G343.75, G328.25, and G335.58 transitions with higher upper level energies are detected.  Most of these transitions can be well fitted with a single Gaussian component.

Furthermore, we observe an intrinsic difference in the line profiles for the G328.25 source that is discussed in detail in \citet{paperI}. While the rest of the sources in our sample exhibits rather narrow line profiles with widths of typically 
2--6\,\kms, toward G328.25 the highest upper level energy transitions of CH$_3$OH $\varv_{\rm t}=0$, CH$_3$OH $\varv_{\rm t}=1$, CH$_3$OCH$_3$, CH$_3$OCHO, and HC(O)NH$_2$ exhibit double peaked line profiles with peak centroids that are offset from the source LSR velocity, \vlsr. This is consistent with kinematic signatures tracing accretion shocks, as first  described in \citet{Csengeri2018} and discussed in detail subsequently in \citet{Csengeri2019}, based on ALMA observations of this source. A complete view on the molecular composition and  kinematics of the material in the accretion shocks was described in \citet{paperI}. 

For a complete analysis of the six sources, we use here the results for G328.25 from \citet{paperI} that includes the complete line analysis for that source.
In contrast to the complex line profiles exhibited by several COMs towards G328.25, a single-component Gaussian fit was typically performed for the rest of the sources. The integrated intensities of each transition were used to build rotational diagrams (Sect.\,\ref{sec:rotational}). 

\subsection{Rotational diagrams}
\label{sec:rotational}
Using the area integrated over the spectral lines, we constructed rotational diagrams for each species. The large frequency coverage of the survey gives us access to a large number of rotational transitions for each molecule. Assuming LTE conditions, this allows us to estimate the total column density, $N$, and the rotational temperature, $T_{\rm rot}$. 
We used rotational diagrams to obtain a first  estimate of $N$ and $T_{\rm rot}$, which then served as an input for a more detailed LTE modelling using Weeds following an approach similar to that described in \citet{paperI}.  
To build these rotational diagrams, we use the integrated intensity, $W$, of each unblended line at a frequency, $\nu$, from the Gaussian fitting that uses one- to two-component fit.  Following \citet{Goldsmith1999}, 
for each optically thin transition of a molecule, the population of the upper rotational level $u$ is given by 
\begin{equation}
    N_u^{thin}=\frac{8\pi k\nu^2W}{hc^3A_{\rm ul}B_{\rm dil}},
\end{equation} 
where $A_{\rm ul}$ is the Einstein coefficient for spontaneous emission, $c$ is the speed of light, $h$ is the Planck constant, $k$ is the Boltzmann constant, and $B_{\rm dil}$ is the beam dilution factor that depends on the size of the emitting region, $\theta_{\rm S}$, and the FWHM beam size, $\theta_{\rm beam}$.
Since we assumed that most of the emission is unresolved within the beam, we 
take into account the beam dilution factor defined as $B_{\rm dil}=\frac{\theta_{\rm S}^2}{\theta_{\rm S}^2+\theta_{\rm beam}^2}$, where a Gaussian shape is assumed both for the source and the beam. Even for a fixed source size, the beam dilution factor shows a significant variation due to the large frequency coverage of our dataset. We estimate the size of the emitting region  based on the rotational diagrams and the LTE modelling using Weeds in an iterative process (see Sect. \ref{sec:LTE}).

We find that from the COMs discussed here, only methanol has a few transitions that are optically thick with $\tau>0.6$. We built our rotational diagrams considering only the optically thin transitions, and therefore, do not discuss optical depth effects here. 

The total column density of each molecule, $N_{\rm tot}$, is given by  
\begin{equation}
    ln\Big(\frac{N_{\rm u}}{g_{\rm u}}\Big)=ln\Big(\frac{N_{\rm tot}}{Z}\Big)-\frac{E_{\rm up}}{kT_{\rm rot}},
\end{equation}
where and $g_{\rm u}$ and $E_{\rm up}$ are, respectively, the statistical weight and the the upper-level energy of each transition; and $Z$ is the rotational partition function at rotational temperature $T_{\rm rot}$ (see, for 
example \citet{Goldsmith1999}). The rotational diagrams are then displayed as $ln\Big(\frac{N_{\rm u}}{g_{\rm u}}\Big)$ versus $E_{\rm up}$, and by fitting a straight line 
one derives $N_{\rm tot}$ from the line's y-intersect and $T_{\rm rot}$ from its slope.
The partition function was interpolated from values listed in the tables given in the CDMS and JPL databases. We performed a linear least-square fit to the rotational diagrams, and from these parameters, we computed the rotational temperature and the total column density of each temperature component of the molecule. We estimated uncertainties on the column density and the rotational temperature using a Monte Carlo approach. We varied each measurement in the rotational diagram within its error bars assuming a uniform distribution and repeated the least-square fit. The final results correspond to the average of the estimated column densities and rotational temperatures, while their uncertainties correspond to the standard deviation of the values determined with our Monte Carlo method. 

\begin{figure}
    \centering
    \includegraphics[width=1\linewidth]{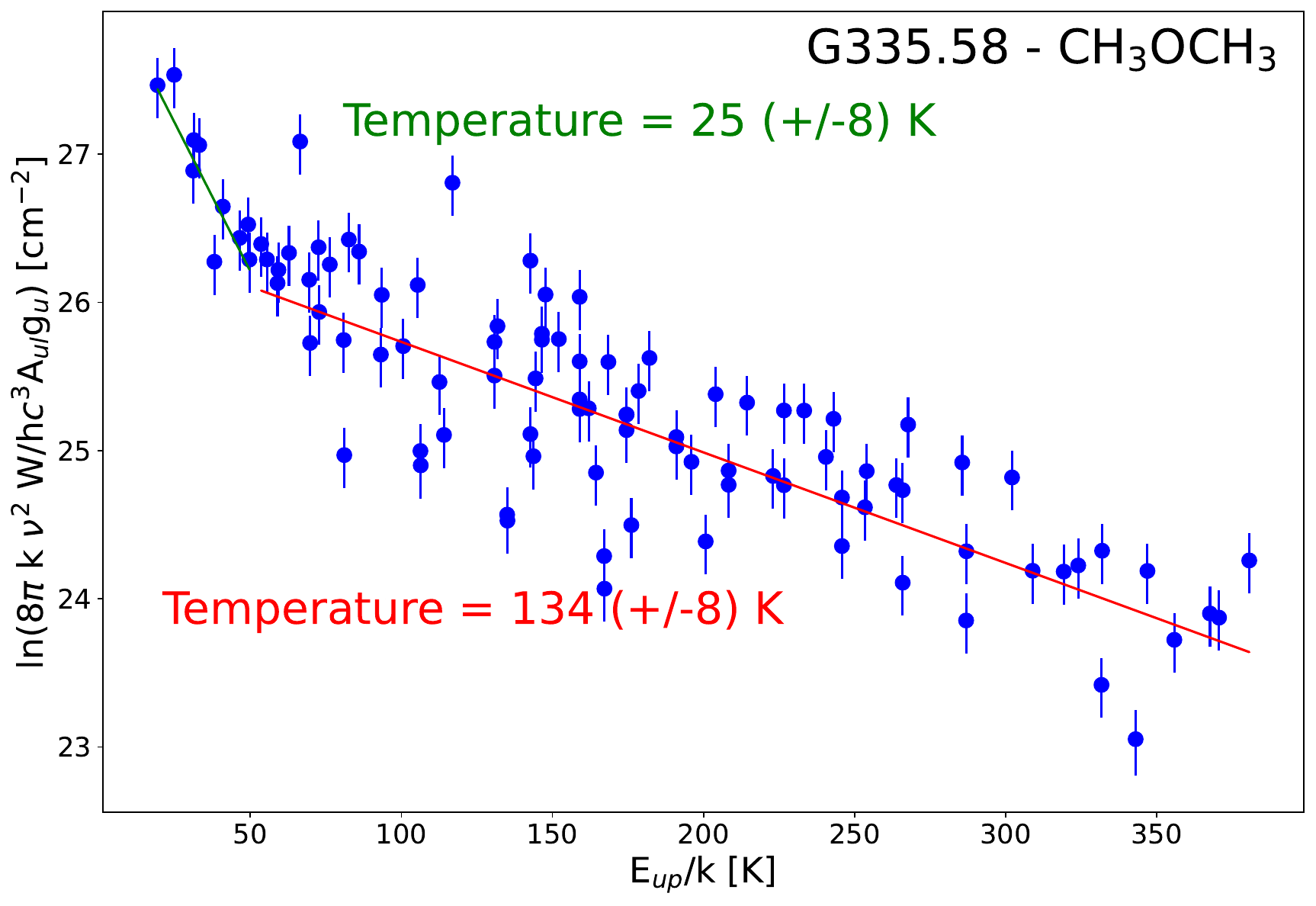}
    \includegraphics[width=1\linewidth]{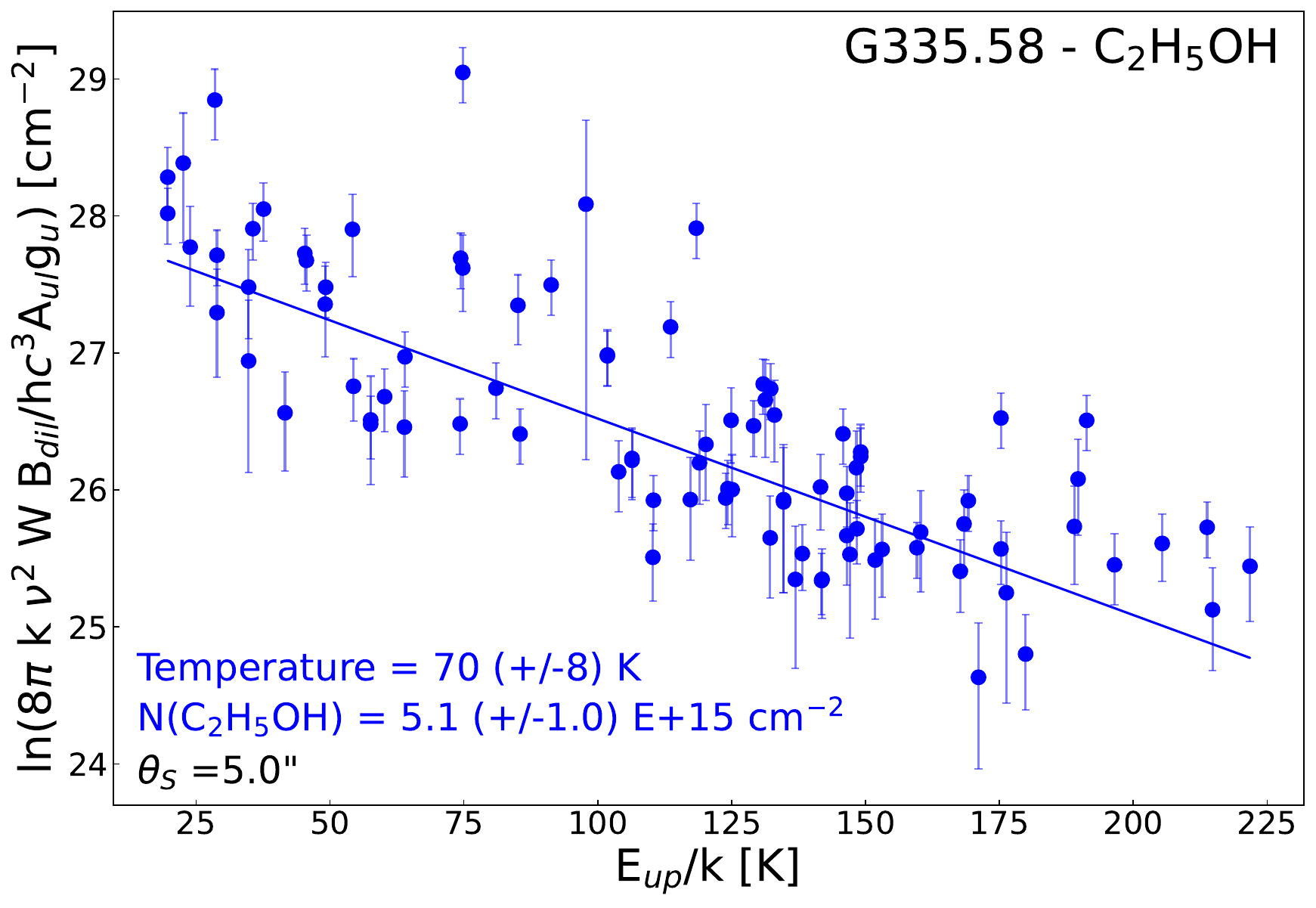}
    \caption{Rotational diagram of dimethyl ether (CH$_3$OCH$_3$) and ethanol (C$_2$H$_5$OH) towards G333.58. The errorbars represent a 1$\sigma$ statistical uncertainty plus a 20\% calibration error. We used the Monte Carlo method and assumed a uniform distribution of the error for each data point to calculate the uncertainties on the rotational temperature and the column density.}
    \label{fig:rotdiag}
\end{figure}

Where the distribution of the measurements in the rotational diagrams is consistent with a single temperature component (Fig.\,\ref{fig:rotdiag}, lower panel), we performed the procedure explained above. However, some rotational diagrams suggest two temperature components (Fig.\,\ref{fig:rotdiag}, upper panel). 
In these cases we performed two linear fits with different slopes, 
assuming different emission sizes, and estimated different column densities and rotational temperatures.
These two temperature components correspond to the warm and cold components of the envelope. Such a two temperature component fit was typically necessary for methanol for all the objects and also for CH$_3$OCH$_3$ towards G328.25, and G335.58 for instance (see App.\,\ref{app:rot_diag} for the rotational diagrams).
The column densities of the cold component typically contribute sufficiently little to the overall column density so that we ignored their contributions to the fit of the higher temperature component.
This approach gives reasonable first guess estimates for the temperature and the column density, however, our LTE modelling (see Sect.\,\ref{sec:LTE}) considers the cold and warm components together.

\subsection{LTE parameters}\label{sec:LTE}
In order to fit the observed spectrum, we performed a LTE modelling with Weeds \citep{Maret2011}. The input parameters for each molecule were the full-width at half maximum (FWHM) size of the emission ($\theta_S$ [$''$]), the excitation temperature ($T_{\rm ex}$ [K]), the column density ($N_{\rm tot}$ [cm$^{-2}$]), the FWHM of the spectral line ($\Delta v$ [\kms]), and the velocity offset ($v_{\rm off}$ [\kms]) with respect to the systemic velocity of the source. For the line width and the velocity offset, we used the average of the line widths obtained from their individual Gaussian fits. In order to identify all the transitions of a given molecule, we first initiated a model with a first guess of the parameters for $T_{\rm ex}$ and $N_{\rm tot}$, which allowed us to identify its unblended transitions of COMs, and derived a rotational diagram (see Sect.\,\ref{sec:rotational}). We used then the $N_{\rm tot}$ and $T_{\rm ex}$ from the rotational diagrams to obtain the final LTE model and corrected the list of the detected transitions if necessary in an iterative fashion. 

\begin{figure*}
    \centering
    \includegraphics[trim=0cm 13.2cm 2cm 2cm,clip,width=1\linewidth]{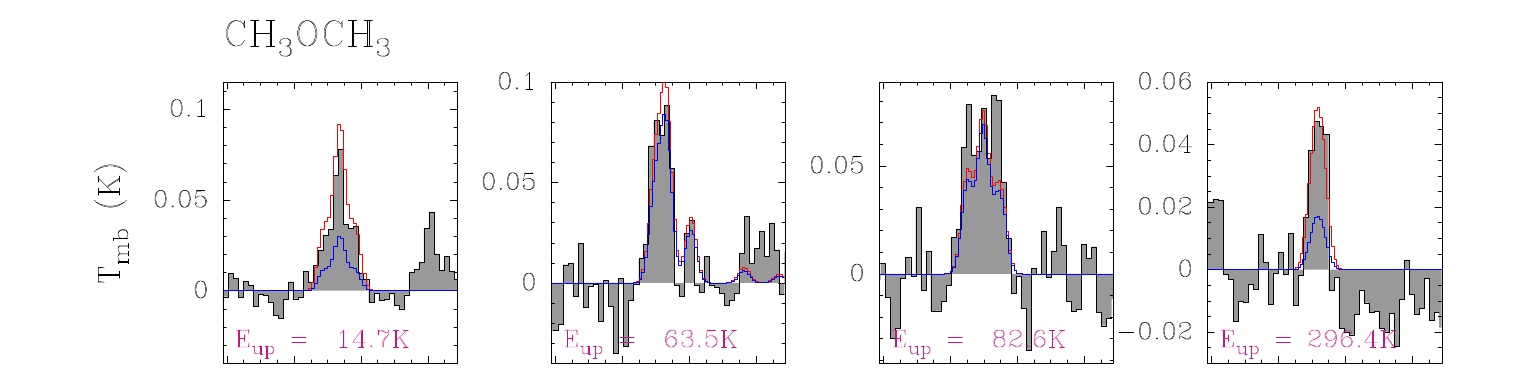}

    \includegraphics[trim=0cm 11cm 2cm 2cm, clip, width=1\linewidth]{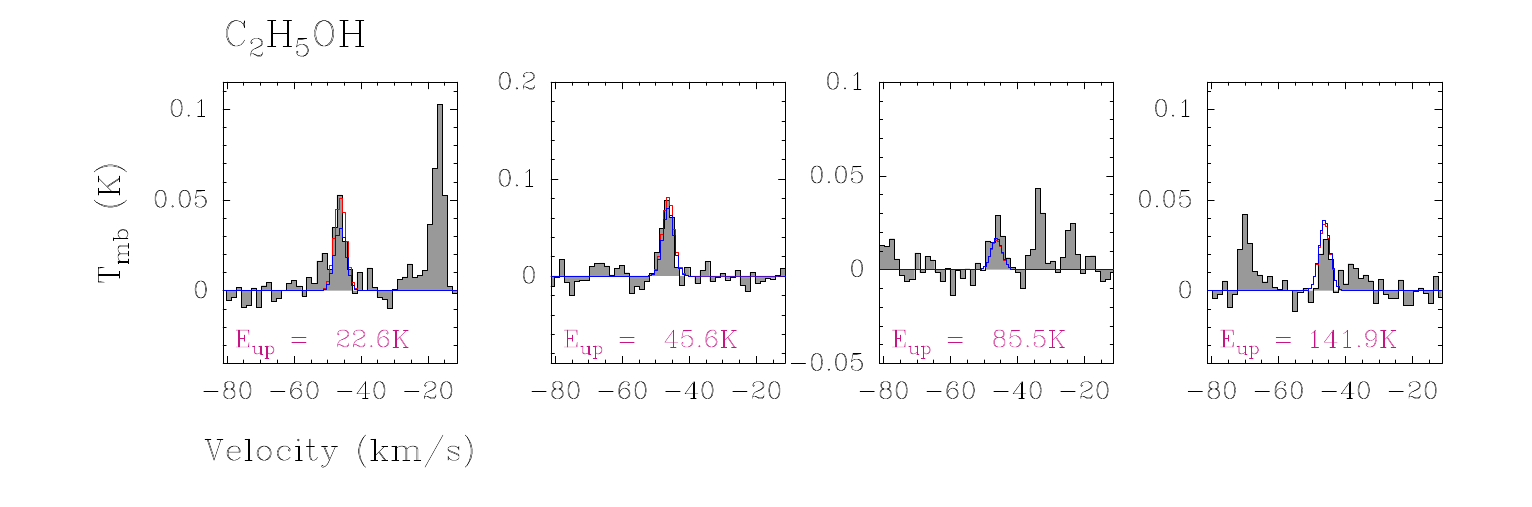}
    \caption{CH$_3$OCH$_3$ (top panel) and C$_2$H$_5$OH (bottom panel) lines (black spectrum) towards G335.58. The blue and red spectra represent the one and two temperature component models, respectively.  We indicate the  upper level energies in the bottom of each panel.  C$_2$H$_5$OH is well fitted with only one temperature component while CH$_3$OCH$_3$ requires a second, higher temperature component in order to fit all lines simultaneously. }
    \label{fig:Mod1vs2comp}
\end{figure*}

We found that some models required two temperature components, a single component could not give a satisfactory fit to the lines. We show an example for this using CH$_3$OCH$_3$ towards G335.58 (Fig.\,\ref{fig:Mod1vs2comp}), where a single temperature component with 70\,K does not fit well the low and high upper level energy transitions and a two temperature component model (red line) is needed to fit the broad range of upper level energies.
On the other hand, we found examples, such as C$_2$H$_5$OH, where a single component at a luke-warm temperature (60\,K) already gives a good fit to the spectra. We tested whether a two temperature component fit with sufficiently distinct temperatures ($\Delta T>100$~K) between the components could also fit the spectra (see Sect.\,\ref{sec:dichotomy_NO} for a discussion). Keeping all the six parameters free ($N$, $T_{\rm ex}$, $\theta_{\rm S}$ for each temperature component), we can indeed obtain satisfactory results, due to the large number of free parameters. These results are, however, degenerate. Therefore, we decided to use the simplest, single temperature component model in these cases. We show an example towards G335.58  (Fig.\,\ref{fig:Mod1vs2comp}) to demonstrate that a single temperature component model gives a satisfactory fit to the spectra for C$_2$H$_5$OH, while for CH$_3$OCH$_3$ the two component model fit gives a noticeably better result.

We list the resulting LTE model parameters for each molecule and source in Table\,\ref{tab:COM_LTE}. There are molecules where only few transitions are unblended, and in some cases we could not obtain a robust estimate of the source size using our iterative method. These exceptions and the adopted assumptions  to obtain $N$ and $T_{\rm ex}$ are discussed for each source in App.\,\ref{sec:app_excpetions}.

\longtab{
\begin{longtable}{l c c c c c c c}
\caption{LTE parameters of COMs towards all six sources of the sample.} \\
\label{tab:COM_LTE} \\
\hline 
\hline
Molecule            & $N$ &   $T$ &  $\theta_S$ & $v_{\rm off}$ & $\Delta_V$ &  N$_X$/N$_{\rm H_2}$ &    Comments\\
 & [cm$^{-2}$] &   [K] &  [$''$] & [\kms] & width[\kms] &   &    \\
\hline
G320.23 \\
\hline
CH$_3$CHO            &     3.2$\times$10$^{13}$ &     18 &      25.0 &      0.3$\pm$      0.3  &      3.4 $\pm$      0.8 & 2.0$\times$10$^{-10}$ &\\
CH$_3$CN             &     8.0$\times$10$^{12}$ &     26 &      30.0 &      0.8$\pm$      0.4  &      5.0 $\pm$      0.8 & 6.0$\times$10$^{-11}$ &\\
$^\dag$C$_2$H$_5$CN         &  $\leq$ 1.0$\times$10$^{13}$ &     26 &      30.0 &      0.8$\pm$      0.0  &      5.0 $\pm$      1.4 & $\leq$ 7.5$\times$10$^{-11}$ & (1) \\
C$_2$H$_3$CN         &  $\leq$ 8.0$\times$10$^{12}$ &     26 &      30.0 &      0.8$\pm$      0.0  &      5.0 $\pm$      0.0 & $\leq$ 6.0$\times$10$^{-11}$ & (1) \\
CH$_3$OH,$\varv_t=0$     &     4.1$\times$10$^{14}$ &     26 &      31.0 &      1.0$\pm$      0.2  &      5.1 $\pm$      0.5 & 3.2$\times$10$^{-9}$ &\\
HC(O)NH$_2$          &  $\leq$ 3.2$\times$10$^{12}$ &     40 &      15.0 &      0.3$\pm$      0.0  &      3.4 $\pm$      0.0 & $\leq$ 1.2$\times$10$^{-11}$ &\\
CH$_3$SH             &  $\leq$ 3.2$\times$10$^{13}$ &     40 &      15.0 &      0.3$\pm$      0.0  &      3.4 $\pm$      0.0 & $\leq$ 1.2$\times$10$^{-10}$ &\\
C$_2$H$_5$OH         &  $\leq$ 1.3$\times$10$^{14}$ &     40 &      10.0 &     -0.8$\pm$      0.0  &      3.2 $\pm$      0.0 & $\leq$ 3.3$\times$10$^{-10}$ & (1) \\
CH$_3$OCHO           &     2.0$\times$10$^{15}$ &     45 &       5.0 &      1.4$\pm$      1.1  &      4.0 $\pm$      0.8 & 2.5$\times$10$^{-9}$ &\\
CH$_3$OCH$_3$        &     3.0$\times$10$^{14}$ &     60 &      20.0 &      1.4$\pm$      0.3  &      5.0 $\pm$      1.1 & 1.5$\times$10$^{-9}$ &\\
CH$_3$OH,$\varv_t=0$     &     4.1$\times$10$^{15}$ &    123 &       6.0 &      1.3$\pm$      0.5  &      6.4 $\pm$      1.0 & 6.2$\times$10$^{-9}$ &\\
CH$_3$CN             &     2.5$\times$10$^{15}$ &    207 &       1.0 &      1.5$\pm$      0.5  &      5.7 $\pm$      1.4 & 6.3$\times$10$^{-10}$ &\\
$^\dag$C$_2$H$_5$CN         &  $\leq$ 9.1$\times$10$^{15}$ &    207 &       1.0 &      1.5$\pm$      0.0  &      5.7 $\pm$      0.0 & $\leq$ 2.3$\times$10$^{-9}$ & (1) \\
C$_2$H$_3$CN         &  $\leq$ 1.0$\times$10$^{16}$ &    207 &       1.0 &      1.5$\pm$      0.0  &      5.7 $\pm$      0.0 & $\leq$ 2.5$\times$10$^{-9}$ & (1) \\
\hline
G333.46 \\
\hline
CH$_3$SH             &     7.5$\times$10$^{13}$ &     18 &      20.0 &     -0.5$\pm$      0.5  &      4.0 $\pm$      1.8 & 3.5$\times$10$^{-10}$ &\\
CH$_3$OH,$\varv_t=0$     &     5.0$\times$10$^{14}$ &     23 &      40.0 &     -0.6$\pm$      0.2  &      4.7 $\pm$      0.6 & 4.7$\times$10$^{-9}$ &\\
CH$_3$CHO            &     6.5$\times$10$^{13}$ &     26 &      20.0 &     -0.5$\pm$      0.3  &      4.4 $\pm$      1.2 & 3.0$\times$10$^{-10}$ &\\
CH$_3$CN             &     1.2$\times$10$^{13}$ &     33 &      35.0 &     -0.7$\pm$      0.3  &      4.7 $\pm$      1.0 & 9.8$\times$10$^{-11}$ &\\
HC(O)NH$_2$          &     7.0$\times$10$^{12}$ &     45 &      15.0 &     -0.7$\pm$      0.4  &      3.5 $\pm$      0.6 & 2.5$\times$10$^{-11}$ & (3)\\
CH$_3$OCH$_3$        &     2.7$\times$10$^{14}$ &     57 &      20.0 &     -0.5$\pm$      0.6  &      4.4 $\pm$      1.3 & 1.3$\times$10$^{-9}$ &\\
$^\dag$C$_2$H$_5$CN         &  $\leq$ 9.3$\times$10$^{12}$ &     57 &      20.0 &     -0.5$\pm$      0.0  &      4.4 $\pm$      0.0 & $\leq$ 4.4$\times$10$^{-11}$ & (2)\\
C$_2$H$_3$CN         &  $\leq$ 9.2$\times$10$^{12}$ &     57 &      20.0 &     -0.5$\pm$      0.0  &      4.4 $\pm$      0.0 & $\leq$ 4.3$\times$10$^{-11}$ & (1)\\
CH$_3$OCHO           &     2.7$\times$10$^{14}$ &     57 &      20.0 &     -0.5$\pm$      0.5  &      4.0 $\pm$      1.8 & 1.3$\times$10$^{-9}$ &\\
CH$_3$COCH$_3$       &  $\leq$ 3.0$\times$10$^{15}$ &     80 &       2.0 &     -0.7$\pm$      0.0  &      5.2 $\pm$      0.0 & $\leq$ 1.4$\times$10$^{-9}$ &\\
C$_2$H$_5$OH         &     2.5$\times$10$^{15}$ &     80 &       3.0 &     -0.8$\pm$      0.8  &      4.2 $\pm$      1.3 & 1.8$\times$10$^{-9}$ &\\
CH$_3$OH,$\varv_t=0$     &     2.0$\times$10$^{16}$ &    121 &       3.0 &     -0.8$\pm$      0.4  &      4.7 $\pm$      1.1 & 1.4$\times$10$^{-8}$ &\\
$^\dag$C$_2$H$_5$CN         &  $\leq$ 2.3$\times$10$^{16}$ &    185 &       1.0 &     -0.7$\pm$      0.0  &      5.2 $\pm$      0.0 & $\leq$ 5.4$\times$10$^{-9}$ & (2) \\
C$_2$H$_3$CN         &  $\leq$ 1.0$\times$10$^{16}$ &    185 &       1.0 &     -0.7$\pm$      0.0  &      5.2 $\pm$      0.0 & $\leq$ 2.3$\times$10$^{-9}$ & (1)\\
CH$_3$CN             &     3.0$\times$10$^{15}$ &    185 &       1.0 &     -0.7$\pm$      0.6  &      5.2 $\pm$      1.5 & 7.0$\times$10$^{-10}$ &\\
\hline
G335.58 \\
\hline
CH$_3$OCH$_3$        &     3.0$\times$10$^{14}$ &     22 &      34.0 &     -0.2$\pm$      0.7  &      4.0 $\pm$      0.7 & 1.7$\times$10$^{-9}$ &\\
HC(O)NH$_2$          &     9.0$\times$10$^{12}$ &     23 &      17.0 &     -0.6$\pm$      0.3  &      4.6 $\pm$      1.3 & 2.5$\times$10$^{-11}$ &\\
CH$_3$CN             &     2.0$\times$10$^{13}$ &     25 &      39.0 &     -0.5$\pm$      0.2  &      5.0 $\pm$      0.5 & 1.3$\times$10$^{-10}$ &\\
CH$_3$OCHO           &     6.0$\times$10$^{14}$ &     28 &      25.0 &     -0.3$\pm$      0.3  &      3.9 $\pm$      0.9 & 2.5$\times$10$^{-9}$ &\\
CH$_3$OH,$\varv_t=0$     &     1.0$\times$10$^{15}$ &     29 &      40.0 &     -0.5$\pm$      0.1  &      4.5 $\pm$      0.4 & 6.5$\times$10$^{-9}$ &\\
CH$_3$CHO            &     8.7$\times$10$^{13}$ &     32 &      25.0 &     -0.4$\pm$      0.4  &      4.1 $\pm$      0.9 & 3.6$\times$10$^{-10}$ &\\
CH$_3$SH             &     5.0$\times$10$^{15}$ &     60 &       2.0 &     -0.6$\pm$      0.8  &      4.5 $\pm$      0.6 & 1.6$\times$10$^{-9}$ &\\
$^\dag$C$_2$H$_5$CN         &     4.3$\times$10$^{14}$ &     70 &       7.0 &     -0.7$\pm$      0.3  &      5.0 $\pm$      0.8 & 4.9$\times$10$^{-10}$ &\\
C$_2$H$_5$OH         &     1.0$\times$10$^{16}$ &     80 &       3.0 &     -0.0$\pm$      0.6  &      3.9 $\pm$      1.4 & 4.9$\times$10$^{-9}$ &\\
CH$_3$COCH$_3$       &     4.2$\times$10$^{16}$ &     85 &       1.0 &     -0.0$\pm$      1.3  &      4.3 $\pm$      1.1 & 6.9$\times$10$^{-9}$ &\\
HC(O)NH$_2$          &     1.1$\times$10$^{14}$ &     90 &       5.0 &     -0.7$\pm$      1.0  &      4.7 $\pm$      1.7 & 9.0$\times$10$^{-11}$ &\\
CH$_3$OCHO           &     3.0$\times$10$^{16}$ &    100 &       3.0 &     -0.3$\pm$      0.3  &      3.9 $\pm$      0.8 & 1.5$\times$10$^{-8}$ &\\
CH$_3$OCH$_3$        &     1.0$\times$10$^{17}$ &    120 &       2.0 &     -0.2$\pm$      0.7  &      4.0 $\pm$      0.7 & 3.3$\times$10$^{-8}$ &\\
C$_2$H$_3$CN         &     1.5$\times$10$^{16}$ &    148 &       1.0 &     -0.8$\pm$      0.6  &      4.7 $\pm$      1.5 & 2.5$\times$10$^{-9}$ & (1) \\
CH$_3$OH,$\varv_t=0$     &     2.5$\times$10$^{16}$ &    150 &       5.0 &     -0.8$\pm$      0.3  &      5.3 $\pm$      1.1 & 2.0$\times$10$^{-8}$ &\\
CH$_3$OH,$\varv_t=1$           &     6.0$\times$10$^{16}$ &    200 &       5.0 &     -0.5$\pm$      0.2  &      4.4 $\pm$      0.6 & 4.9$\times$10$^{-8}$ &\\
$^\dag$C$_2$H$_5$CN         &     3.6$\times$10$^{16}$ &    223 &       1.0 &     -0.7$\pm$      0.4  &      4.7 $\pm$      1.1 & 5.9$\times$10$^{-9}$ &\\
CH$_3$CN             &     3.6$\times$10$^{15}$ &    260 &       2.0 &     -0.5$\pm$      0.6  &      5.2 $\pm$      1.0 & 1.2$\times$10$^{-9}$ &\\
\hline
G335.78 \\
\hline
CH$_3$OH,$\varv_t=0$     &     5.0$\times$10$^{14}$ &     25 &      39.0 &     -1.0$\pm$      0.2  &      5.1 $\pm$      0.5 & 4.1$\times$10$^{-9}$ &\\
CH$_3$CN             &     1.2$\times$10$^{13}$ &     26 &      39.0 &     -0.8$\pm$      0.2  &      5.0 $\pm$      0.9 & 9.8$\times$10$^{-11}$ &\\
$^\dag$C$_2$H$_5$CN         &  $\leq$ 1.2$\times$10$^{13}$ &     26 &      39.0 &     -0.8$\pm$      0.0  &      5.0 $\pm$      0.0 & $\leq$ 9.8$\times$10$^{-11}$ & (2)\\
C$_2$H$_3$CN         &  $\leq$ 1.2$\times$10$^{13}$ &     26 &      39.0 &     -0.8$\pm$      0.0  &      5.0 $\pm$      0.0 & $\leq$ 9.8$\times$10$^{-11}$ & (2) \\
CH$_3$CHO            &     5.5$\times$10$^{13}$ &     27 &      23.0 &     -0.6$\pm$      0.5  &      4.4 $\pm$      1.1 & 2.6$\times$10$^{-10}$ &\\

CH$_3$OCH$_3$        &     3.5$\times$10$^{14}$ &     46 &      15.0 &     -0.9$\pm$      1.0  &      5.5 $\pm$      1.6 & 1.1$\times$10$^{-9}$ &\\
HC(O)NH$_2$          &  $\leq$ 1.0$\times$10$^{13}$ &     46 &      15.0 &     -0.9$\pm$      0.0  &      5.5 $\pm$      1.4 & $\leq$ 3.1$\times$10$^{-11}$ & (2)\\
CH$_3$SH             &  $\leq$ 1.0$\times$10$^{14}$ &     46 &      15.0 &     -0.9$\pm$      0.0  &      5.5 $\pm$      1.4 & $\leq$ 3.1$\times$10$^{-10}$ & (2)\\
CH$_3$OH,$\varv_t=0$     &     1.8$\times$10$^{16}$ &    130 &       4.0 &     -1.3$\pm$      0.5  &      6.4 $\pm$      1.0 & 1.5$\times$10$^{-8}$ &\\
CH$_3$CN             &     4.2$\times$10$^{15}$ &    220 &       1.0 &     -0.7$\pm$      0.7  &      5.7 $\pm$      1.8 & 8.8$\times$10$^{-10}$ &\\
$^\dag$C$_2$H$_5$CN         &  $\leq$ 1.8$\times$10$^{16}$ &    220 &       1.0 &     -0.7$\pm$      0.0  &      5.7 $\pm$      0.0 & $\leq$ 3.8$\times$10$^{-9}$ & (2)\\
C$_2$H$_3$CN         &  $\leq$ 8.0$\times$10$^{15}$ &    220 &       1.0 &     -0.7$\pm$      0.0  &      5.7 $\pm$      0.0 & $\leq$ 1.7$\times$10$^{-9}$ & (2)\\
\hline
G343.75 \\
\hline
CH$_3$OH,$\varv_t=0$     &     4.5$\times$10$^{14}$ &     31 &      39.0 &     -1.0$\pm$      0.2  &      5.1 $\pm$      0.5 & 2.9$\times$10$^{-9}$ &\\
CH$_3$CHO            &     5.0$\times$10$^{13}$ &     34 &      29.0 &     -0.6$\pm$      0.6  &      5.4 $\pm$      1.5 & 2.4$\times$10$^{-10}$ &\\
CH$_3$CN             &     1.5$\times$10$^{13}$ &     35 &      35.0 &     -0.8$\pm$      1.5  &      5.3 $\pm$      0.5 & 8.8$\times$10$^{-11}$ &\\
CH$_3$OCHO           &     5.5$\times$10$^{14}$ &     42 &      17.0 &     -0.4$\pm$      0.8  &      6.7 $\pm$      1.4 & 1.6$\times$10$^{-9}$ &\\
HC(O)NH$_2$          &     1.0$\times$10$^{14}$ &     52 &       6.0 &     -1.0$\pm$      0.9  &      6.0 $\pm$      2.0 & 1.0$\times$10$^{-10}$ &\\
CH$_3$SH             &     1.5$\times$10$^{14}$ &     70 &      13.0 &     -0.2$\pm$      0.4  &      5.8 $\pm$      1.9 & 3.3$\times$10$^{-10}$ &\\
CH$_3$OCH$_3$        &     9.0$\times$10$^{15}$ &     80 &       5.0 &     -0.2$\pm$      0.5  &      4.8 $\pm$      1.0 & 7.5$\times$10$^{-9}$ &\\
$^\dag$C$_2$H$_5$CN         &     5.9$\times$10$^{15}$ &    110 &       2.0 &     -1.3$\pm$      0.7  &      6.9 $\pm$      1.3 & 2.0$\times$10$^{-9}$ &\\
CH$_3$OCHO           &     9.5$\times$10$^{15}$ &    112 &       3.0 &     -0.6$\pm$      0.9  &      6.3 $\pm$      1.3 & 4.8$\times$10$^{-9}$ &\\
CH$_3$OH,$\varv_t=0$     &     8.0$\times$10$^{15}$ &    130 &       9.0 &     -1.3$\pm$      0.5  &      6.4 $\pm$      1.0 & 1.2$\times$10$^{-8}$ &\\
C$_2$H$_3$CN         &  $\leq$ 1.0$\times$10$^{15}$ &    170 &       2.0 &     -0.7$\pm$      0.0  &      5.7 $\pm$      0.0 & $\leq$ 3.3$\times$10$^{-10}$ & (2)\\
CH$_3$OH,$\varv_t=1$           &     2.5$\times$10$^{17}$ &    195 &       2.0 &     -1.0$\pm$      0.8  &      7.0 $\pm$      1.5 & 8.4$\times$10$^{-8}$ &\\
CH$_3$CN             &     2.0$\times$10$^{15}$ &    200 &       2.0 &     -1.3$\pm$      0.6  &      6.6 $\pm$      1.1 & 6.7$\times$10$^{-10}$ &\\
\hline
\end{longtable}
\tablefoot{$^\dag$ The column densities of these molecules have been corrected by the vibrational factor to account for the contribution of torsionally or vibrationally excited states to the partition function. All parameters are calculated for the given $T_{\rm ex}$. The vibrational factors for C$_2$H$_5$CN are: 1.01 at 57\,K, 1.04 at 70\,K, 1.18 at 110\,K, 1.78 at 185\,K, and 2.28 at 185\,K.  
(1) Upper limit is based on a 3$\sigma$ noise level. (2): Upper limit is based on a tentative detection. We fixed the following parameters: size, excitation temperature, line width, and velocity offset.  (3): We used the temperature and the size determined for CH$_3$OCH$_3$. 
}
}

\subsection{H$_2$ column densities}
\label{sec:H2}
We use the APEX/LABOCA measurements at 870~$\mu$m from the ATLASGAL survey \citep{Csengeri2014}, and estimate the H$_2$ column density using the following expression:
\begin{equation}
\label{eq:NH2}
    N_{\rm {H_2}} [{\rm{cm}^{-2}}]= \frac{F_{\nu} R}{B_{\nu}(T_{\rm d})\Omega \kappa_{\nu} \mu_{\rm H_2} m_H}
\end{equation}
where $F_{\nu}$ is the beam averaged flux density from \citet{Csengeri2014}, $B_{\nu}(T_{\rm d})$ the Planck function, $\Omega$ the solid angle of the beam using a HPBW of 19\rlap{.}{\arcsec}2, $\mu_{\rm H_2}$ the mean molecular weight per hydrogen molecule for which we use 2.8 \citep{Kauffmann2008}, and $m_{\rm H}$ the mass of a hydrogen atom. R is the gas to dust ratio for which we assume 100. For the dust absorption coefficient, $\kappa_{\nu}$, we use 1.85~cm${^2}$~g${^{-1}}$ following \citet{Schuller2009} (see also references therein).
We adopt the dust temperature from \citet{Urquhart2018} who fit the spectral energy distribution using Herschel data, as listed in Table~\ref{tab:source_coord}.

\section{Properties of COMs in the sample}
\label{sec:COMs_properties}
\subsection{Detection statistics}
\label{sec:det_stats}
We perform a statistical assessment of the eleven COMs detected over  the sample (Table\,\ref{tab:COM_detection}). We firmly identify five COMs (CH$_3$OH, CH$_3$CN, CH$_3$OCHO, CH$_3$OCH$_3$, CH$_3$CHO) towards all sources suggesting that these COMs may become abundant already in the infrared-quiet phase. The second most frequent detections are HC(O)NH$_2$ and CH$_3$SH detected towards four sources, while C$_2$H$_5$CN is detected towards only half of the sample. Rotational transitions from vibrationally excited states are only detected for CH$_3$OH for three sources. We discuss here only the complex organic molecules with at least six atoms and an atom of oxygen or nitrogen.

The most line-rich source is G335.58 which has 11 COMs detected, followed by G328.25 and G333.46 with nine COMs detected, and then G343.75 with eight COMs detected. The most line-poor sources are G320.23 and G335.78 with a detection of five COMs. Overall this implies that the molecular richness and the presence of COMs can be characteristic of the early evolutionary stages accompanying high-mass star and cluster formation.

However, in order to discuss these detection rates of COMs, we need to consider the distance range of our sample (1.8--4.2\,kpc, see Table\,\ref{tab:source_coord}, \citealt{Csengeri2017}). Observationally, we aimed to achieve the same noise level for all sources, however, considering the same physical size of the emitting region of a given molecule, a factor of 2.3  in distance range may imply a factor of 5 difference in line intensities (for spatially unresolved emission). Using a detection threshold of 3$\sigma$ implies that any detection below 15$\sigma$ for the closest sources may become undetectable for the most distant ones, biasing our detection statistics.
The brightest COM lines of CH$_3$OH and CH$_3$CN are always detected above this threshold of 15$\sigma$ and are hence not impacted by the distance effect.

The first noticeable difference of the detection statistics among the sample is that rotational transitions from the first torsionally excited state of methanol ($v_t$=1) are only detected towards G328.25 and G335.58. Since these transitions are bright, and G335.58 is among the most distant sources, we would expect these transitions to be detected towards all sources if originating from regions with similar sizes. Consequently, the non-detection of CH$_3$OH $v_t$=1 transitions towards G320.23, G333.46, and G335.78 suggests intrinsic differences among the sample. 

Considering the distance effect, we would also expect complex cyanides to be detected towards all sources based on the detection of these molecules towards G328.25 and G335.58. Consequently, their non-detection towards G320.23, G333.46, and G335.78 suggests intrinsic differences in the molecular composition of the sample. Therefore, the different detection rates of COMs (Table\,\ref{tab:COM_detection}) likely correspond to intrinsic differences among the sources such as the column density, the luminosity, or the mass of the central object rather than an observational bias due to the distance range of the sample. 

The molecular composition and the detection rates of individual lines are, however, not identical in all the sources. Table\,\ref{tab:COM_Eup} shows the range of upper level energies of the transitions covered and detected by our spectral survey. 
We find that the three sources, G320.23, G333.46, and G335.78, exhibit detections with typically lower upper level energies than G328.25, G335.58, and G343.75. We can also notice that the cyanides are detected where we detect the highest upper level energies. 
Overall, this allows us to conclude that the COMs exhibit intrinsic differences among the sample that could be linked to chemical history, age sequence, or the population of embedded protostars within the clump (Sect.\,\ref{sec:dichotomy_NO}).

\subsection{Temperatures of COMs}

We find that for CH$_3$OH, $v_{\rm t}$=0 and CH$_3$CN we typically need two temperature components, corresponding to a cold and a warm gas phase (see Sect.\,\ref{sec:discuss_temp}). The highest temperatures are traced by the CH$_3$CN molecule for all sources.
We find an extended cold gas phase with temperatures between 22 and 35\,K and a compact warm gas phase with a source size ($\theta_S$) of $\sim$1--2\arcsec\ and a temperature reaching between 185 and 260\,K. Such high temperatures are exclusively traced by CH$_3$CN, and the source size of this component corresponds to the most compact sizes we can extract from the fitting procedure. 
The heavier complex cyanides, ethyl and vinyl cyanide also trace a warm and compact phase with a somewhat lower temperature between 90\,K and 200\,K, although they are not ubiquitously detected.
Interestingly, these two species have the largest dispersion in temperature among the detected molecules. 

On the contrary, methanol traces a cold extended gas phase and its hot compact component has a maximum temperature of 130\,K towards G333.46. Although, the upper level energies covered by our spectral survey is similar for methanol and methyl cyanide, methanol does not trace the compact hottest region in the vicinity of the protostar, and it exhibits up to 100~K lower temperatures. Similarly, other O-bearing COMs, when detected in the warm gas phase, are found to have a temperature of $\sim$80\,K (for instance G335.58) significantly lower than that of N-bearing COMs with a --CN functional group. 

The rest of the molecules trace cold gas below 100\,K. Overall, we measure an average temperature between 60\,K and 90\,K and a median temperature between 43 and 81\,K considering all COMs (Table\,\ref{tab:temp_stats}). The cold gas phase has a range of temperatures between 20\,K and 55\,K. This is a clear indication that the bulk of the gas towards infrared quite clumps remains cold or luke-warm. Towards G343.75, a warm component at 1$\arcsec$ emerges (corresponding to a physical size scale of 1800\,au). Temperatures close to $\sim$100~K are typically traced by COMs appearing in the warm phase. Overall, we find that while some sources have a pronounced warm component, others remain mainly cold. 

While our spectral survey is sensitive enough to detect several COMs, we find that towards half of the sample we cannot separate the cold and warm components of the envelope based on COMs. As discussed in  Sect.\,\ref{sec:LTE}, and shown in Fig.\,\ref{fig:Mod1vs2comp}, the spectra can be equally well fitted with a cold and a warm temperature component (e.g., 30\,K and 90\,K) but also with a single temperature component at $\sim$50\,K. Since the parameters of the two temperature component fit cannot be robustly constrained, we choose the single component model to fit the spectra. However, this may suggest that a compact warm component can be present, although it may not have yet a significant contribution to the overall molecular emission. We conclude that in these cases an emerging warm gas phase may simply be too compact to be robustly detected.

For example, the cold and warm gas components of CH$_3$OCH$_3$ could be clearly identified towards both G328.25 and G335.58. 
However, for the other sources, G333.46, G320.23, G335.78, and G343.75, the spectra could be equally well fitted with a single or a two temperature component model. Since in such cases, the parameters of the multiple components are poorly constrained, we consider the single component model as the best fit. 

\subsection{Source structure}

Considering the velocity profiles of all spectral features in the band, we notice broad and narrow line profiles, as well as outflow wings. This suggests that although it remains spatially unresolved in our observations, the protostellar envelope has an internal structure with physical parameters that are not uniform. From the sample, only G328.25 has been studied in detail. Its envelope structure has been revealed by ALMA observations in \citet{Csengeri2018}, and has been modeled for its temperature and density profiles in \citet{paperI}. Such a detailed modeling for all sources is beyond the scope of this paper, we discuss, however the change of physical parameters throughout the envelopes.

\begin{figure*}
    \centering
    \includegraphics[width=1\linewidth]{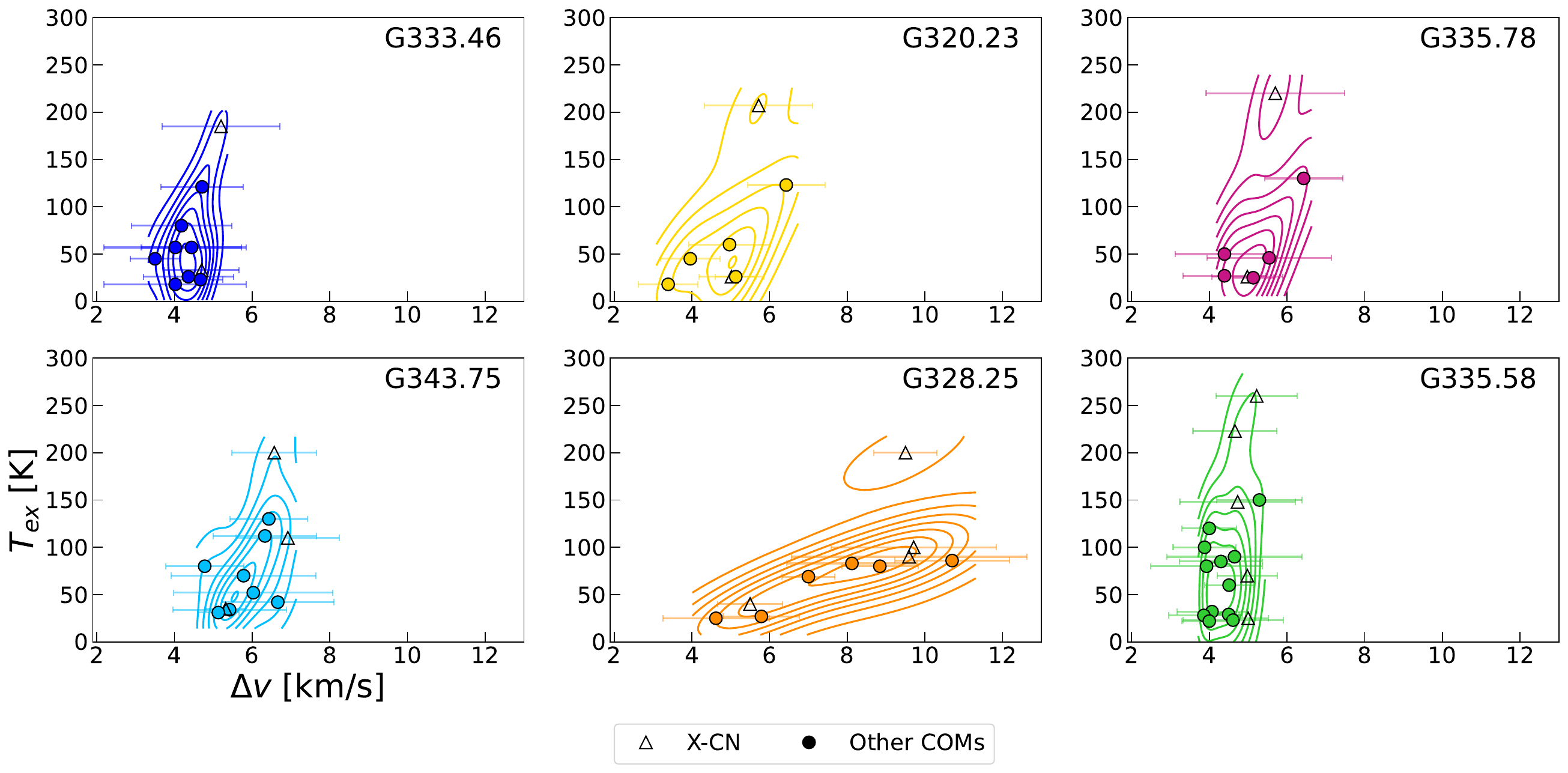}
    \caption{Excitation temperature ($T_{\rm ex}$) versus mean line width ($\Delta {v}$) of each COM in the cold and warm components. Each data point represents the line width of a given molecule averaged over all its transitions. The contours show the Gaussian Kernel distribution of the datapoints. 
Transitions fitted with multiple velocity components are considered as individual measurements in this figure. The errorbars represent the dispersion between the line width measurements for all the transitions. The triangles represent the complex cyanides and the circles represent the other COMs.}
    \label{fig:TexVsLineWidth}
\end{figure*}

We show the distribution of line widths versus $T_{\rm ex}$ (Fig.\,\ref{fig:TexVsLineWidth}), where  $T_{\rm ex}$ is the result of the LTE fitting and the line width corresponds to the average line widths from the Gaussian line profile fitting. We find that the line widths towards all objects (except G328.25) range between 2~\kms\ and 7~\kms. Only towards G328.25 we find lines as broad as 10.7~\kms. The larger velocity dispersion observed towards G328.25 is due to the accretion shocks found at $\pm4.2$~\kms\ offset from the \vlsr\ \citep{paperI}. In \citet{paperI} we show that the accretion shocks have relatively narrow line widths of $\sim$3~\kms, and as shown in Fig.\,\ref{fig:CH3OH_lineprofile} 
they can blend with the envelope component.

\begin{figure*}
    \centering
    \includegraphics[width=1\linewidth]{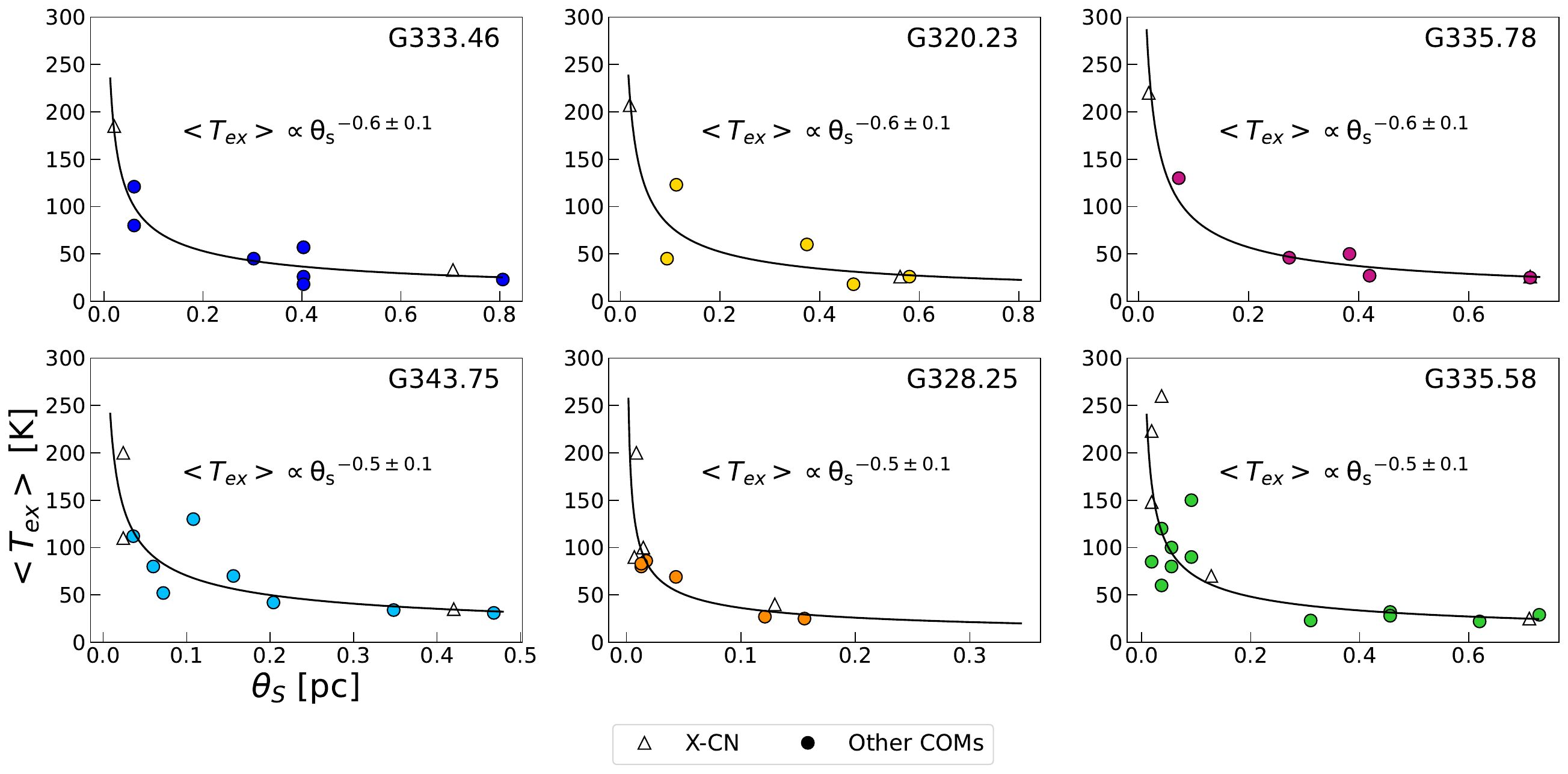}
    \caption{Excitation temperature of COMs from the cold and warm gas vs their size of emission region. The black curve represents the power-law fit of the data. The triangles represent the cyanides and the circles represent the other COMs. We show here the results of both the warm and cold gas phases.}
    \label{fig:TexVsSize}
\end{figure*}

In Fig.\,\ref{fig:TexVsSize}, we show $T_{\rm ex}$ as a function of the size of the emitting region ($\theta_{\rm S}$), both parameters were obtained from the LTE modelling. We find that this can be well described with a power law function in the form of $<T_{\rm ex}>\propto \theta_{\rm S}^{-\beta}$, where $\beta$ is between 0.5 and 0.6. Our fit suggests a gradual increase towards all sources in the excitation temperature spatially averaged over the innermost regions.
In \citet{paperI} we discussed the same figure for G328.25 including all species identified from the entire $\sim$206\,GHz frequency range of the survey. Extracting here only the information for COMs, we find that the exponent of the power law fit to the data is consistent within the errors with the results of \citet{paperI}. 

The relatively small dispersion in the correlation between $T_{\rm ex}$  and $\theta_{\rm S}$ suggests that the obtained $T_{\rm ex}$ may correspond to an averaged kinetic temperature over the given source size ($\theta_{\rm S}$). This would be consistent with a stratified envelope, where the different molecules originate from different regions. 
We notice that the temperature distribution of the warmest component in G320.20, G333.46, and G335.78 is not as well defined as for the other sources (Fig.\,\ref{fig:TexVsSize}), because only
CH$_3$CN and CH$_3$OH trace a compact warm component and the rest of the species are at lower temperatures.

\begin{table*}
    \centering
    \caption{Statistics of the excitation temperatures 
    of 
    COMs towards the sample. }
    \label{tab:temp_stats}
    \begin{tabular}{l c c c c c c}
    \hline
    \hline
    & G333.46 & G320.23 & G335.78 & G343.75 & G328.25 & G335.58\\
    \hline
    Mean temperature [K] & 61 & 72 & 75 & 83 & 80 & 90\\
    Median temperature [K] & 43 & 45 & 46 & 70 & 82 & 65\\
    Max temperature [K] & 185 & 207 & 220 & 200 & 200 & 260 \\
    Mean cold gas [K] & 22 & 22 & 22 & 27 & 23 & 25\\
    Median cold gas [K] & 20 & 20 & 20 & 30 & 23 & 25\\
    \hline
    \end{tabular}
\end{table*}

\begin{figure*}
    \centering
    \includegraphics[width=0.49\linewidth]{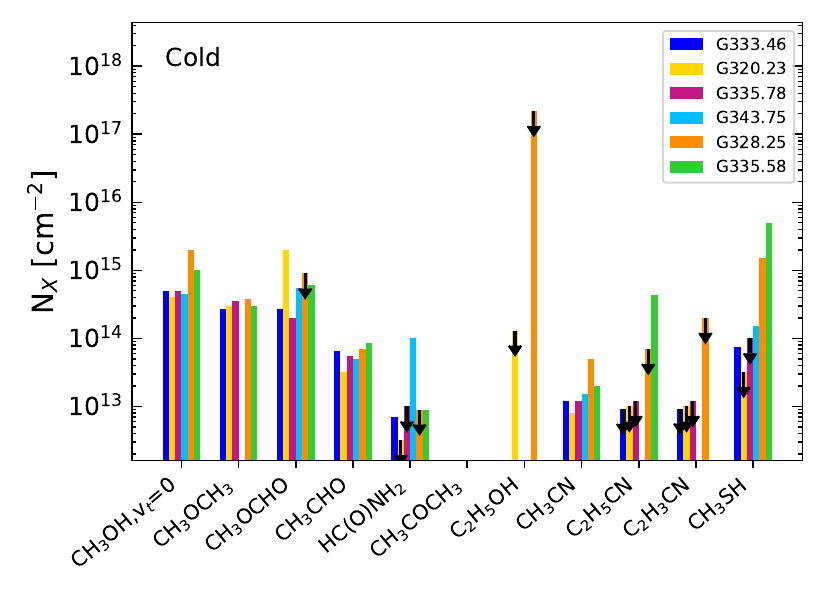}
    \includegraphics[width=0.49\linewidth]{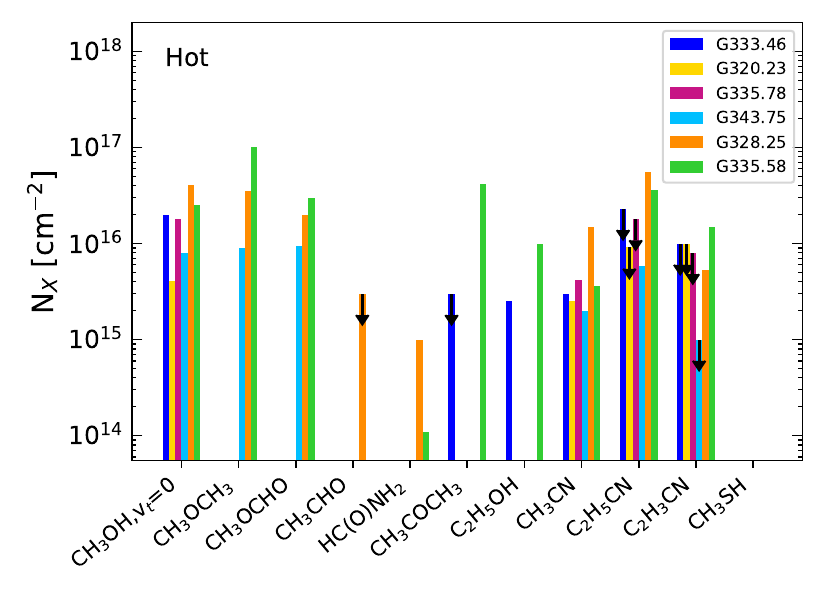}
    \caption{Column densities of the COMs in the cold (left panel) and warm (right panel) gas components. Black arrows represent upper limits. Each source is represented by a different color.}
    \label{fig:histo_dens_COMs}
\end{figure*}

\subsection{Column densities of COMs}
\label{sec:abund_COMs}

Figure\,\ref{fig:histo_dens_COMs} and Table\,\ref{tab:COM_LTE} show the column densities of COMs for each source estimated based on the LTE assumption  described in Sect.\,\ref{sec:LTE}. Using the large range of upper-level energies for each COM, we were able to recognize COMs originating from the warm and cold gas, respectively. We also discuss these column density estimates considering the different source sizes in Sects.\,\ref{sec:mol_abundances} and \ref{sec:dichotomy_NO}.

The O-bearing COM with the highest column density in the cold gas phase (except for G320.23) is CH$_3$OH, $v_{\rm t}$=0 with values 
between $4\times10^{14}$~\unidens\ and $2\times10^{15}$~\unidens.
It exhibits less than an order of magnitude dispersion over the sample (Fig.\,\ref{fig:histo_dens_COMs}, left panel). 
CH$_3$OCH$_3$ has column densities between 2.7 and 3.2~$\times10^{14}$~\unidens, and is relatively constant for the sample, with a variation of a factor of less than 2. CH$_3$CHO has the lowest column density among the sample in the cold gas phase with values between 3.2 and 8.7$\times10^{13}$~\unidens. 
Interestingly, CH$_3$OCHO shows the largest variation among the sample, and it is not detected in the cold gas phase towards G328.25, where we only estimate an upper limit for its column density. For the other sources we estimate  a column density between 2.7$\times10^{14}$ and 2.0$\times10^{15}$\unidens, which corresponds to a somewhat larger variation by a factor of $\sim$10. 

CH$_3$CN is detected in the cold gas phase towards all sources and shows a large variation between 8.0$\times10^{12}$\,\unidens\ and 5.0$\times10^{13}$\unidens. 
Heavier complex cyanides remain undetected in the cold gas phase, and only upper limits can be inferred (except for C$_2$H$_5$CN towards G335.58 where we can fit a cold component).

The molecule HC(O)NH$_2$ is detected towards three sources (G333.46, G343.75, G335.58) with a column density of 7-9$\times10^{12}$\,\unidens\ for G333.46 and G335.58, for G343.75 we estimate 1$\times10^{14}$\,\unidens\ but within a size of 6\arcsec and an excitation temperature of 52~K. This emission is more compact than for the other sources, and could thus originate from the lukewarm layers.

CH$_3$SH is only detected in the cold gas towards G333.46, G343.75, G328.25, and G335.58 with column densities between 7.5$\times10^{13}$ and 5.0$\times10^{15}$\unidens. It shows thus the largest variation in column density among the sample.

It is clear that the cold gas phase is richer in the variety of COMs compared to the hot gas phase (Fig.\,\ref{fig:histo_dens_COMs}, right panel). Due to the more compact emitting regions, COMs in the warm gas exhibit column densities of up to 3 orders of magnitude higher compared to the cold gas, and this jump in column density is the most significant for the complex cyanides. CH$_3$CN exhibits values between 2.5 and $2.1\times10^{15}$~\unidens\ for all sources except G328.25, and a peak of $\sim10^{16}$~\unidens\ for G328.25. For C$_2$H$_5$CN we estimate column densities between 5.9$\times 10^{15}$ and 3.6$\times 10^{16}$~\unidens, except for G328.25 where a peak value of 5.5$\times 10^{16}$~\unidens\ is observed.
The highest column densities of C$_2$H$_3$CN reach 1.5$\times 10^{16}$~\unidens\ towards G335.58.

Methanol, on the other hand shows a less significant jump between the cold and warm gas phase. It exhibits column densities between 4.1$\times$10$^{15}$\unidens\ and 4.1$\times$10$^{16}$\unidens. 
For G320.23, we only find one order of magnitude higher column densities in the hot gas phase compared to the cold gas.
We also detect emission from the CH$_3$OH $v_{\rm t}$=1 state towards three objects: G328.25, G335.58, and G343.75 that we fitted independently. Towards G328.25, CH$_3$OH $v_{\rm t}$=1 is proposed to originate from the accretion shocks \citep{Csengeri2018}. Due to blending of the $v_{\rm t}$=1 transitions, for the other two sources we have a higher uncertainty on the column densities of CH$_3$OH in the $v_{\rm t}$=1 state, and estimate values on the order of 10$^{16}$~\unidens. For G335.38, our column density estimates exceed that of the $v_{\rm t}$=0 state by a factor of 2.4. Typically the $v_{\rm t}$=1 excited state is expected to be a minor contributor to the overall methanol column density, this suggests therefore that our overall column density estimates for methanol in the hot gas phase could be uncertain for this source by a factor of a few. 

In the hot gas phase, we estimate column densities between 9.0$\times$10$^{15}$\unidens\ and 1$\times$10$^{17}$\unidens\ for CH$_3$OCH$_3$ (G343.75, G335.58, and G328.25), and of 9.5$\times$10$^{15}$\unidens\ and 3.6$\times$10$^{16}$\unidens\ for CH$_3$OCHO towards G343.75 and G335.58, respectively. These column densities are higher compared to that of methanol, however, they originate from a more compact region. We compare these column density estimates considering the different sizes in more detail in  Sects.\,\ref{sec:mol_abundances} and \ref{sec:dichotomy_NO} (see also Fig.\,\ref{fig:add_col_dens}), where we show that after correcting for the different emission sizes methanol is the most abundant O-bearing COM in the gas phase.

\begin{figure*}
    \centering
    \includegraphics[width=0.49\linewidth]{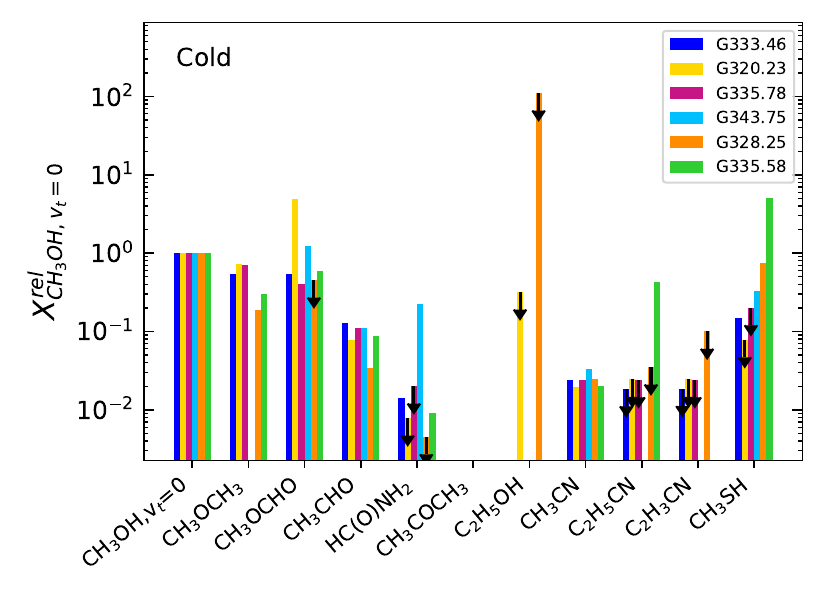}
    \includegraphics[width=0.49\linewidth]{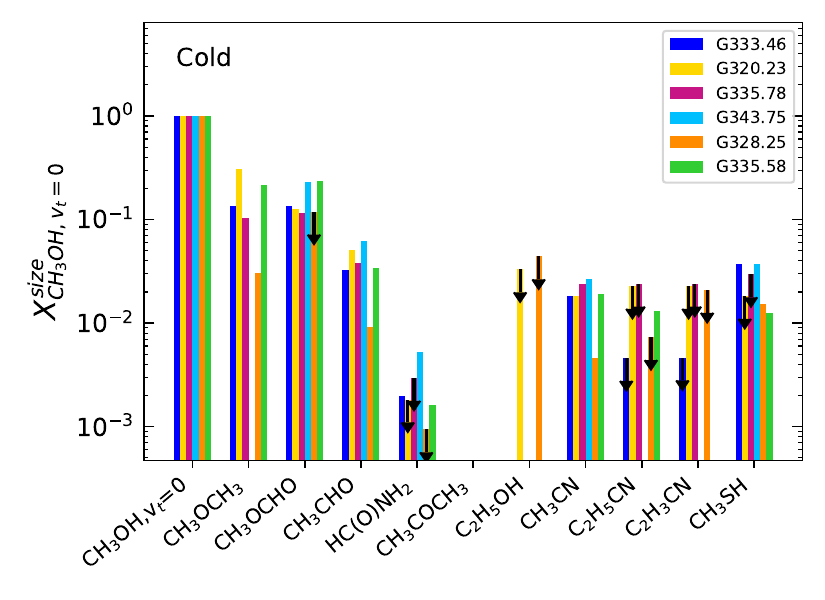}
    \includegraphics[width=0.49\linewidth]{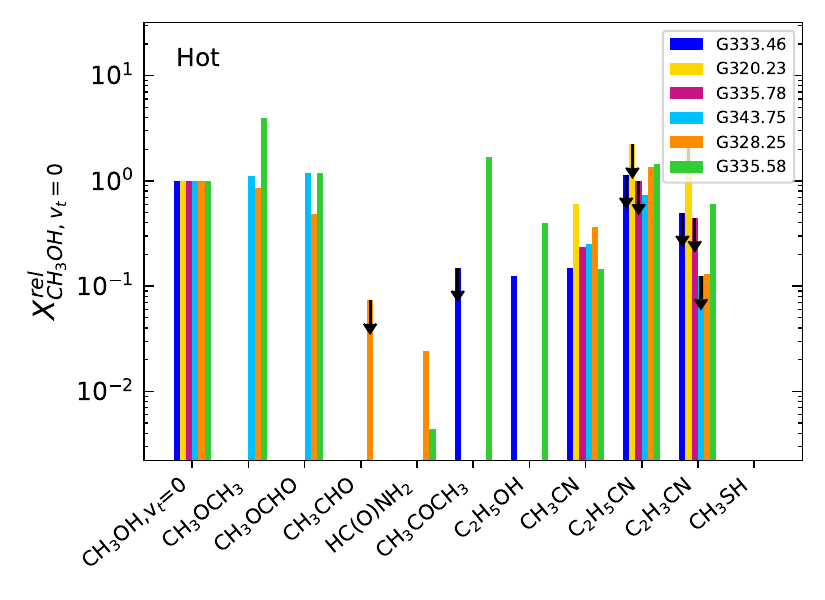}
    \includegraphics[width=0.49\linewidth]{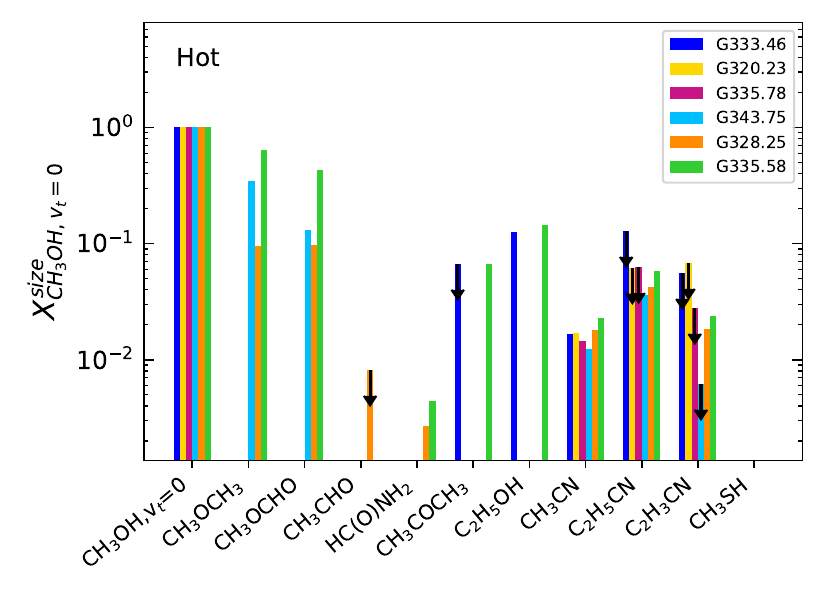}
    \caption{Abundances of COMs relative to CH$_3$OH, $v_t$=0 in the cold and warm components of the envelope. The first row represents the cold component of the envelope with the relative abundances to CH$_3$OH, $v_t$=0 without size correction (top left panel) and including the correction by the size of the emitting region (top right panel). The bottom row represents the hot component of the envelope with the relative abundances to CH$_3$OH, $v_t$=0 without size correction (bottom left panel) and with size correction (bottom right panel). Each source is indicated in a different color. Black arrows indicate upper limits.}
    \label{fig:COMs_relCH3CHO}
\end{figure*}

\begin{figure*}
    \centering
    \includegraphics[width=0.49\linewidth]{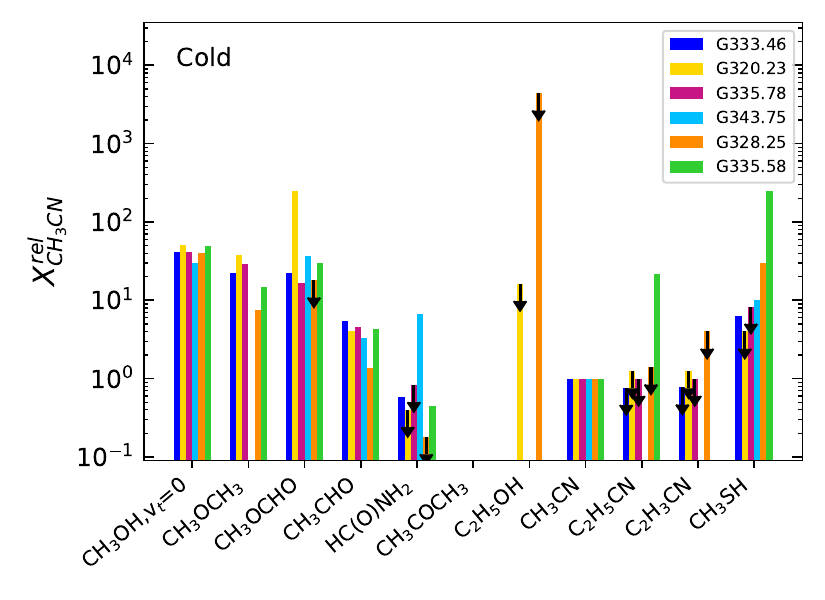}
    \includegraphics[width=0.49\linewidth]{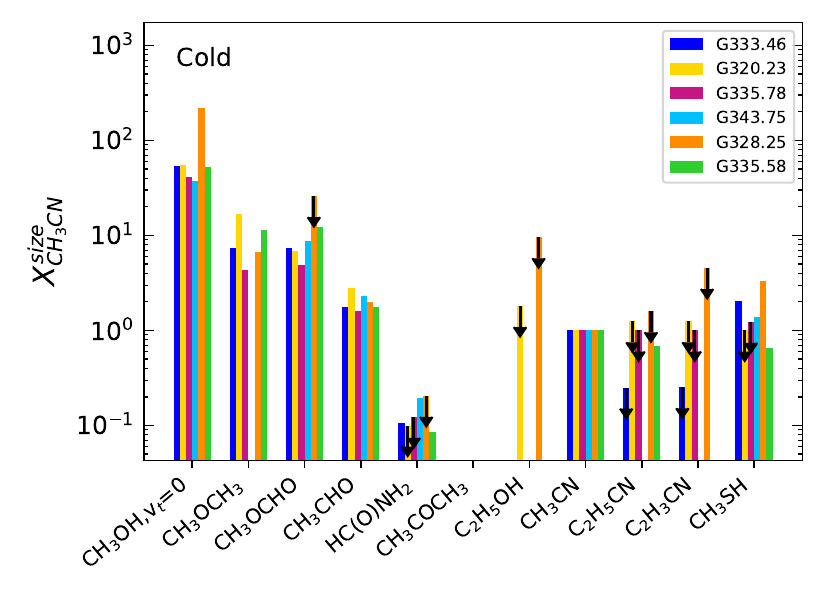}
    \includegraphics[width=0.49\linewidth]{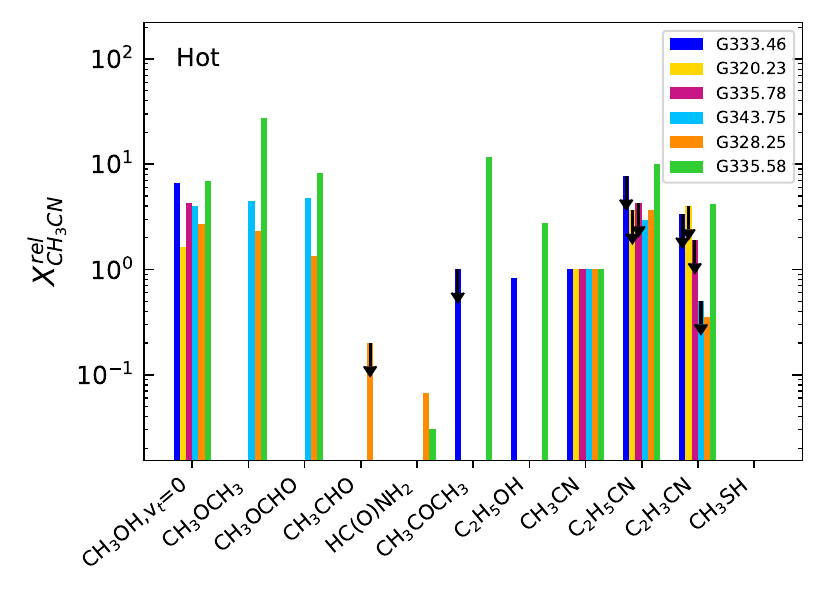}
    \includegraphics[width=0.49\linewidth]{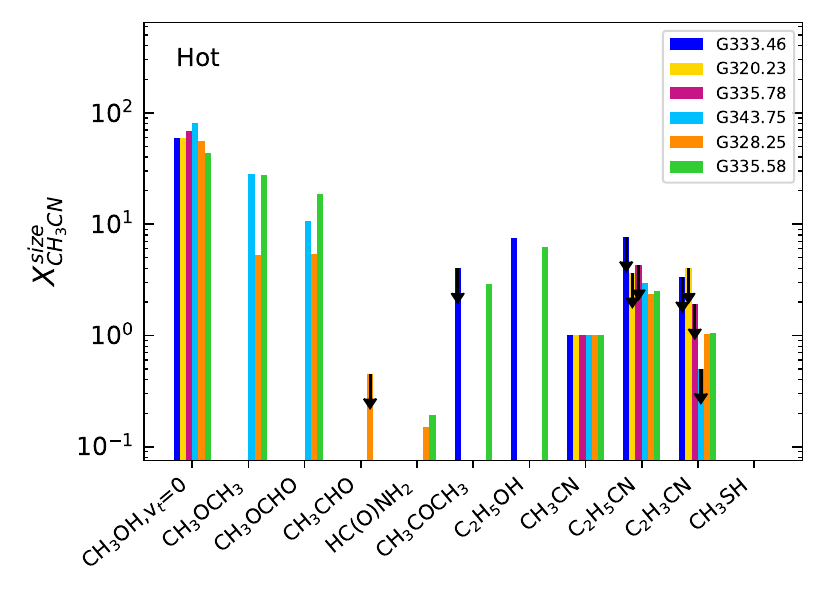}
    \caption{Abundances of COMs relative to CH$_3$CN in the cold and warm components of the envelope. The first row represents the cold component of the envelope with the relative abundances to CH$_3$CN without size correction (top left panel) and including the correction by the size of the emitting region (top right panel). The bottom row represents the hot component of the envelope with the relative abundances to CH$_3$CN without size correction (bottom left panel) and with size correction (bottom right panel). Each source is indicated in a different color. Black arrows indicate upper limits.}
    \label{fig:COMS_relCH3CN}
\end{figure*}

\subsection{Molecular abundances}
\label{sec:mol_abundances}
Molecular abundances are computed relative to another species, where the choice is frequently H$_2$. However, since the H$_2$ column density is not straightforward to obtain, and column density estimates of COMs suffer from less bias, it became a common practice in observational astrochemistry to use other abundant species as a reference, for example methanol, methyl formate, methyl cyanide \citep[e.g][]{Drozdovskaya2019, Jorgensen2020}.
Here we define $X_{ref}^{rel}(mol)=\frac{N(mol)}{N(ref)}$ as the abundance of a molecule ($mol$) relative to that of another species ($ref$), where any molecule from the same observational dataset can be taken as a reference. Since the emission from various molecules may arise from different regions within the envelope, we can also consider to correct for the difference in the size of the emitting regions using  $N(mol)\times \theta_S^2(mol)$.  We also introduce  
$X_{ref}^{size}(mol)=\frac{N(mol)\times \theta_S^2(mol)}{N(ref)\times \theta_S^2(ref)}$ the relative abundances to a molecule corrected for the different sizes of the emitting regions. 

For this analysis, ideally we would require  
a molecule to be detected towards all the sources, and both in the cold and warm gas components. In this study methanol is well detected towards all the objects and in both the warm and cold gas components, and therefore we could use it to compare the relative molecular abundances of the sample (Fig.\,\ref{fig:COMs_relCH3CHO}).
 However, due to its high molecular abundance
methanol may become optically thick, which introduces higher uncertainties on the methanol temperature and column density compared to the other COMs. Furthermore, 
it may often exhibit multiple velocity components that complicate its analysis.
Therefore, when  comparing to measurements from the literature (Sect.\,\ref{sec:sec_comp}), we use dimethyl ether as a reference for O-bearing COMs because it is expected to be less affected by optical depth effects, \citep[e.g.,][]{Bisschop2007, Agundez2019, Widicus2017}
 and CH$_3$CN for N-bearing COMs (Fig.\,\ref{fig:COMS_relCH3CN}), because it is the most frequently detected
N-bearing COM.  
Towards our sample, these species (CH$_3$OCH$_3$, CH$_3$CN) are detected in both components of the envelope.

We use molecular abundances relative to H$_2$ in Sect.\,\ref{sec:comp_models} for the compact hot gas phase with the purpose of comparing our relative abundance estimates to predictions from chemical models. For this, we use the formula for $N_{\rm H_2}$ given in Sect.\,\ref{sec:H2}. However, to infer H$_2$ column density at scales well below the 19\rlap{.}{\arcsec}2 resolution of the ATLASGAL survey, we need to scale the measured $F_{\rm{\nu}}$ flux densities and the H$_2$ column density to the size of the emitting region of COMs (c.f.\,Table~\ref{tab:COM_LTE}). Comparing the ATLASGAL flux densities to measurements at a similar central frequency of 345~GHz obtained with ALMA at 3--4\arcsec resolution from \citet{Csengeri2017b}, we find that they are broadly consistent with a source density profile of $\rho(r)\sim r^{-2}$ (see as an example the discussion on source structure for G328.25 in \citealp{paperI}). Therefore, to account for such a source structure we scaled $N_{\rm H_2}$ as $\sim r^{-1}$ using the values from Table\,\ref{tab:source_coord} as references.

\section{Discussion}
\label{sec:discussion}

\subsection{Emerging hot gas component}
\label{sec:dichotomy_NO}

Using an unbiased spectral survey performed towards our sample of six infrared-quiet  clumps, we derived  column densities and temperatures of COMs.  
Our main finding is that COMs are detected towards all objects and their excitation temperatures are typically low, they originate from cold gas ($T\sim$20--30\,K). Methanol is the major contributor of the COM column densities in the cold gas.

We find, however, several examples where another temperature component corresponding to warm or hot gas is necessary. This is consistent with our earlier result (see Sect.\,\ref{sec:det_stats}) that sources with a higher mean temperature have lines detected over a larger range of upper level energies  (Table\,\ref{tab:COM_Eup}). The warm component typically exhibits temperatures of $\sim$90\,K for O-bearing molecules while cyanides trace temperatures between 90 and 250\,K (see Sect.\,\ref{sec:discuss_temp} for further discussion). Additionally, the average temperature of the COMs in the envelope of our objects is between 61 and 90\,K (Table\,\ref{tab:temp_stats}).   We see an 
emerging hot and compact gas phase, while the bulk of the gas remains cold.

\begin{figure*}
    \centering
    \includegraphics[trim=0cm 3.7cm 0cm 0cm, clip, width=1\linewidth]{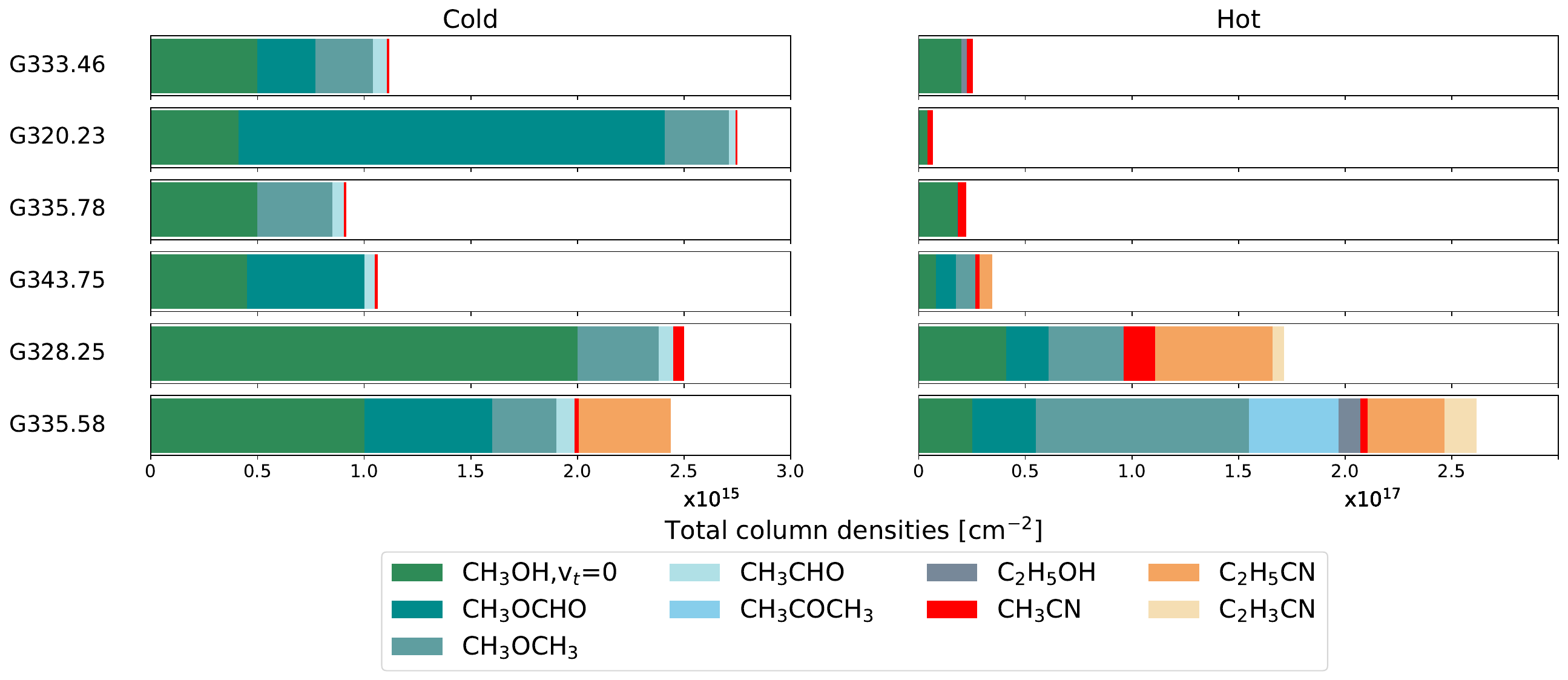}
    \includegraphics[width=1\linewidth]{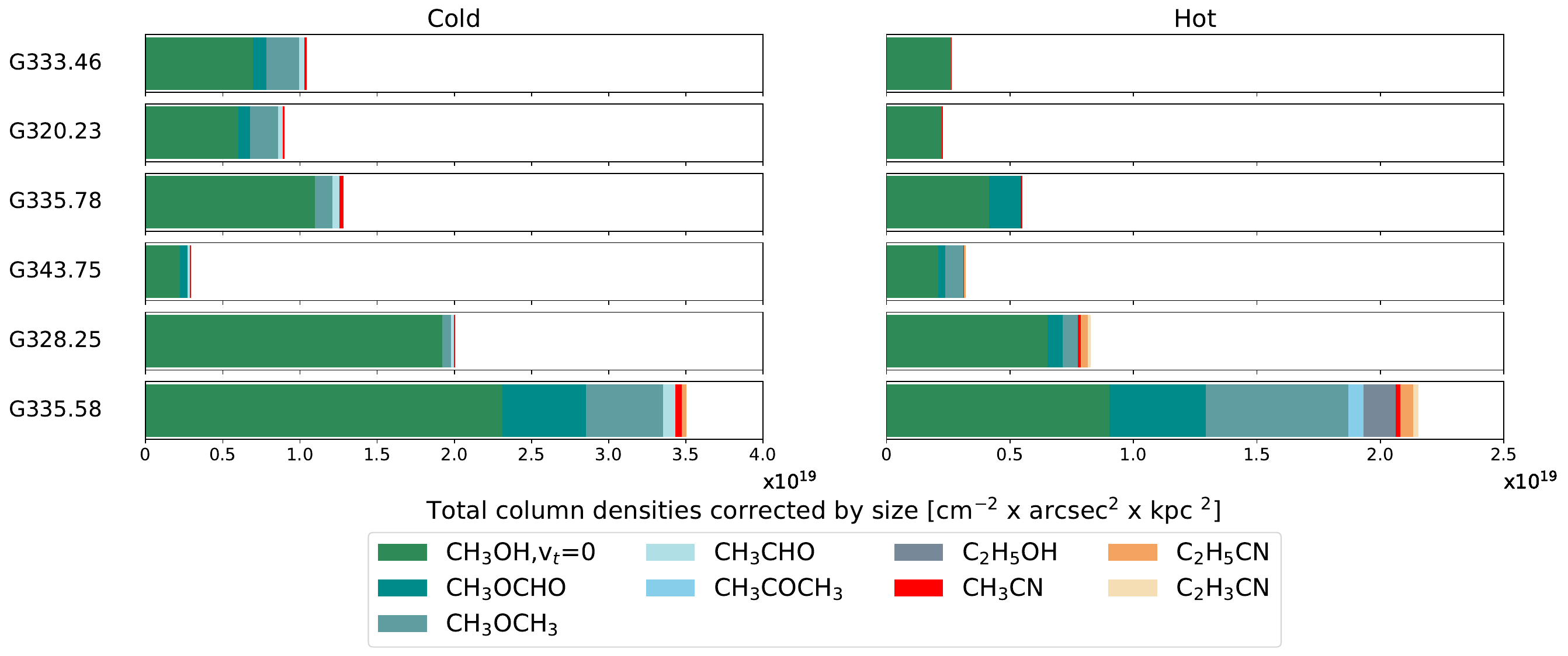}
    \caption{{\it Top panel:} Column densities of the detected COMs in the cold (left column) and hot (right column) gas. {\it Bottom panel:} Column densities normalized by the physical size of the emitting region of the COMs in the cold (left column) and hot (right column) gas. Each molecule is represented by one colour.}
    \label{fig:add_col_dens}
\end{figure*}

While Figs.\,\ref{fig:TexVsLineWidth} and \ref{fig:TexVsSize} suggest a decreasing temperature profile as a function of radius within the envelope, in order to discuss  the impact of radiative heating on the molecular composition of the envelope,  we need to separate contribution from the cold, quiescent gas phase from that of the hot component impacted by the radiative heating. For this we adopt a temperature threshold of 80\,K corresponding to the onset of 
the \textit{thermal} desorption of H$_2$O found from experiments of \citet{Collings2004}. Above this temperature threshold, we assume that COMs thermally co-desorb together with water ice. In this picture, we define the cold gas phase as the quiescent gas with a temperature below 80\,K.
Figure\,\ref{fig:add_col_dens} (top row) shows the sum of the column densities of the detected COMs in the cold and hot gas for each source. 
Our sample exhibits  H$_2$ column densities that vary by less than a factor 2 (Table\,\ref{tab:source_coord}), therefore, it is relevant to compare the column densities of COMs of the sample.

The total molecular content of COMs in the cold gas phase is rather similar over the sample. It is characteristic of all sources that the contribution of complex cyanides is minimal in the total amount of COMs.
The sum of the column density of COMs in the cold gas ranges between 1.1$\times$10$^{15}$cm$^{-2}$ and 2.8$\times$10$^{15}$cm$^{-2}$. 
The broadly similar  amount of detected COMs in the cold gas suggests that the chemical processes at the origin of these COMs are similar for all sources. 

In contrast to the cold gas phase, the warm gas shows a considerably larger diversity in COMs. 
The column density of COMs towards is larger from the two sources G328.25 and G335.58. These two sources also show  a higher mean temperature compared to the rest of the sample (Table\,\ref{tab:temp_stats}). G343.75 exhibits overall lower column densities of all COMs compared to G328.25 and G335.58, although its molecular composition is similar to theirs with the detection of ethyl cyanide and methyl formate. Particularly apparent is that the ratio of O-bearing COMs versus -CN bearing COMs shows a diversity that is more pronounced in the cold gas phase (Fig.\,\ref{fig:COMs_relCH3CHO}). We suggest that the hot gas phase reflects the impact of the embedded source, which, compared to other sources, is either more luminous or more evolved in G328.25 and G335.58 (see also Sect.\,\ref{sec:lum}).

In order to account for the different 
distances of the sample, we multiply the obtained column densities by the area of the emitting region for each species ($\theta^2$) and the square of distance of the source (in kpc$^2$). 
Fig.\,\ref{fig:add_col_dens} (bottom row) and find that the sum of the column density of detected COMs 
shows variations. In this representation, G343.75 resembles G335.78 with a larger molecular diversity and a better defined warm phase.
Motivated by this, we define the first group of sources composed of G333.46, G320.23, and G335.78, in which the warm gas is mainly represented by O-bearing COMs, mostly methanol. The second group of sources composed of G343.75, G328.25, and G335.58, exhibits a more pronounced warm gas phase, where COMs other than methanol start to appear, and complex cyanides are present. The warm component of the envelope towards the second group of objects shows a richness in COMs that suggests an increasing chemical complexity at higher temperatures.

Overall, we conclude that the sources of our sample of infrared quiet clumps have large column densities of COMs in the cold gas phase. Three sources (G333.46, G320.23, and G335.78) remain  mostly cold and host a low total column density of COMs in the warm gas (2.5$\times$10$^{16}$\unidens). The rest of the sample (three sources) shows a well defined ($T > 80$\,K), higher column density warm gas phase. The molecular composition of the hot gas phase shows differences among the sample, especially the amount of O-bearing COMs compared to heavy \mbox{-CN} bearing COMs (such as C$_2$H$_5$CN and C$_2$H$_3$CN) shows a diversity.
Sites of high-mass star and cluster formation are known to be associated with a cold gas phase, where infrared-quiet clumps have been found to exhibit lower bulk temperatures compared to more evolved massive star forming sites \citep[c.f][]{Wienen2012,Giannetti2014,Molinari2016,Wienen2018}. One a few  high-mass star forming regions are known to exhibit COM emission in the cold gas phase. COMs have been identified, for example towards Sgr~B2 in the cold gas phase by \citet{Li2017} over an extended region, and by \citet{Busch2022} at reaching as low as about 30-60~K. Extended cold methanol emission has been reported towards the IRDC G028.23-00.19 by \citet{Sanhueza2013}. However, no systematic studies explored heavier COMs in the cold gas reaching as low as 30~K, other than CH$_3$OH and CH$_3$CN \citep{Oberg2013, Sanhueza2013, Giannetti2014}. This is because observational studies of COMs towards high-mass star forming regions mainly focused on more evolved stages and their warm gas phase \citep[e.g.][]{Bisschop2007,Widicus2017}. On the other hand, nearby dark clouds, and pre-stellar cores associated with cold gas, reaching temperature below 10~K, have been found to exhibit COMs  \citep{Bacmann2012, Scibelli2021}, challenging theoretical models (\citealp{Garrod2022}, see however \citealp{Balucani2015}). In this context our results suggest that COMs in the cold, pristine gas phase are also present towards high-mass star forming regions. Two regimes of COM desorption, at low and high temperatures, have also been reported by \citet{Busch2022}.

We compare and discuss the properties of the emerging warm gas phase to a larger number of hot cores and hot corino sources from the literature in Sect.\,\ref{sec:sec_comp}.

\subsection{Origin of COMs: thermal desorption?}
\label{sec:discuss_temp}
Although our sample of young massive clumps have their bulk of the gas at cold temperatures, we also detect emission from COMs that is compact and at an elevated temperature, thus potentially impacted by radiative heating of the protostar. Current chemical models explain the formation of COMs mainly on the surface of dust grains \citep{Garrod2006}, followed by thermal or non-thermal desorption to the gas phase. Gas phase reactions become important once the gas becomes hot \citep{Charnley1997, Garrod2006}, although some reactions may proceed at lower temperatures to form COMs \citep{Balucani2015, Vasyunin2017}.

\begin{figure*}
    \centering
    \includegraphics[width=1\linewidth]{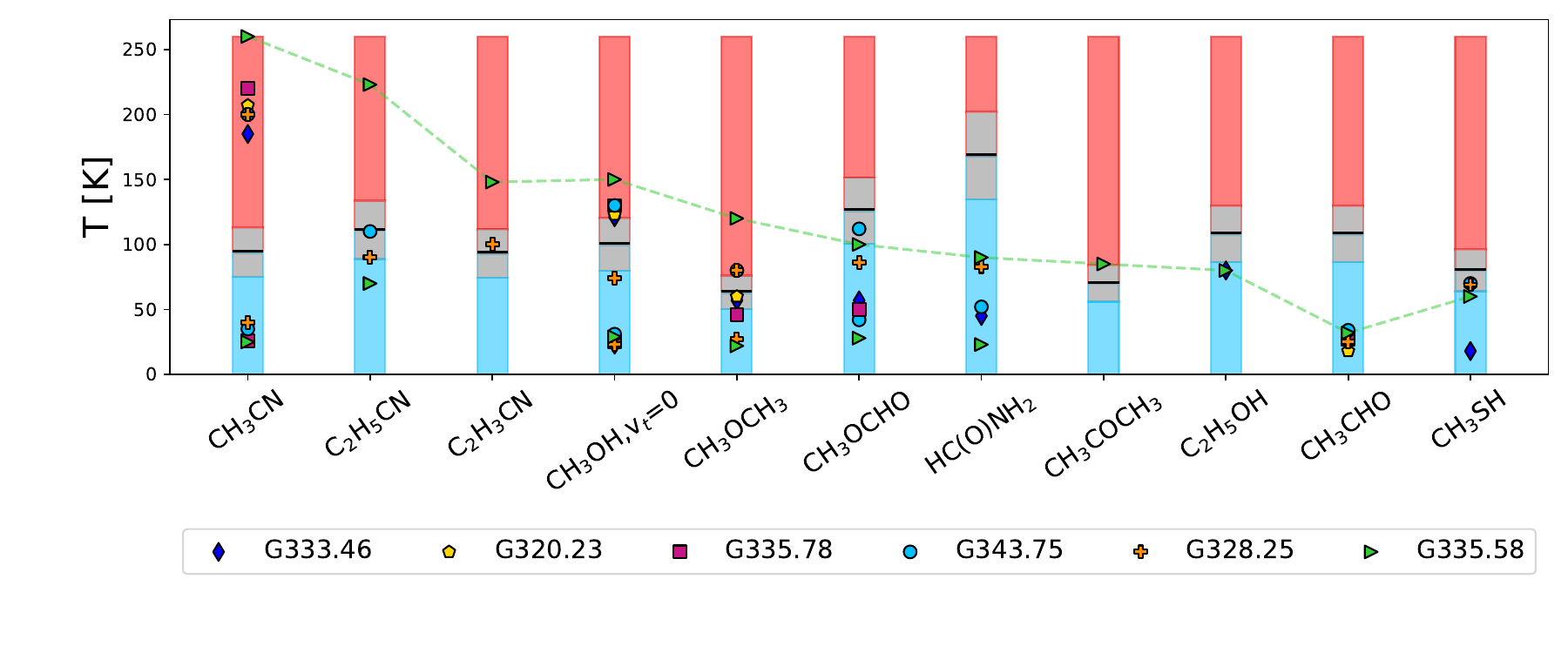}
    \caption{Excitation temperatures of the COMs. Each source is represented by a different color and symbol. The black line shows the desorption temperature for each molecule (references are given in Table\,\ref{tab:desorption_temperature}). The blue and red histograms represent the temperatures below and above the desorption temperature, respectively. The grey area shows a 20\% error on the desorption temperature. The green line represents the variation of temperature between different molecules towards G335.58.}
    \label{fig:scatter_temp_COMs}
\end{figure*}

In Fig.\,\ref{fig:scatter_temp_COMs} we show the distribution of the measured excitation temperatures compared to the estimated desorption temperatures ($T_{\rm des}$) that we compute as $T_{\rm des}$=0.02$\times E_{\rm des}$ following \citet{Hollenbach2009}, and where $E_{\rm des}$ is the desorption energy (in K). The desorption temperature should be considered as a range around this temperature, since its value depends on the binding energies that may vary  depending on the molecular content of the ices \citep[e.g][]{Tielens1991, Collings2004, Ferrero2020}. 
We list the values of desorption energy and the corresponding desorption temperature in Table\,\ref{tab:desorption_temperature}.

We identify three groups of molecules: for the first one (CH$_3$CN, C$_2$H$_5$CN, and C$_2$H$_3$CN) the excitation temperature is clearly higher than the desorption temperature suggesting that desorption can happen through thermal processes. The second group (CH$_3$CHO, C$_2$H$_5$OH, HC(O)NH$_2$, and CH$_3$SH) exhibits  excitation temperatures considerably below the corresponding thermal desorption temperature, suggesting that these molecules are liberated to the gas phase through non-thermal desorption processes if formed on the grain surfaces. From this group of molecules only for HC(O)NH$_2$ the excitation temperature is close to the water desorption temperature, which means that it could possibly co-desorb with water ice. This is clearly not the case for CH$_3$CHO. 
Chemical models predict CH$_3$CHO and C$_2$H$_5$OH to form on the grain surfaces while a region is cold. Hence 
non-thermal desorption mechanisms are needed to 
explain their presence in the cold gas phase. The formation routes for HC(O)NH$_2$ are still debated, and both gas phase and grain surface mechanisms have been invoked \citep[e.g][]{Barone2015, Douglas2022}, and consequently we cannot conclude on its origins. Our results for HC(O)NH$_2$ are different from the results of \citet{Nazari2022} who, based on high angular resolution observations with ALMA towards a sample of hot cores, suggest that formamide traces the hottest gas phase. Towards our sample, formamide does not follow the complex cyanides that are better tracers of the hottest regions, instead it resembles more the O-bearing COMs, similarly as imaged for the hot core precursor G328.25 by ALMA in \citet{Csengeri2019}.

Finally, the third group (CH$_3$OH, CH$_3$OCH$_3$, CH$_3$OCHO) exhibits  excitation temperatures that are comparable to the desorption temperature, where both thermal and non-thermal desorption can contribute to releasing these molecules to the gas phase.

Since the temperatures of the warm gas are still moderate for our sample except for the complex cyanides, it is unclear whether the other COMs may fully thermally desorb, given 
the uncertainties about the thermal desorption temperature 
\citep{Wakelam2017, Ferrero2022}.  For instance, \citet{Collings2004} find that H$_2$O and CH$_3$OH, the most abundant molecules on the grains, start desorbing at temperatures of 120\,K and have a peak desorption efficiency at 160\,K suggesting the COMs might desorb only later, at higher temperatures. 

However, based on high-angular resolution observations with ALMA towards SgrB2(N1), \citet{Busch2022} suggest that COMs may sublimate at lower temperatures, together with water ice at about 100\,K. Lower desorption temperatures in water-poor layers could explain the presence of COMs in the gas phase below $<$100~K. 
Overall, a mixture of non-thermal desorption and an early onset of thermal desorption could explain the globally lower COM abundances in the warm gas phase compared to hot cores (see Sect.\,\ref{sec:comp_hotcorinoscores}).

\begin{table*}
    \centering
    \caption{Desorption energies and temperatures of the COMs and lighter molecules.}
    \label{tab:desorption_temperature}
    \begin{tabular}{l c c c}
    \hline
    \hline
    Molecule & Desorption energy [K] & Desorption temperature\tablefootmark{a} [K] & Reference \\
    \hline
    CH$_3$OH & 5000& 100 & \citet{Wakelam2017} \\
    CH$_3$OCH$_3$ & 3150 & 63 & \citet{Garrod2006}\\
    CH$_3$OCHO & 6300 & 126 & \citet{Garrod2006} \\
    CH$_3$COCH$_3$ & 3500 & 70 & \citet{Garrod2006}\\
    C$_2$H$_5$OH & 5400$\pm$1620 & 108$\pm$32 & \citet{Wakelam2017}\\
    CH$_3$CHO &5400$\pm$1620 & 108$\pm$32 & \citet{Wakelam2017}\\
    HC(O)NH$_2$ & 8420$\pm$960 & 168$\pm$20 & \citet{Chaabouni2018} \\
    CH$_3$CN & 4680 & 94 & \citet{Garrod2017}\\
    C$_2$H$_5$CN & 5537 & 111 & \citet{Garrod2017}\\
    C$_2$H$_3$CN & 4637 & 93 & \citet{Garrod2017}\\
    CH$_3$SH & 4000$\pm$1200 & 80$\pm$24 & \citet{Wakelam2017}\\
    \hline
    \end{tabular}
    \tablefoot{The uncertainties are given when available in the literature.
        \tablefoottext{a}{Obtained with the relation $T_{\rm des}=0.02E_{\rm des}$ \citep{Hollenbach2009}.}
    }
\end{table*}

\subsection{Chemical differentiation}
Observations suggest a dichotomy between O- and N-bearing molecules both at low \citep{Caselli1993} and high angular resolution \citep{Brouillet2013, Csengeri2019, Qin2022, Busch2022}. N-bearing molecules show different spatial distributions compared to O-bearing molecules, moreover  
molecules with a CN functional group were found at elevated temperatures compared to O-bearing molecules \citep{Widicus2017}.  We observe a similar trend in our sample where the complex cyanides trace a more compact and warmer gas component compared to the O-bearing molecules suggesting they may not originate from the same gas and may trace different physical regions.

Figures\,\ref{fig:TexVsLineWidth} and \ref{fig:TexVsSize} show that the complex cyanides are located in high temperature regions and may be exclusive tracers of the gas originating from the vicinity of the protostar unlike the O-bearing COMs, which are tracing the more extended components. 
From the complex cyanides in our sample, CH$_3$CN traces the hottest component, while C$_2$H$_5$CN and C$_2$H$_3$CN trace lower temperatures.

We compare the molecular abundances of the O-bearing molecules relative to methyl cyanide in the cold and warm components of the envelope (Fig.\,\ref{fig:COMS_relCH3CN}). 
If the formation and desorption mechanisms of the complex cyanides and the O-bearing molecules were similar in both gas components, then the ratio of these two types of molecular families would be similar both in the cold and warm gas. Our results suggest that the column density ratios between O-bearing COMs and methyl cyanide are similar within a factor of a few when comparing the cold and warm gas phase with the correction by the size, and exhibit a variation of one order of magnitude when ignoring the different size of the emission. In the cold gas phase, they are about 10-40 for CH$_3$OH, CH$_3$OCH$_3$ and CH$_3$OCHO, except for G320.23, which has a peak in the CH$_3$OCHO/CH$_3$CN column density ratio of about 300. This is due to the largely different sizes, since this peak is not apparent when considering the different source sizes. However, in the hot gas phase these values are considerably lower, and range between 1 and 40. The column density ratios between the heavy complex cyanides and CH$_3$CN are an order of magnitude smaller, and are between 1--3 in the warm gas and below 0.7 in the cold gas, except for C$_2$H$_5$CN towards G335.58 in the cold gas phase, that has a peak of about 20. Similarly as above, this peak is due to the different sizes, when correcting for the size, this ratio is about 0.7 (Fig.\,\ref{fig:COMS_relCH3CN}).
This confirms that the cold gas is poorer in heavy complex cyanides compared to the warm gas. We have an increase of the complex cyanides
towards the warmest region suggesting that complex cyanides have no or less efficient formation pathways in the cold gas and/or are not as easily desorbed. We can conclude that the complex cyanides and the O-bearing molecules trace two different layers of the envelope, where the complex cyanides are more likely to trace the immediate surroundings of the emerging protostar. On the other hand, the O-bearing molecules are more abundant compared to the heavy cyanides in the cold gas.

Another argument in favor of this interpretation is that we detect acethaldehyde only in the cold component of the envelope with an excitation temperature of 
25\,K. CH$_3$CHO is commonly observed in dark clouds at low temperature \citep{Gratier2016} and at higher temperature in hot corinos and hot cores \citep{Widicus2017}. However, we find here that it shows the opposite behaviour compared to the cyanides, which are detected only in the warm gas.  
Towards the hot core of SgrB2(N1) \citet{Busch2022} find that the abundance of CH$_3$CHO increases above 100~K, while the abundances of other O-bearing COMs drop. It is also possible that our sensitivity is insufficient to detect a hot component of CH$_3$CHO.

\subsection{Molecular composition of the envelope}\label{sec:molecular_composition}

\begin{figure*}
    \centering
    \includegraphics[width=1\linewidth]{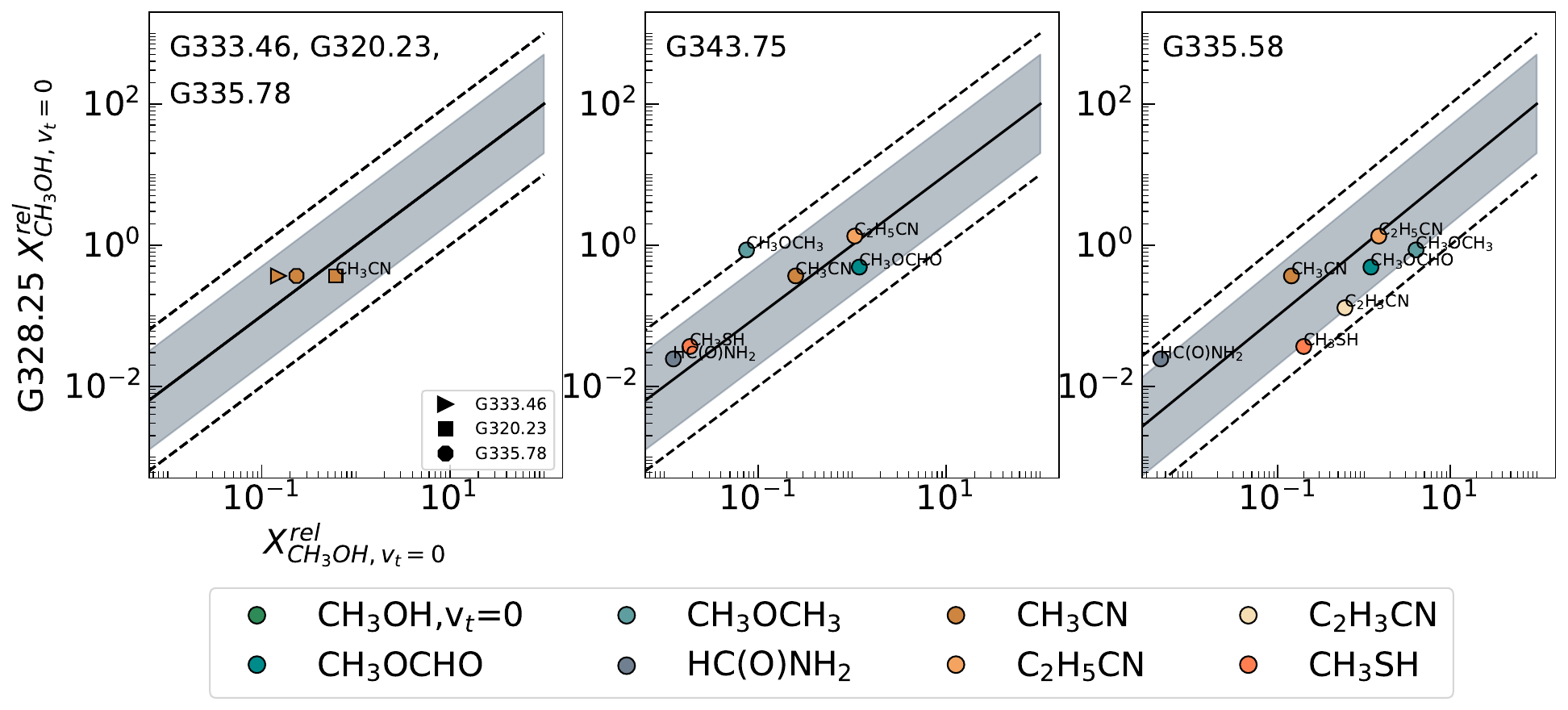}
    \caption{Comparison of molecular abundances relative to methanol (CH$_3$OH, $v_t$=0) in the warm gas phase of our sample where G328.25 is chosen as a reference. The left panel shows G333.46, G320.23, and G335.78, the middle panel G343.75, and the right panel G335.58. The black solid line represents the one-to-one relation. The grey shaded region corresponds to a factor 5 scatter from the one-to-one relation. The dashed lines represent a factor 10 scatter from the one-to-one relation. Each molecule is displayed in a different color.}
    \label{fig:comp_G328p25}
\end{figure*}

We compare the COMs in the warm gas phase of the sample using molecular abundances relative to methanol in Fig.\,\ref{fig:comp_G328p25}. We use  G328.25 as a reference, and display G333.46, G320.23 and G335.78 on the same panel since these sources have the fewest molecules detected in the warm gas. 
As discussed in Sect.\,\ref{sec:abund_COMs}, there are differences among the sample in terms of detection of the different COMs in the warm component of the envelope.
However, the molecular abundances relative to CH$_3$OH show strong similarities. The relative molecular abundances are similar for the sample within a factor 5 for all the molecules, except for CH$_3$OCH$_3$ towards G343.75, for which we find a factor 10 difference compared to G328.25. This suggests that the chemistry at the origin of COMs in the warm gas phase is similar for all the sources, which is in line with results obtained on larger samples of COMs in high-mass star forming regions \citep{Coletta2020, Nazari2022}. However, our results point out noticeable variations  in  the overall quantity of the COMs. The dispersion of relative abundances within a factor of a few is consistent with chemical models by \citet{Garrod2022}; in fact these models have a larger dispersion in the peak predicted molecular abundances when considering different mechanisms, than their final model using different warm-up time-scales.
Their final models considering varying warm-up time-scales  produce peak molecular abundances for the COMs discussed here that vary only by a factor of 5 for each molecule, respectively (see Sect.\,\ref{sec:comp_models} for a more detailed comparison to models).

Overall, we found that the observed differences in terms of COM abundances in the warm gas are unlikely to be explained only by sensitivity limitations (see Sect.\,\ref{sec:COMs_properties}). Instead they are more likely to point to intrinsic variations in the physical conditions and evolutionary stage of the sample.

\subsection{COMs tracing an evolutionary sequence?}

Our sample consists of infrared quiet massive clumps that are expected to form high-mass stars and clusters \citep[c.f.][]{Csengeri2017}. They exhibit  $L_{\rm bol}\lesssim2\times10^4$\,\lsol\ that corresponds to at least one order of magnitude lower bolometric luminosities compared to Galactic {\hii} regions, suggesting that our sample represents an earlier evolutionary stage.  This is consistent with the lack of strong radio continuum emission from ionised gas \citep{Csengeri2017}, and our findings confirm that the bulk of the gas is cold (see Sect.\,\ref{sec:dichotomy_NO}). 
The low bolometric luminosity 
of our sample suggests that protostellar embryos have masses at most of $\lesssim$16--20\,\msol\ according to models of protostellar evolution \citep{Hosokawa2009}, because more massive protostellar embryos would exhibit higher luminosities.

In Sect.\,\ref{sec:molecular_composition} we found that while the molecular reservoir of COMs in the cold gas phase is similar for all sources, there is significant difference, in particular, in the quantity of complex cyanides in the warm gas. 

Consequently, variations in the molecular emission 
are expected to be intrinsic to the sources reflecting their physical conditions related to age, protostellar content or chemical history. 
A more massive protostar would imply higher 
temperatures in its vicinity leading to a larger extent of COM emission, and likely more elevated mass weighted temperatures. Alternatively, an age  difference would imply longer warm-up phase, potentially leading to more efficient sublimation of ices and gas phase reactions to enrich the COM content of the warm gas phase. While the age and mass of the protostar are related to the accretion history, both of these effects may manifest as larger amount of COMs in the gas phase.  
We propose an evolutionary sequence for the sample based on the quantity of complex cyanides in the hot gas phase (Fig.\,\ref{fig:add_col_dens}, lower right panel) and considering the mean excitation temperature: 
\begin{multline}
    \rm{G333.46 \rightarrow G320.23 \rightarrow G335.78 \rightarrow} \\ \rm{\rightarrow G343.75 \rightarrow G328.25 \rightarrow G335.58}
\end{multline}

This sequence could also equally well reflect an increasing trend in the current protostellar mass. In this picture,
G333.46 is the coldest, and hence probably the youngest object of the sample, and G335.58 the warmest and most evolved source. In fact, we find that such a purely temperature and COM based sequence corresponds well to the estimated bolometric luminosities (see more details in Sect.\,\ref{sec:lum}). A purely empirical evolutionary stage is often established using the $L_{\rm bol}$/$M$ ratio as an indicator \citep[e.g.][]{Molinari2008}. Our sample has an $L_{\rm bol}$/$M$ range between 2.5 and 20.3, consistent with a globally young evolutionary stage. 

\subsection{Comparison to chemical models}
\label{sec:comp_models}
\begin{figure}
    \centering
    \includegraphics[width=1\linewidth]{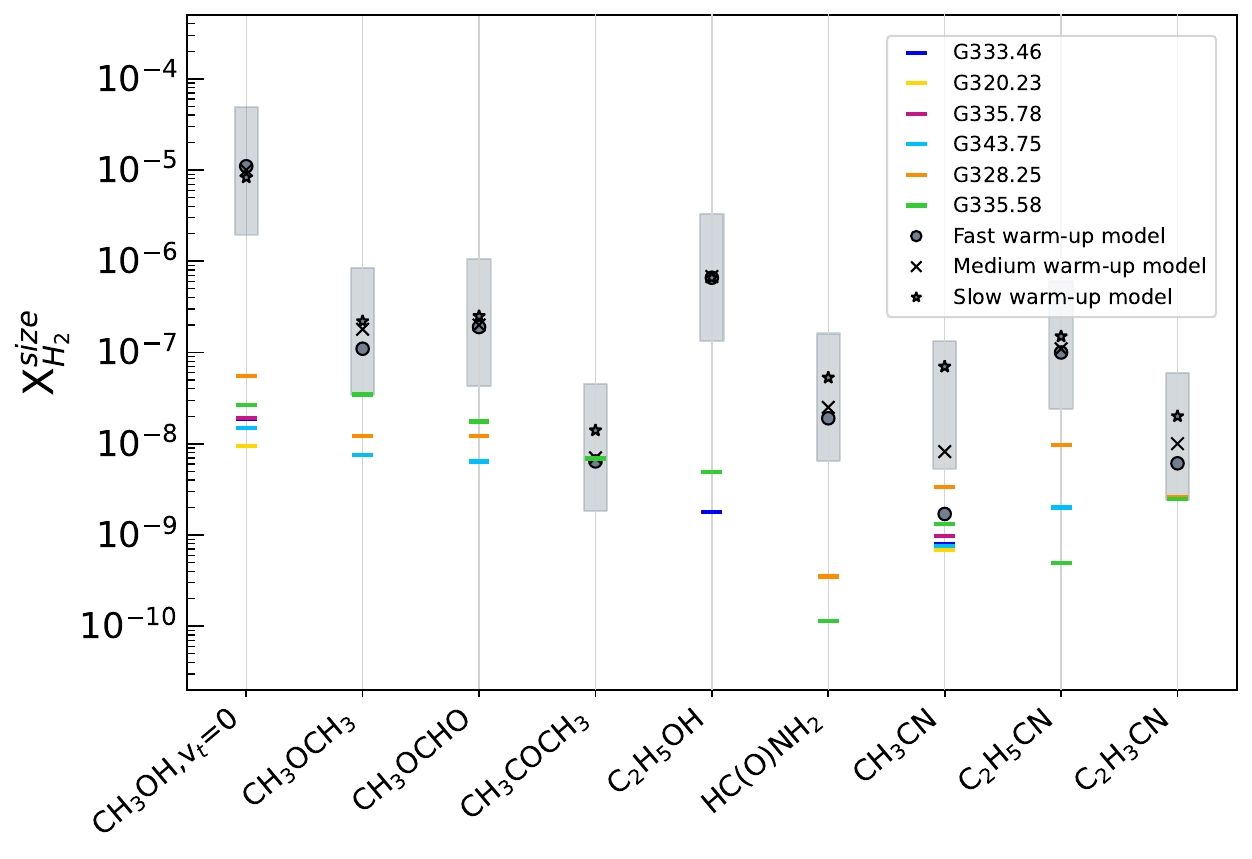}
    \includegraphics[width=1\linewidth]{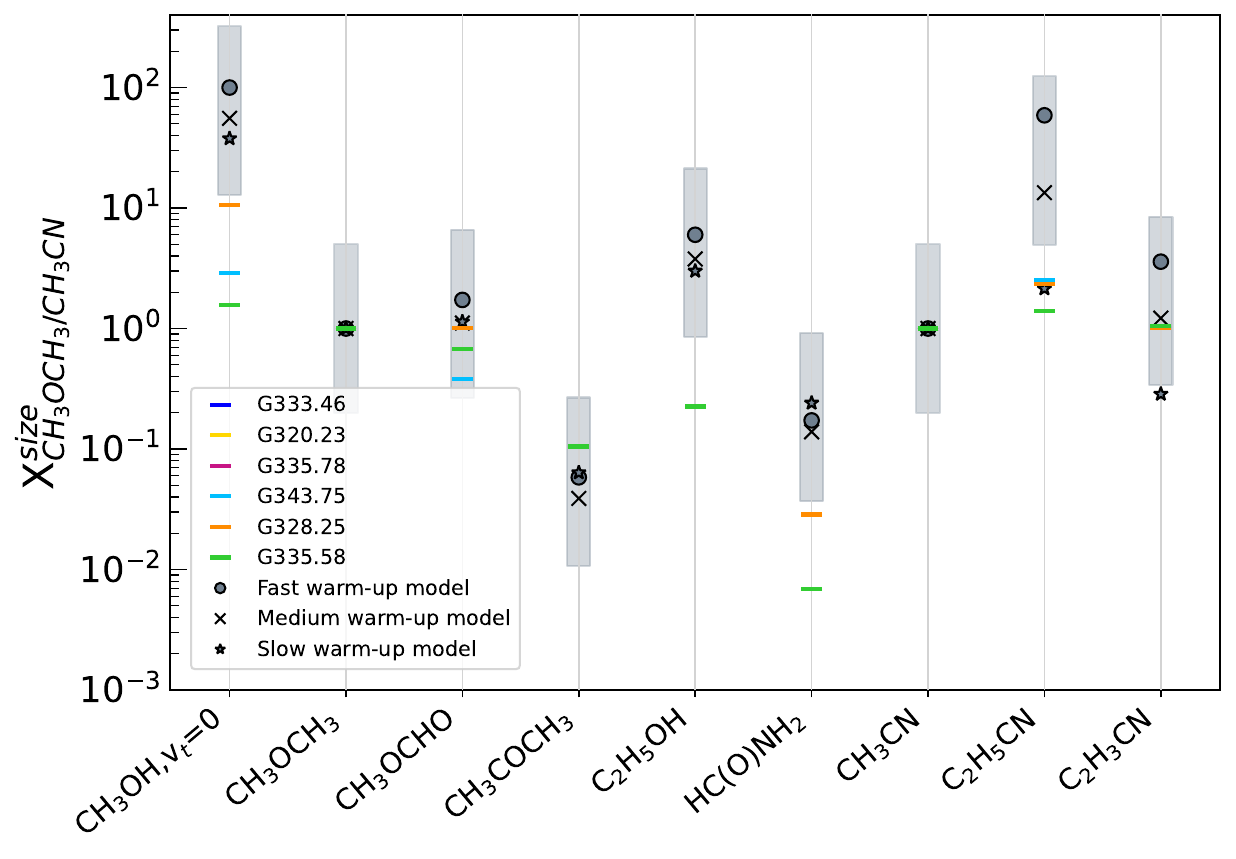}
    \caption{{\it Upper panel:} Comparison between the abundances relative to H$_2$ in our objects and the peak surface abundances in the models from \citealt{Garrod2022} (Table 17). {\it Lower panel:} Comparison between the abundances relative to CH$_3$OCH$_3$ for O-bearing molecules and CH$_3$SH and relative to CH$_3$CN for cyanides. Each source is represented by a different color. The models are represented by three symbols. The grey area represents a factor 5 scatter from the average model values.}
    \label{fig:comp_models}
\end{figure}

Models of the chemical evolution of the gas and grains can predict the expected molecular abundances as a function of physical parameters that evolve with time. The time evolution is typically linked to the emerging heating source leading to a gradual increase in temperature. Comprehensive models of the chemical evolution of hot cores treat the gas-grain chemistry, as well as chemical reactions taking place in the bulk-ice. We use as a reference the latest models of \citet{Garrod2022} that also include non-diffusive reaction processes in  the ice on grain surfaces and compare our observational results from the hot gas to these model calculations in Fig.\,\ref{fig:comp_models}.

We indicate a dispersion by a factor of five of the average stage 2 peak gas-phase molecular abundances of the three models (slow, medium and fast warm-up time-scale) and notice that they show less dispersion compared to our observational results. We compare 
our observed molecular abundances relative to H$_2$, where $N_{H_2}$ is a beam averaged column density from Table~\ref{tab:source_coord} based on the 870~$\mu$m ATLASGAL observations, and is scaled to the size of the molecular emission discussed here (Fig.\,\ref{fig:comp_models}, upper panel; see Sects\,\ref{sec:H2} and \ref{sec:mol_abundances} for more details on the estimates of H$_2$ column densities). We discuss formamide together with the O-bearing COMs due to their similar behaviour.  
We immediately notice a significant discrepancy between the model predictions and the estimated CH$_3$OH abundances from our observations that are almost about two orders of magnitude. Similarly large discrepancies of two to three orders of magnitude are found for C$_2$H$_5$OH and HC(O)NH$_2$ as well. For the other O-bearing COMs, CH$_3$OCH$_3$ and CH$_3$OCHO we find that their molecular abundances  are typically one order of magnitude below the model predictions. However, we find that CH$_3$COCH$_3$ is the only O-bearing COM well reproduced by the models. For complex cyanides the observational estimates are  about an order of magnitude or more below the model values. Only CH$_3$CN can be reproduced by the fast warm-up model.
Overall this picture is consistent with a partial thermal desorption scenario discussed in Sect.\,\ref{sec:discuss_temp}. The differences between the models and the observational results could be explained by the difference in the temperature range, the estimated gas phase temperatures of our sample are clearly below the temperature corresponding to the peak gas phase abundances used from the model predictions. This is because in the considered temperature range, between 80 and 110\,K, the model predictions vary up to 4 orders of magnitude for most of the COMs in the models of \citet{Garrod2022} that prevents us from making more precise comparisons at the corresponding temperatures. Furthermore, our estimation of $N(H_2)$ may not be robust at the scales discussed here.

While discussing the molecular abundances relative to H$_2$ gives an estimate about the quantity of molecules produced, comparing abundances relative to other molecules provides a hint on the chemical processes. For this purpose, in
Fig.\,\ref{fig:comp_models} (lower panel) we show the comparison of the  abundances of O-bearing molecules relative to CH$_3$OCH$_3$ and complex cyanides relative to CH$_3$CN. We choose CH$_3$OCH$_3$ over CH$_3$OH due to the large discrepancy between the models and observational estimates discussed above. For O-bearing molecules,
we find that the observed CH$_3$OH abundance is still by a factor of $\sim$60 lower compared to the model results, suggesting that we lack CH$_3$OH in the warm gas phase of our objects. However, 
for other O-bearing COMs, such as CH$_3$OCH$_3$, CH$_3$OCHO, and CH$_3$COCH$_3$, where the estimated gas temperature is similar to the desorption temperature, we are very close to the model predictions within a factor of 5. We still notice a discrepancy between the model and the observed values for C$_2$H$_5$OH and HC(O)NH$_2$ (corresponding to the molecules where the gas temperature is below the desorption temperature, see Sect\,\ref{sec:discuss_temp}). Overall, we find the most noticeable differences for the R--OH molecular families, and formamide. 
 
For complex cyanides here we use CH$_3$CN as a reference molecule and find that the relative abundance of C$_2$H$_3$CN corresponds well to that of the model predictions. The discrepancy for C$_2$H$_5$CN is larger than a factor of five, however we see a larger dispersion in the model predictions for this molecule. Overall, our comparison suggests that for species where thermal desorption is already started (CH$_3$OCH$_3$, CH$_3$OCHO, and CH$_3$COCH$_3$, and complex cyanides) their chemistry is rather well reproduced by the models of \citet{Garrod2022}.  Concerning methanol, ethanol and formamide, either the chemical network is incomplete, or we simply trace a physical/chemical stage where significant gas phase abundance variations do not allow us to perform the appropriate comparison to the model predictions.

\section{Comparison to hot cores and hot corinos: a new evolutionary stage preceding the emergence of hot cores} \label{sec:sec_comp}

Chemically rich star forming regions are often characterised as hot cores corresponding to massive cores and clumps \citep[e.g.][]{Kurtz2000, Cesaroni2005}, or hot corinos for their low-mass counterparts \citep{Ceccarelli2000,Bottinelli2004}. Hot cores and hot corinos are rich in saturated hydrocarbons, and while both types of object exhibit significant diversity in their overall molecular content \citep{Widicus2017,Lopez-Sepulcre2017, Belloche2020}, the recent study by \citet{Jorgensen2020} suggests however that they may be similar in terms of their molecular reservoirs for certain COMs. 
In Sect.\,\ref{sec:discussion} we discussed the in-depth characterisation of COMs towards our sample, and here we aim to compare their COM properties to that of hot cores and hot corinos. 
Our targets being infrared quiet massive clumps correspond to an early evolutionary stage with a low bolometric luminosity ($<$2$\times10^4$L$_\odot$) that is 1--2 orders of magnitude below that of the most prominent Galactic hot cores (see Table~\ref{tab:literature} for examples). The mass reservoir of our sample of clumps is sufficient to suggest that they may evolve into hot cores forming high-mass stars and clusters \citep[c.f.][]{Csengeri2017}.
Their luminosities are, however, at least 2--3 orders of magnitude higher compared to that of hot corinos, which is likely due to higher accretion rates and more massive deeply embedded protostars.

\subsection{Size and temperature of the warm component}
The physical characteristics of our sample corresponds to that of massive clumps, and \citet{Csengeri2017b} demonstrates that they host highly concentrated single massive dense cores. Comparing the properties of their heated regions to hot cores and hot corinos, we first focus on discussing the size (corresponding to the emitting region of COMs) 
 and temperature of the warm gas phase. For example, early observations of hot cores discovered massive clumps with a significant amount of hot gas with temperatures above $150$\,K and sizes $>$0.2\,pc \citep{Kurtz2000, Cesaroni2005}. High angular resolution studies, however, reveal more compact massive cores rich in COMs, in particular recent ALMA observations identified compact hot cores, with sizes $> $3000-5000\,au \citep{SanchezMonge2014, Ginsburg2018, Bonfand2017, Colzi2021}.  
At these size-scales such objects are expected to correspond to individual protostellar envelopes, typically forming high-mass (proto)stars, although fragmentation at even smaller scales has also been observed \citep[e.g.][]{Maud2017, Olguin2021}.

Hot molecular cores have been explained by the impact of radiative feedback from the protostellar object leading to elevated temperatures and thus sublimation of ices and subsequent gas phase reactions that lead to a molecular complexity represented by a large amount of COMs \citep{Charnley1997, Garrod2006}. Considering their physical size, hot corinos are scaled down versions of hot cores, corresponding to at least one order of magnitude smaller regions of 100-300\,au \citep[e.g.][]{Bianchi2020, Belloche2020} that is broadly consistent with their lower bolometric luminosities typically corresponding to low- to intermediate mass protostars. 

As shown in Sect.\,\ref{sec:dichotomy_NO}, molecules may trace different layers of the heated gas phase, and hence 
provide different values for the temperature estimates of hot cores and hot corinos. 
In this context, our sample exhibits maximum estimates of gas temperatures of 70 to 150\,K on 6000\,au scales with a median of $\sim$100~K, while the most compact and warmest regions are traced by CH$_3$CN reaching 200 and 250\,K temperatures (see Table\,\ref{tab:temp_stats}) at scales of 2000-6000\,au. These high temperatures at the smallest scales are consistent with recent results for gas temperatures of hot cores that are $>$150\,K \citep{Bonfand2017, Colzi2021,Nazari2022}.  
Similarly, hot corinos exhibit temperatures typically $>$100~K on small scales \citep[e.g][]{Jorgensen2016, Belloche2020}. 
Overall, this suggests that the bulk of the warm gas phase is somewhat less hot towards our objects compared to hot cores and hot corinos, while the temperatures of complex cyanides are similar for both hot cores, hot corinos and our sample.

Considering
the size of the emitting region of COMs, we estimate 
more compact sizes of $< 6000$~au for our sample compared to hot cores although 
recent ALMA observations reveal rather compact hot core sizes reaching few thousands of au \citep{Allen2018, Bonfand2017, Ginsburg2018}. \citet{Bonfand2023} discuss a sample of hot core candidates from the ALMA-IMF large program \citep{Motte2022} using CH$_3$OCHO as a tracer with multiplet transitions of $E_{\rm up}/k=109$~K, and find a large dispersion of sizes with the largest sources extending to 13000 au.  
As shown in \citet{Bonfand2023}, for hot cores with $L_{\rm bol}\sim10^4$~\lsol\ the 120~K radius is around 800~au following the calculations of \citet{Wilner1995}. Our $FWHM$ COM source size are typically by a factor of 2--3 larger that is a reasonably good agreement considering the uncertainties of the models, and our temperature and size estimates. 

Considering that our average temperatures are lower, and the sizes are more compact than that of hot cores we suggest that our sample of infrared-quiet clumps corresponds to an evolutionary stage prior to the emergence of hot cores. From the chemistry point of view, 
we argued that our sources exhibit emission from several molecules of COMs below the thermal desorption temperature  suggesting that these COMs may still be sticking to the ices on grain surfaces. This stage could therefore be different from radiatively heated hot cores, where a substantial fraction of ices has already evaporated and gas-phase reactions may have taken over the chemical evolution of the gas.

\subsection{Molecular composition: hot core or hot corino-like?} 
\label{sec:comp_temp}
We compare the molecular composition of our sample to that of hot cores and hot corinos from the literature, and
discuss differences in terms of the presence or absence of COMs. 
We list the physical properties of the sources and the corresponding column densities used for the comparison in Table\,\ref{tab:ref_comp}.

\subsubsection{The total amount of detected COMs}
\begin{figure}
    \centering
    \includegraphics[width=1\linewidth]{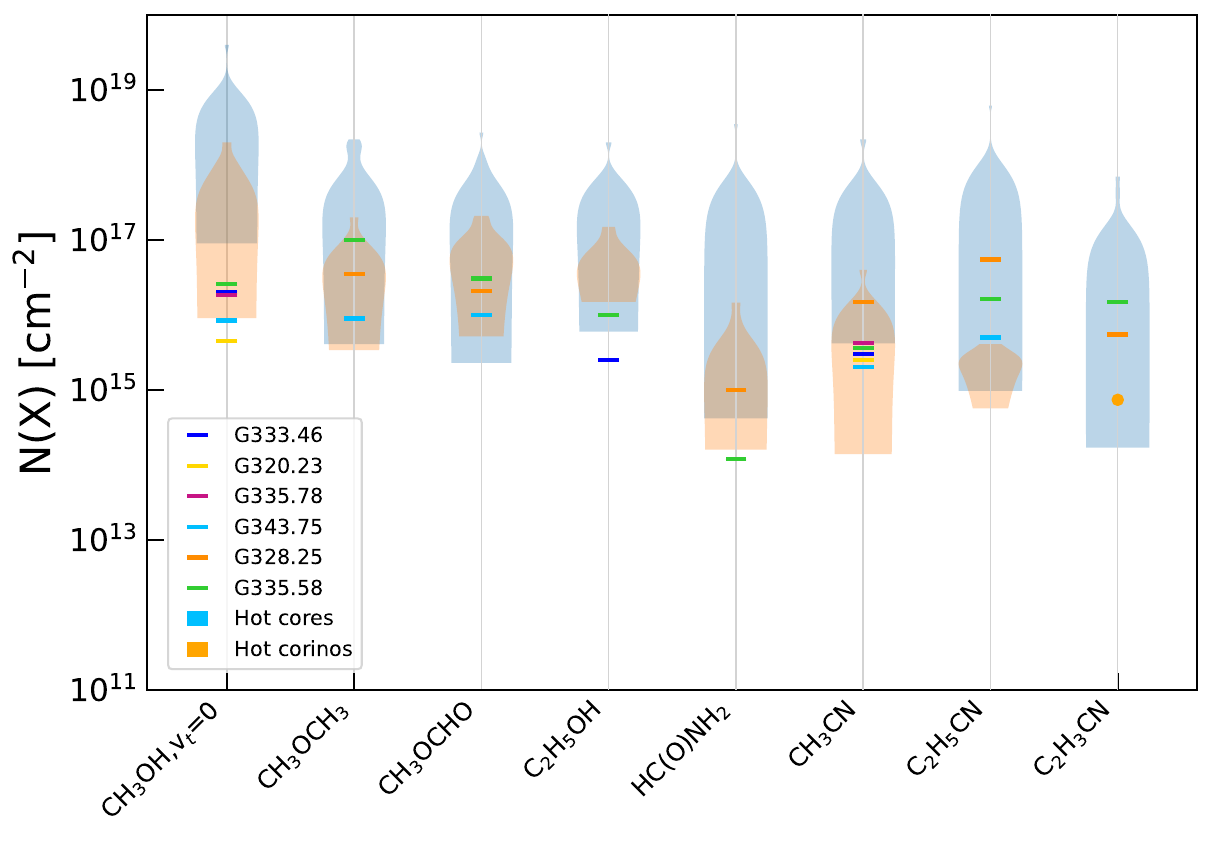}
    \caption{Comparison of the column densities in the warm gas between our sources and a sample of hot corinos \citep{Jorgensen2016, Belloche2020, Yang2021} and hot cores \citep{Belloche2016, Belloche2017, Rolffs2011, Allen2017, Bonfand2017, Law2021, Bogelund2019, Baek2022}. The blue and orange areas represent the hot core and hot corino distributions (respectively).}
    \label{fig:Comp_Dens}
\end{figure}

We first discuss the  peak column density of COMs in the compact warm gas phase from the literature and our sample (Fig.\,\ref{fig:Comp_Dens}), since it is a measure of the quantity of COMs in the gas phase. Inferring the column density strongly depends on the size of the emitting region, and its ratio to the beam size. Since the heated region in our sample is compact, where available, we use interferometric measurements from the literature for both hot corinos and hot cores (see more in App.\,\ref{tab:ref_comp}).

Due to the differences in the physical conditions (i.e. density and surface density), hot cores and hot corinos exhibit substantial differences in the peak column densities of their COMs,  hot corinos exhibit about two orders of magnitude lower peak column densities for almost all COMs discussed here. 
While the diversity in the molecular composition of hot cores has been found to be significant \citep{Widicus2017}, systematic studies of hot corinos exhibit only an order of magnitude dispersion in the column densities of O-bearing COMs \citep{Belloche2020} that suggest similarities in their molecular composition. 
We show in Fig.\,\ref{fig:Comp_Dens} that the column densities of certain COMs in the warm gas phase of our sample (CH$_3$OH, CH$_3$OCH$_3$, HC(O)NH$_2$ and CH$_3$CN) are more similar to that of hot corinos, and are on average one order of magnitude below that of hot cores. However, the column density estimates of C$_2$H$_5$CN and C$_2$H$_3$CN resemble that of hot cores. In particular, detection of C$_2$H$_3$CN towards hot corinos remains scarce and has only been reported towards IRAS16293B from our Table\,\ref{tab:ref_comp}. 

\subsubsection{Molecular abundances}
\label{sec:comp_hotcorinoscores}
Due to the differences in the physical conditions of 
hot cores and hot corinos,
and because the H$_2$ volume and column densities of hot cores are a few orders of magnitude higher compared to hot corinos, the peak column density measurements of molecules, although directly measured quantities, may not reflect real chemical similarities or differences. For that we also need to discuss here the relative molecular abundances of the observed COMs.
For this purpose we use here molecular abundances relative to  dimethyl ether for the O-bearing COMs, where we expect that the column densities are more reliable than that of CH$_3$OH that may suffer from optical depth effects treated in different ways in the literature. Since we only have a few complex cyanides detected, we can only use CH$_3$CN as a reference, although that may also be affected by optical depth issues.
We show these results towards three sources of our sample, G343.75, G328.25 and G335.58 with the largest number of COM detections in the warm gas phase.

\begin{figure*}[!ht]
    \centering
    \includegraphics[trim=0cm 0cm 0cm 10cm, clip, width=0.9\linewidth]{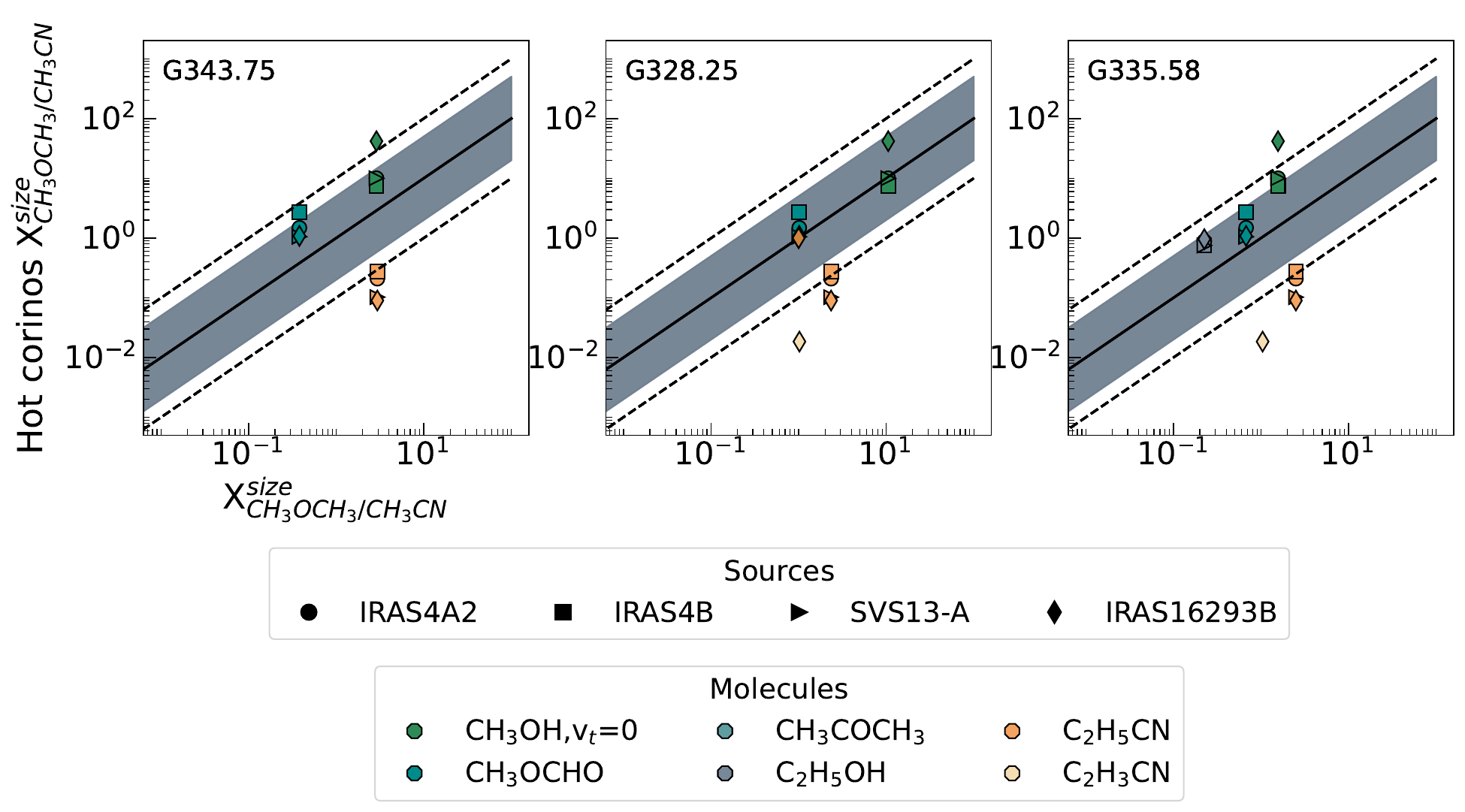}
    \caption{Comparison between the molecular abundances relative to CH$_3$OCH$_3$ for O-bearing molecules and CH$_3$CN for cyanides of our objects and four hot corinos: IRAS4A2, IRAS4B, SVS13-A, and IRAS16293B. The molecular abundances are corrected by the size of the emitting region. Each source is indicated with a different symbol. The black solid line represents the one-to-one relation. The grey shaded region corresponds to a factor 5 scatter from the one-to-one relation. The dashed lines represent a factor 10 scatter from the one-to-one relation.}
    \label{fig:comphotcorino}
\end{figure*}

Figure\,\ref{fig:comphotcorino} shows the comparison of relative COM abundances of our sample to hot corinos from \citet{Belloche2020}. 
We find a relatively good correlation within roughly one order of magnitude between our objects and hot corinos for the O-bearing COMs. However, we notice a strong difference for the N-bearing COMs, 
the complex cyanides. Towards those objects of our sample (G328.25, G335.58 and G343.75), where ethyl and vinyl cyanides are detected in the warm phase, they exhibit higher relative abundances by at least an order of magnitude compared to hot corinos.

\begin{figure*}[!ht]
    \centering
    \includegraphics[trim=0cm 0cm 0cm 10cm, clip, width=0.9\linewidth]{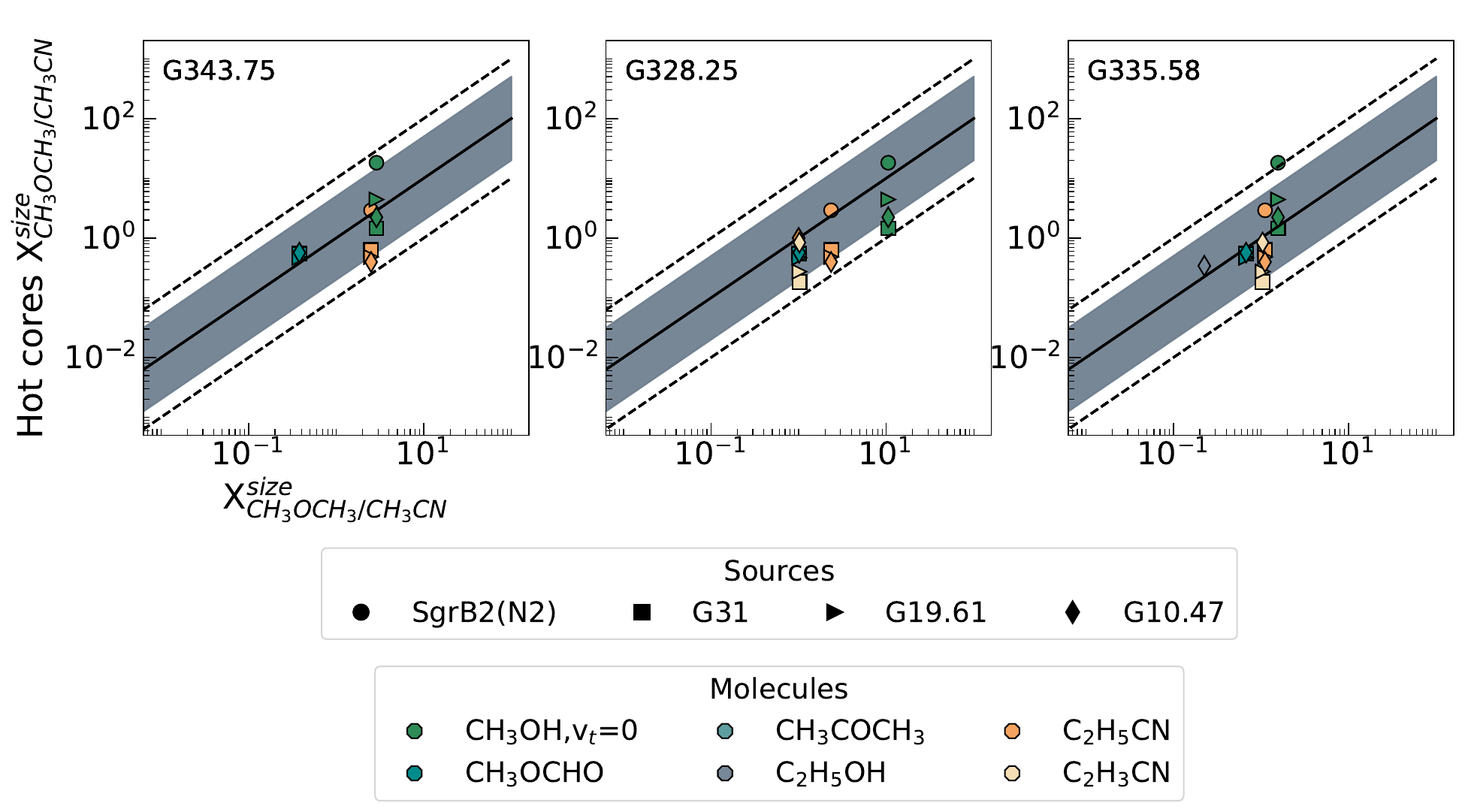}
    \caption{Comparison between the molecular abundances relative to CH$_3$OCH$_3$ for O-bearing molecules and CH$_3$CN for cyanides of our objects and four hot corinos: G19.61, G10.47, G31, and SgrB2(N2). The molecular abundances are corrected by the size of the emitting region. Each source is indicated with a different symbol. The black solid line represents the one-to-one relation. The grey shaded region corresponds to a factor 5 scatter from the one-to-one relation. The dashed lines represent a factor 10 scatter from the one-to-one relation.}
    \label{fig:comphotcore}
\end{figure*}

Concerning the sample of hot cores, we find that they exhibit a dispersion in O-bearing COMs comparable to the hot corinos and within a factor 5 (Fig.\,\ref{fig:comphotcore}). When comparing to our sample, we find about an order of magnitude dispersion  in the relative molecular abundances. 
Nevertheless, we find a generally similar trend in the relative molecular abundances of O-bearing COMs that is consistent with the results of \citet{Jorgensen2020}. For hot cores, we have a larger sample of complex cyanides detected, and they also show a similar scaling like the O-bearing COMs that is markedly different for the comparison with the hot corino sample.

Hot cores, hot corinos and our sample exhibit similar relative molecular abundances of O-bearing COMs within two orders of magnitudes, as found for SgrB2(N) and IRAS16293 by \citet{Jorgensen2020}. We note, however, a difference for complex cyanides, where our sample, similarly to hot cores, exhibit  higher molecular abundances compared to hot corinos. Overall this suggests similar chemical origin for O-bearing COMs over all mass ranges, consistent with the results of \citet{Colzi2021}. On the other hand, complex cyanides may reflect the difference physical conditions in the immediate vicinity of protostars.

\subsubsection{The early warm-up phase}

 The evolutionary sequence for the emergence of high-mass stars and clusters is debated due to the fact that typically multiple stages of the process are observed simultaneously in forming clusters. Nevertheless major characteristic stages have been identified and characterised, corresponding to massive dense cores \citep[c.f.][]{Motte2007}, hot cores and \uchii\ regions \citep[c.f.][]{Kurtz2000}, the earliest phases being the most poorly studied.
 
\begin{figure*}
    \centering
    \includegraphics[width=0.6\linewidth]{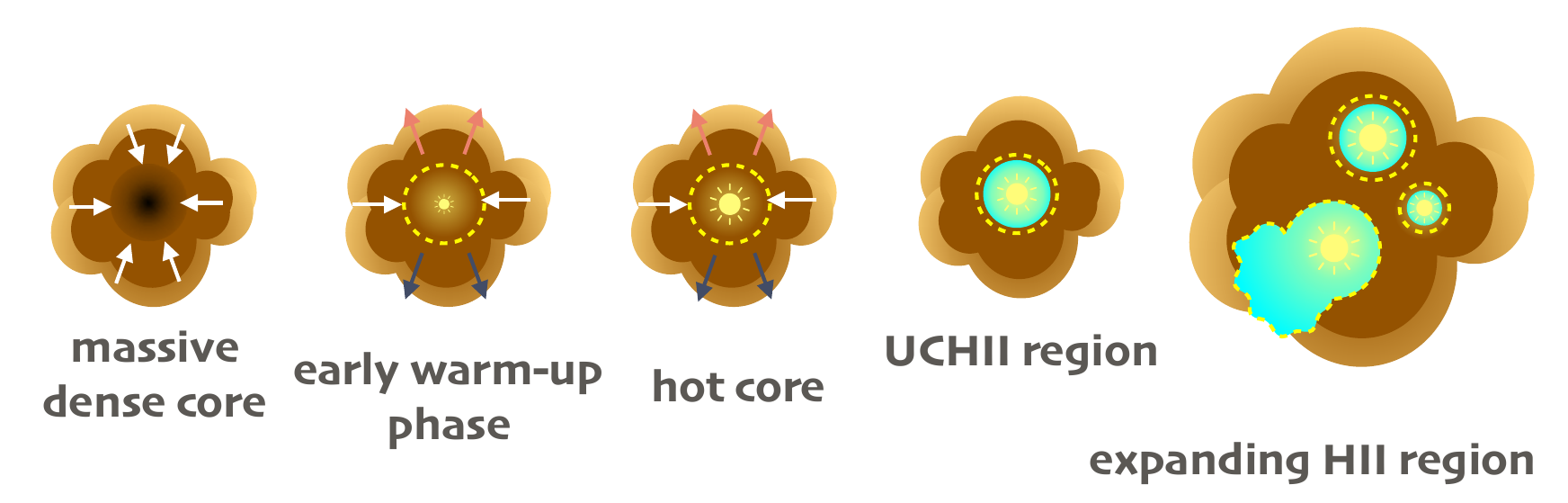}
    \caption{Observationally constrained evolutionary sequence of high-mass star formation, depicting  from: cold massive dense cores, early warm-up phase, hot cores, and {\uchii} regions to {\hii} regions representing the most evolved stages.}
    \label{fig:evolution_sequence}
\end{figure*}

Our sample represents
  low luminosity infrared-quiet clumps dominated by single massive envelopes, all of which drive outflows \citep[c.f.][for G328.25 as an example]{Csengeri2018}. In this picture of evolutionary stages, our objects are  between the cold gas dominated massive dense core stage and hot cores that contain already a significant amount of hot gas characterised by COMs (Fig.\,\ref{fig:evolution_sequence}).  We suggest that they represent an evolutionary stage characterized by the early emergence of a hot core, where the bulk of the gas is still cold, and a compact warm component rich in COMs emerges. The compact heated part of the core is associated with warm temperatures, while the molecular composition of the gas originates from a combination of non-thermal desorption processes, partial thermal desorption or gas-phase reactions. As discussed in Sect.\,\ref{sec:comp_hotcorinoscores} this stage is chemically similar to hot cores and hot corinos with similar molecular abundances for O-bearing COMs. Complex cyanides have, however, similar molecular abundances to hot cores.  
 We can thus conclude that there is a chemically active phase preceding the emergence of hot cores, and suggest that this corresponds to the early warm-up phase  (Fig.\,\ref{fig:evolution_sequence}). This stage is well characterized by luke-warm O-bearing COMs
with a potentially similar chemistry as in hot corinos and hot cores, whereas complex cyanides could represent a distinguishing tracer between hot corinos and high-mass star forming envelopes. Recent results from the ALMA-IMF large program using CH$_3$OCHO lines (with $E_{\rm up}/k$) identify a large sample of chemically active cores, many of them could correspond to this early warm-up phase chemistry \citep{Bonfand2023}.

\subsection{Correlation with luminosity?}\label{sec:lum}
\begin{figure}
    \centering
    \includegraphics[trim=0cm 4.5cm 0cm 0cm, clip, width=0.95\linewidth]{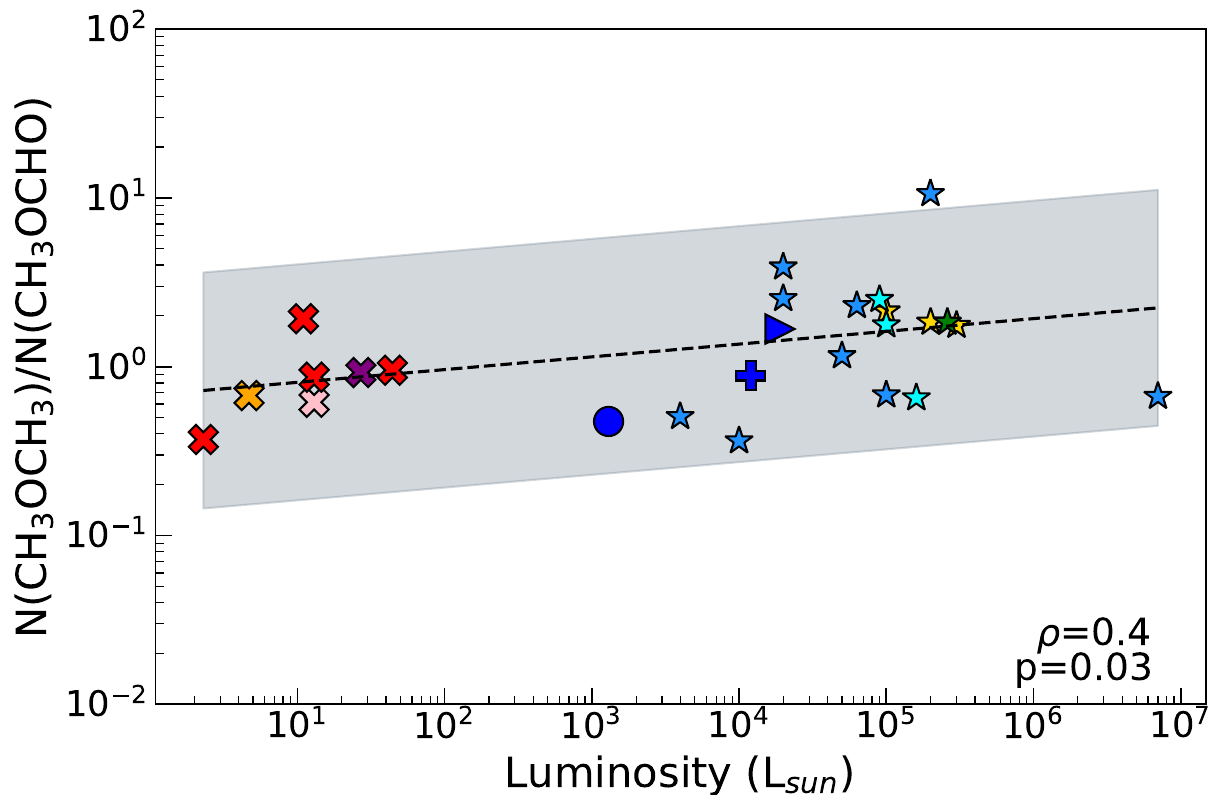}
    \includegraphics[trim=0cm 4.5cm 0cm 0cm, clip, width=0.95\linewidth]{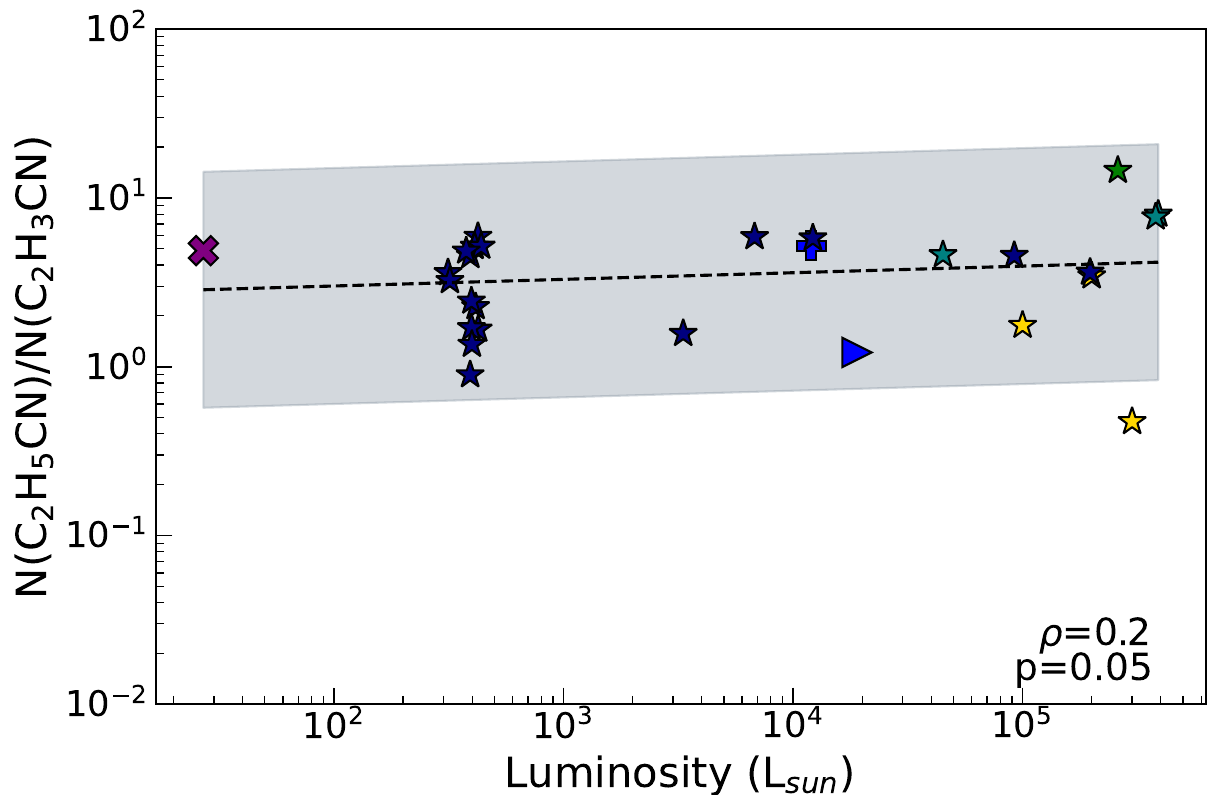}
    \includegraphics[width=0.95\linewidth]{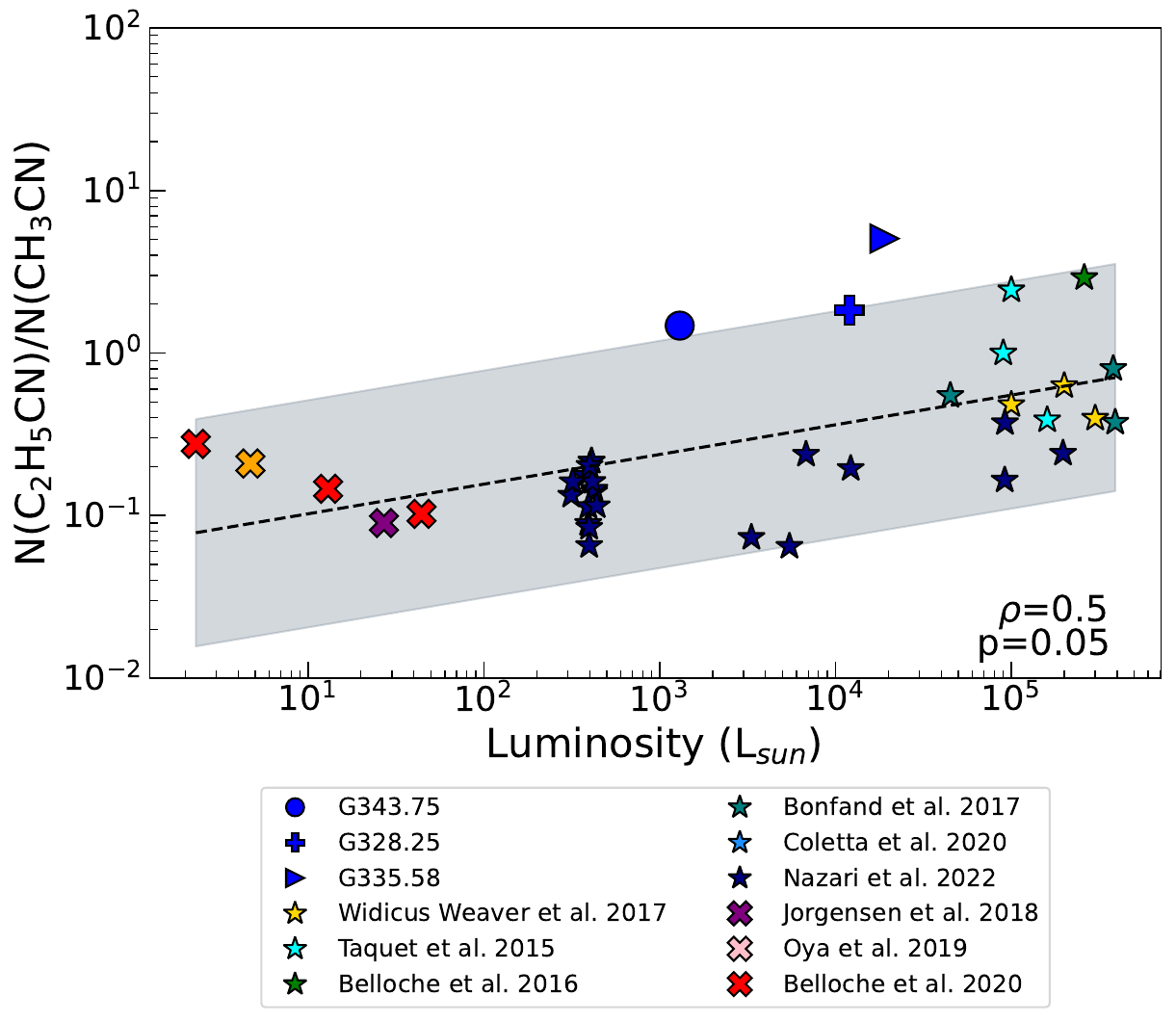}
    \caption{{\it Top panel:} Abundances of CH$_3$OCH$_3$ relative to CH$_3$OCHO versus the bolometric luminosity. {\it Middle panel:} Abundances of C$_2$H$_5$CN relative to C$_2$H$_3$CN versus the bolometric luminosity. {\it Bottom panel:} Abundances of C$_2$H$_5$CN relative to CH$_3$CN versus the luminosity. The dashed black line represents the fit to the data and the grey shaded area represents a factor 5 scatter from the fit. Our data points are represented in blue, the high-mass protostars \citep{Belloche2016,Widicus2017,Coletta2020,Nazari2022} are represented with a star. The low-mass protostars are presented with crosses \citep{Oya2017,Jorgensen2018,Belloche2020}.}
    \label{fig:luminosity}
\end{figure}

In order to elucidate the origin of the observed similarities and differences for our sample and hot cores, hot corinos, we compare the relative molecular abundances 
to their bolometric luminosity ($L_{\rm bol}$) estimates.  
In Fig.\,\ref{fig:luminosity}, we show molecular column density ratios of several pairs of COMs as a function of $L_{\rm bol}$ 
in order to investigate potential chemical variations. 
As discussed in Sect.\,\ref{sec:full_line_prof}, not all types of COMs are detected towards the sample in the warm component, and here we focus only on correlations with the warm gas phase tracing the immediate vicinity of the protostar. We compare our measurements to values from the literature, primarily relying on low-mass objects using the CALYPSO sample \citep{Belloche2020}, IRAS16293B \citep{Jorgensen2020}, L483 \citep {Oya2017}) and hot cores from \citealp{Widicus2017, Coletta2020, Nazari2022} and \citealt{Belloche2016, Bonfand2019} (for Sgr B2(N2), SgrB2(N3), SgrB2(N4), and SgrB2(N5)).

We first look at the $N$(CH$_3$OCH$_3$)/$N$(CH$_3$OCHO) ratio versus $L_{\rm bol}$ and compute a Pearson correlation coefficient ($\rho$) of 0.4 with a $p$ value of 0.03 that is statistically not robust, nevertheless it shows a trend of an increasing ratio
as a function of luminosity (Fig.\,\ref{fig:luminosity}, top panel).
The ratio of CH$_3$OCH$_3$ and CH$_3$OCHO has been found to be relatively  independent of luminosity for low-, intermediate and high-mass envelopes by \citet{Ospina-Zamudio2018, Coletta2020}, albeit with a larger scatter towards high-mass sources. A flat distribution of the CH$_3$OCH$_3$ and CH$_3$OCHO ratio over orders of magnitude in  luminosity \citep{Coletta2020} is  interpreted as an indication for similar formation mechanisms for these species. 
This is in line with the results of chemical models that predict CH$_3$OCH$_3$ and CH$_3$OCHO to be chemically linked \citep[e.g][]{Garrod2006}.
Our results are in line with these results, although they suggest that the production of CH$_3$OCH$_3$ or the destruction of CH$_3$OCHO is more efficient towards the higher luminosity objects.

We also investigated the column density ratio of complex cyanides ($N$(C$_2$H$_5$CN)/$N$(C$_2$H$_3$CN)) versus $L_{\rm bol}$. First we notice the scarcity of C$_2$H$_3$CN detections towards hot corino objects that limits the luminosity range for our comparison, as we only have one hot corino, IRAS16293B detected in C$_2$H$_3$CN.  We find only a weak correlation for this ratio (Fig.\,\ref{fig:luminosity}, middle panel)  with $\rho$ of 0.2 and a $p$-value of 0.05. Such a constant ratio of these molecules may suggest that they originate from the same gas throughout various conditions and stages for star formation, or may suggest common chemical precursor or a chemical link \citep{Coletta2020}.  In fact, C$_2$H$_5$CN and C$_2$H$_3$CN are expected to be partially chemically related molecules \citep{Garrod2022}, as in the cold gas phase C$_2$H$_5$CN is produced primarily on the grain surfaces, and is the daughter molecule of C$_2$H$_3$CN. In the hot gas phase their ratio has been proposed to trace the chemical age of the gas as C$_2$H$_5$CN forms from C$_2$H$_3$CN in the gas phase \citep[c.f.][]{Allen2018}, although this is less clear in the models of  \citet{Garrod2022}. 

The $N$(C$_2$H$_5$CN)/$N$(CH$_3$CN) ratio (Fig.\,\ref{fig:luminosity}, bottom panel) also suggests an increasing trend as a function of luminosity with $\rho$ of 0.5 and $p$-value of 0.05. 
This reflects well our findings in Sect.\,\ref{sec:comp_hotcorinoscores} where we observed higher abundances relative to CH$_3$CN of complex cyanides towards our objects compared to hot corinos, and could also reflect intrinsic chemical differences 
for heavy complex cyanides between hot corinos and hot core like objects.
Comparing our results to that of \citet{Nazari2021}, who find a constant ratio of $N$(C$_2$H$_5$CN)/$N$(CH$_3$CN) for their sample of high-mass sources, we find that our estimates and that of Sgr~B2(N) are mostly above most of their measurements, although the data points show a considerable scatter. An increasing trend of these ratios would suggest a more efficient formation of C$_2$H$_5$CN compared to CH$_3$CN (or destruction of CH$_3$CN) at higher luminosities.

\section{Conclusions}
We performed a spectral survey with the APEX telescope between 159\,GHz and 374\,GHz towards six infrared quiet massive clumps, and focused here on the analysis of COMs. COMs are widely detected among the sources, we detected between five to eleven COMs (plus isotopologues). Using an iterative process relying on rotational diagrams and LTE modelling with Weeds, we estimate the physical conditions (such as column density, excitation temperature and size of the emitting region) and infer relative molecular abundances. For several of the COMs we are able to separate two temperature components. Our main results are:

\begin{itemize}

\item We detect \textit{a minima} five COMs (CH$_3$OH,  CH$_3$CN, CH$_3$OCHO, CH$_3$OCH$_3$, CH$_3$CHO) towards all sources, and detect furthermore HC(O)NH$_2$, C$_2$H$_5$OH, C$_2$H$_5$CN, C$_2$H$_3$CN, CH$_3$SH, and CH$_3$COCH$_3$ in the most line rich source G335.58. 
\item The cold gas phase is richer  in terms of the type of COMs detected compared to the warm gas phase, while the heavier complex cyanides are only present in the warm gas phase. The cold gas is deficient in heavy complex cyanides and is abundant in O-bearing COMs including formamide.

\item We find that the excitation temperatures of O-bearing COMs and formamide are close to, or below their respective thermal desorption temperatures. This suggests that their abundances have contribution from non-thermal desorption processes. 

\item We compared the COMs in the warm gas phase of our objects to that of a sample of hot corinos and hot cores from the literature. A comparison of their physical properties suggests that the COM emission towards our objects are somewhat more compact compared to hot cores, and more extended than hot corinos by almost an order of magnitude. The comparison of molecular abundances reveals that the warm gas phase of our objects is similar to both hot cores and hot corinos except for the heavy complex cyanides that show a deficit for hot corinos compared to our sample and the hot cores. 
\item Column density ratios of  CH$_3$OCH$_3$/CH$_3$OCHO, C$_2$H$_5$CN/C$_2$H$_3$CN, and C$_2$H$_5$CN/CH$_3$CN as a function of luminosity do not reveal strong correlations, and mostly exhibit a flat distribution. Our results suggest, however, an increasing trend of $N$(CH$_3$OCH$_3$)/$N$(CH$_3$OCHO) and $N$(C$_2$H$_5$CN)/$N$(CH$_3$CN)
potentially pointing to a link between the physical properties of the emerging protostar and the chemical evolution of the gas. 

\item We suggest that there is an evolutionary stage preceding the emergence of hot cores that is characterised by COMs at warm temperatures originating from a combination of non-thermal desorption processes, partial thermal desorption or gas-phase reactions. This stage is chemically similar to hot cores and hot corinos with similar molecular abundances for O-bearing COMs. 
Complex cyanides have, however, similar molecular abundances to hot cores. This stage may correspond to the early warm-up phase chemistry, where the full molecular composition of the radiatively heated hot inner component is about to emerge.

\end{itemize}

\bibliographystyle{aa}
\bibliography{Full_Sample_COMs}

\begin{acknowledgements}

This publication is based on data acquired with the Atacama Pathfinder Experiment (APEX). APEX is a collaboration between the Max-Planck-Institut für Radioastronomie, the European Southern Observatory, and the Onsala Space Observatory. 
T. Cs. has received financial support from the French State in the framework of the IdEx Universit\'e de Bordeaux Investments for the future Program.
This work was supported by the
  Programme National “Physique et Chimie du Milieu Interstellaire” (PCMI) and “Programme National de Physique Stellaire” (PNPS)
  of CNRS/INSU with INC/INP co-funded by CEA and CNES.  
The authors wish to thank Dr. Arnaud Belloche for his valuable insights on the analysis of emission from COMs and Dr. Rolf Güsten for help with the observations.
\end{acknowledgements}

\begin{appendix}
\section{Specific cases of the LTE modelling}
\label{sec:app_excpetions}

\subsection{G333.46}
For {\bf dimethyl ether (CH$_3$OCH$_3$)}, the size is difficult to constrain with the detected lines, however the non-detection of the line at 358\,GHz, constrains the size of the emission rather well. The size has to be large: about 20$''$.

{\bf Formamide} is detected in our spectral survey with two transitions above 3$\sigma$. However these two transitions have similar energy levels (85\,K and and 91\,K). This range is not sufficient to determine with precision the excitation temperature, the total column density and the size of the emission of HC(O)NH$_2$. Therefore, we took the size of the emission of CH$_3$OCH$_3$.

{\bf Ethyl cyanide} is not detected in this object. We determined an upper limit for the column density assuming the same size and the same temperature as CH$_3$CN. Although we do not detect C$_2$H$_5$CN with lines above 3$\sigma$, our upper limit relies on several potential signals coming from C$_2$H$_5$CN at a 2$\sigma$ level.

\subsection{G335.58}

Several {\bf CH$_3$OH,$v_{\rm t}$=1} lines were detected and fitted. The errorbars are large and some lines are not well fitted. Therefore the estimate value relies on a visual inspection.

For {\bf C$_2$H$_5$CN}, we fit two temperature components with one temperature of 70\,K and another one at 250\,K.

\subsection{G343.75}
For {\bf CH$_3$C}N, similarly to G328.25, the $K$ ladders were found to be sub-thermally excited. Therefore, we determined the column density, the temperature and the size of the cold component using the $K$=0 ladder. However, this model was not found to be satisfactory and we added a warm compact component to fit the excited $K$ ladders. This fit was found to be satisfactory.
Several lines of vibrationally excited methano, {\bf CH$_3$OH,$v_{\rm t}$=1}, were detected and fitted. The errorbars are large and some lines are not well fitted. 

\section{Rotational diagrams}
\label{app:rot_diag}
\subsection{G320.23}
\begin{figure*}
    \centering
    \includegraphics[width=0.45\linewidth]{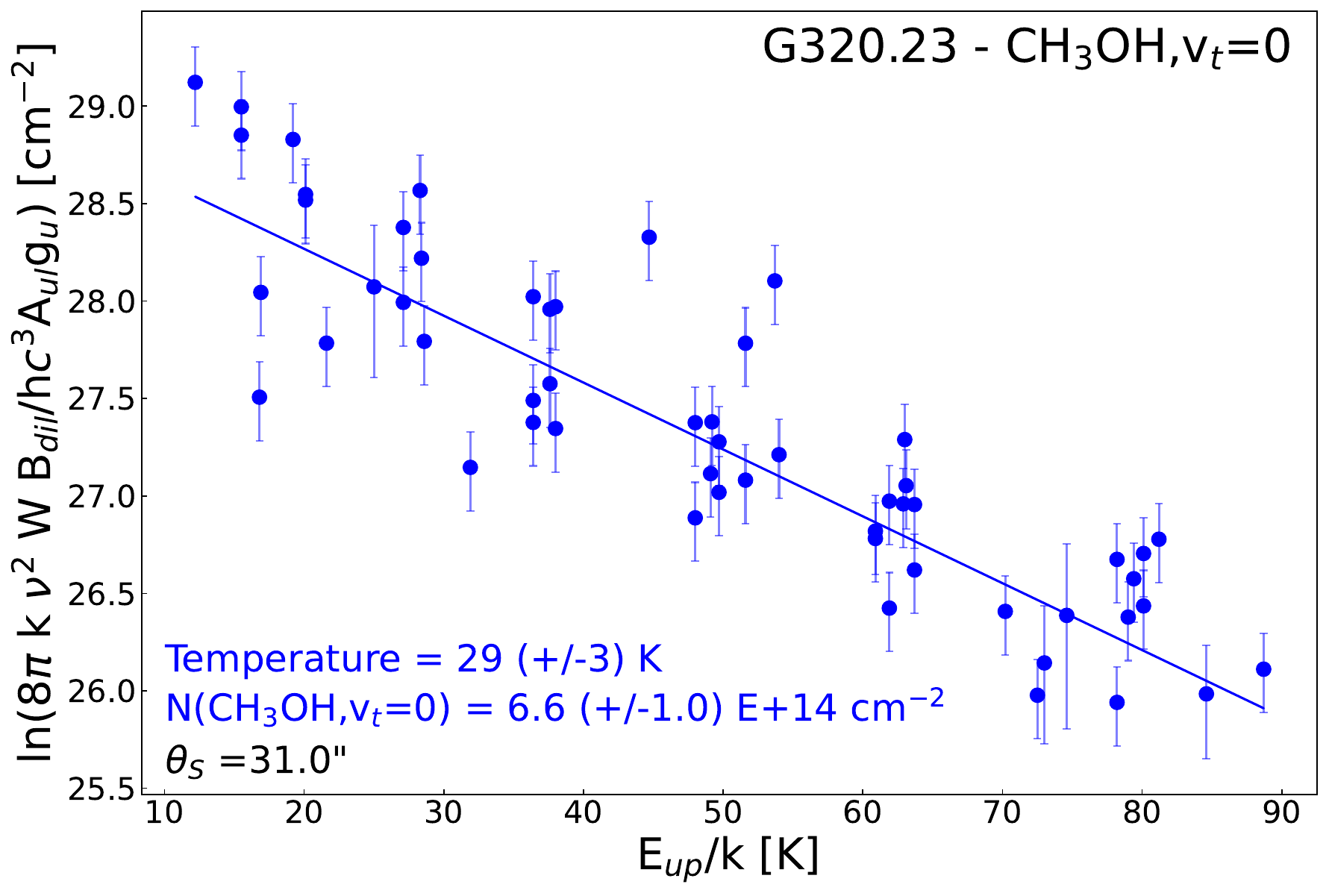}
    \includegraphics[width=0.45\linewidth]{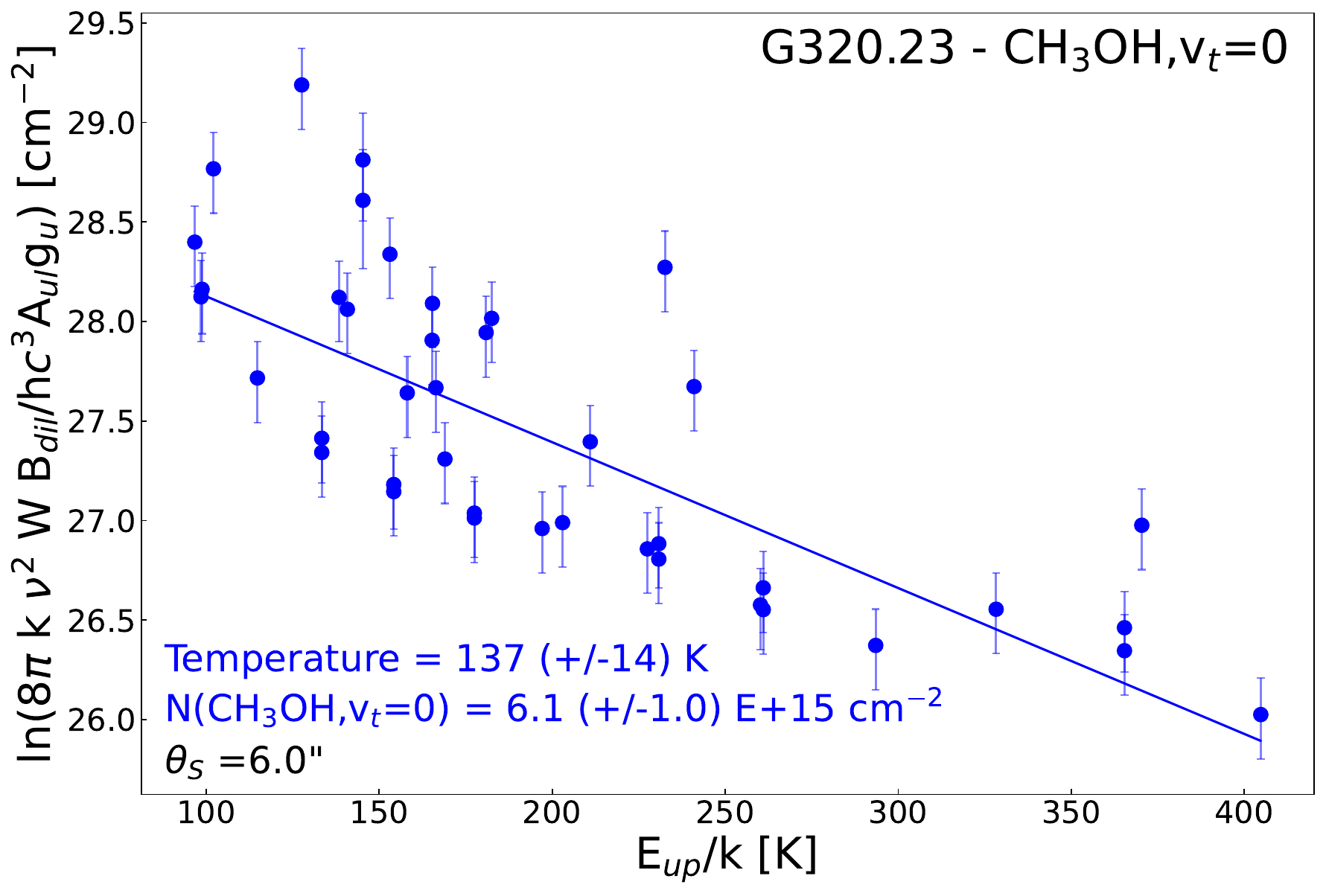}
    \includegraphics[width=0.45\linewidth]{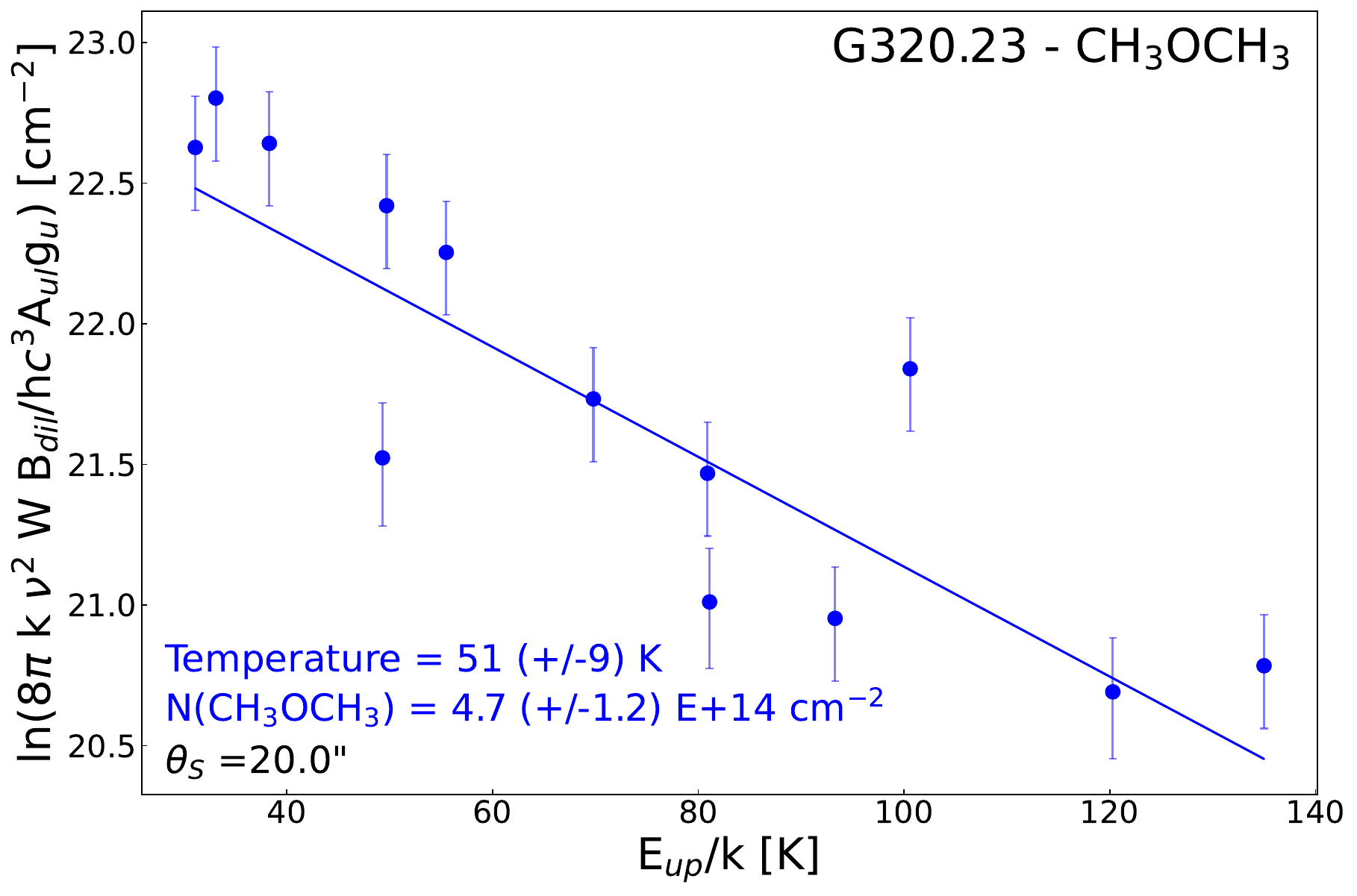}
    \includegraphics[width=0.45\linewidth]{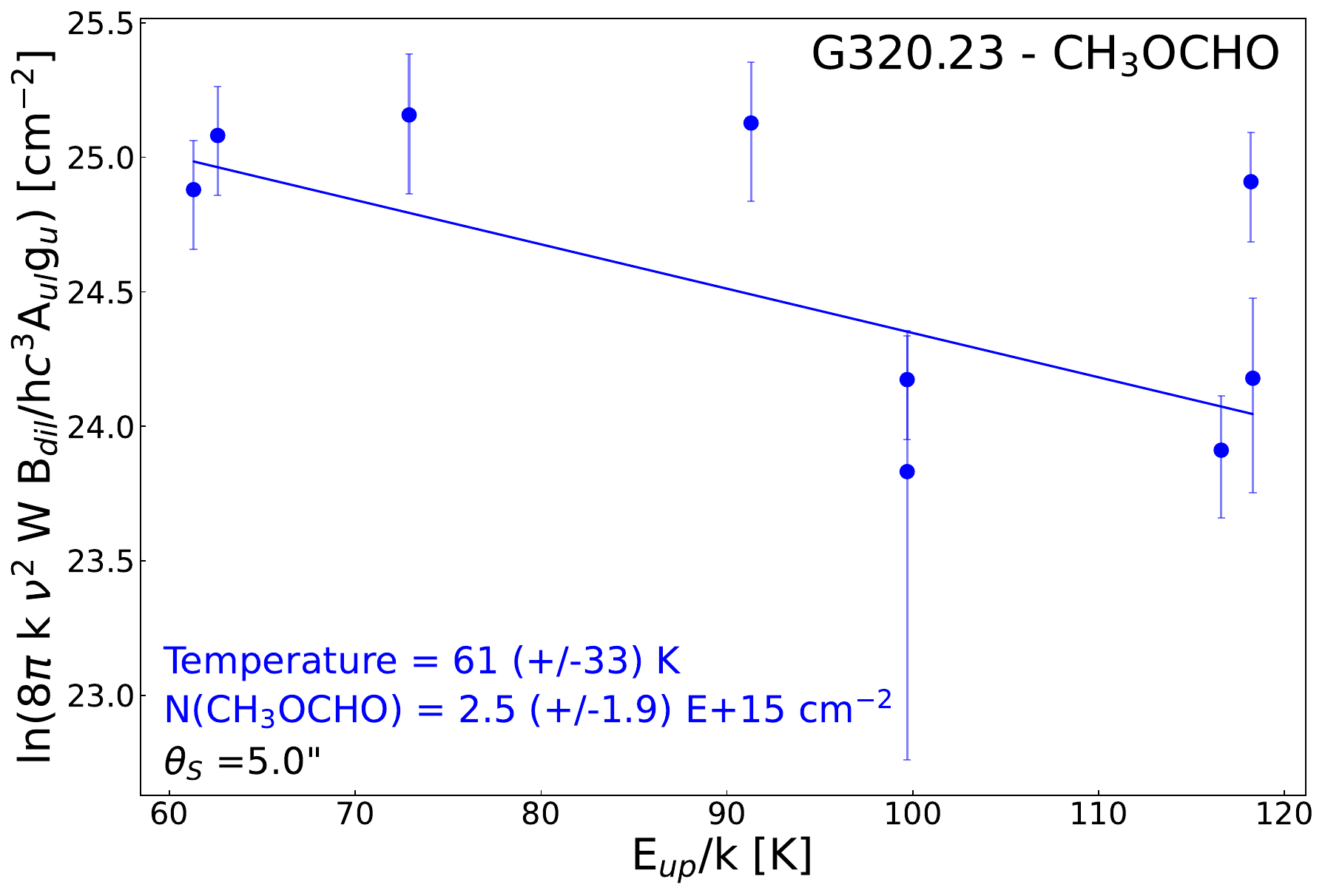}
    \includegraphics[width=0.45\linewidth]{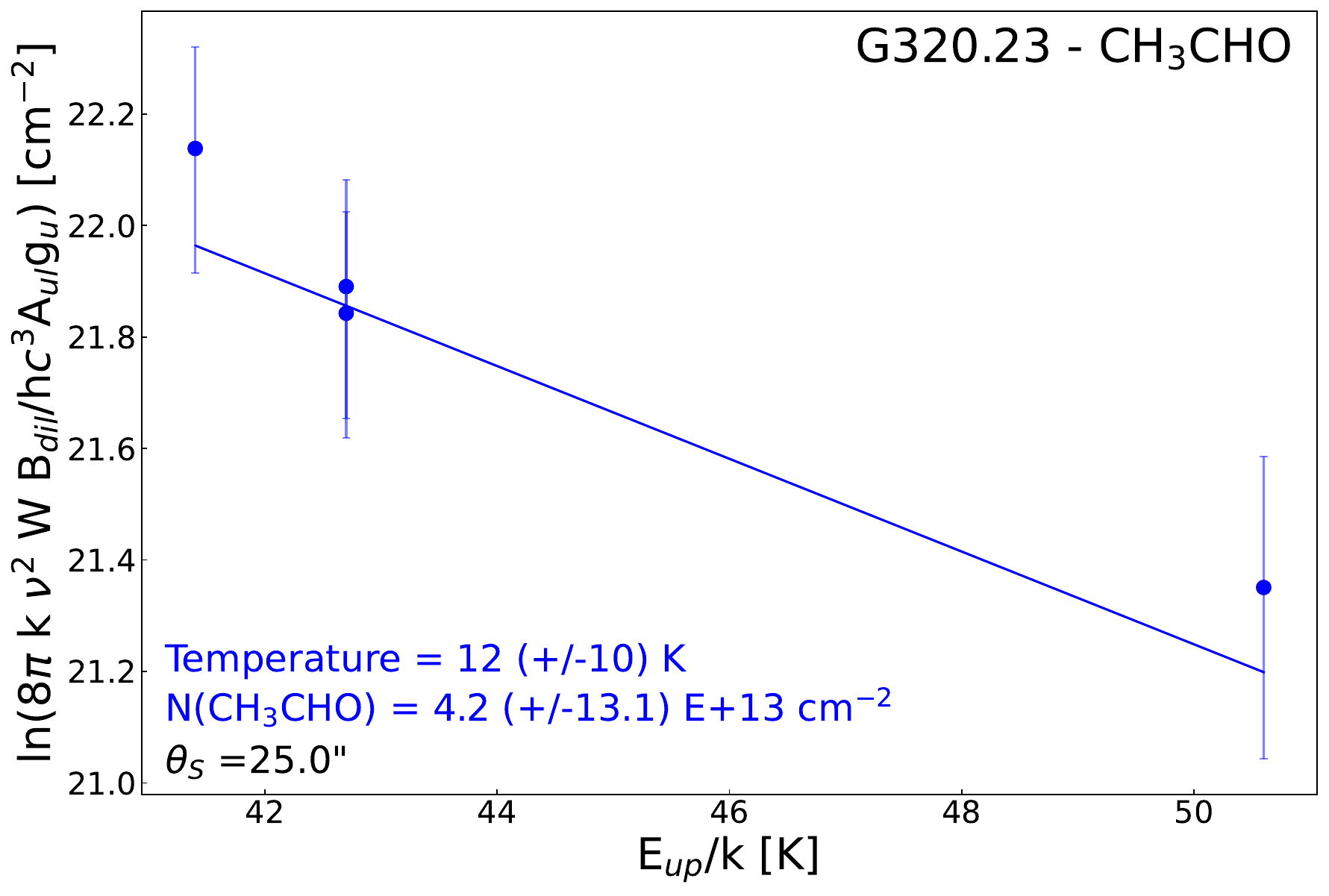}
    \includegraphics[width=0.45\linewidth]{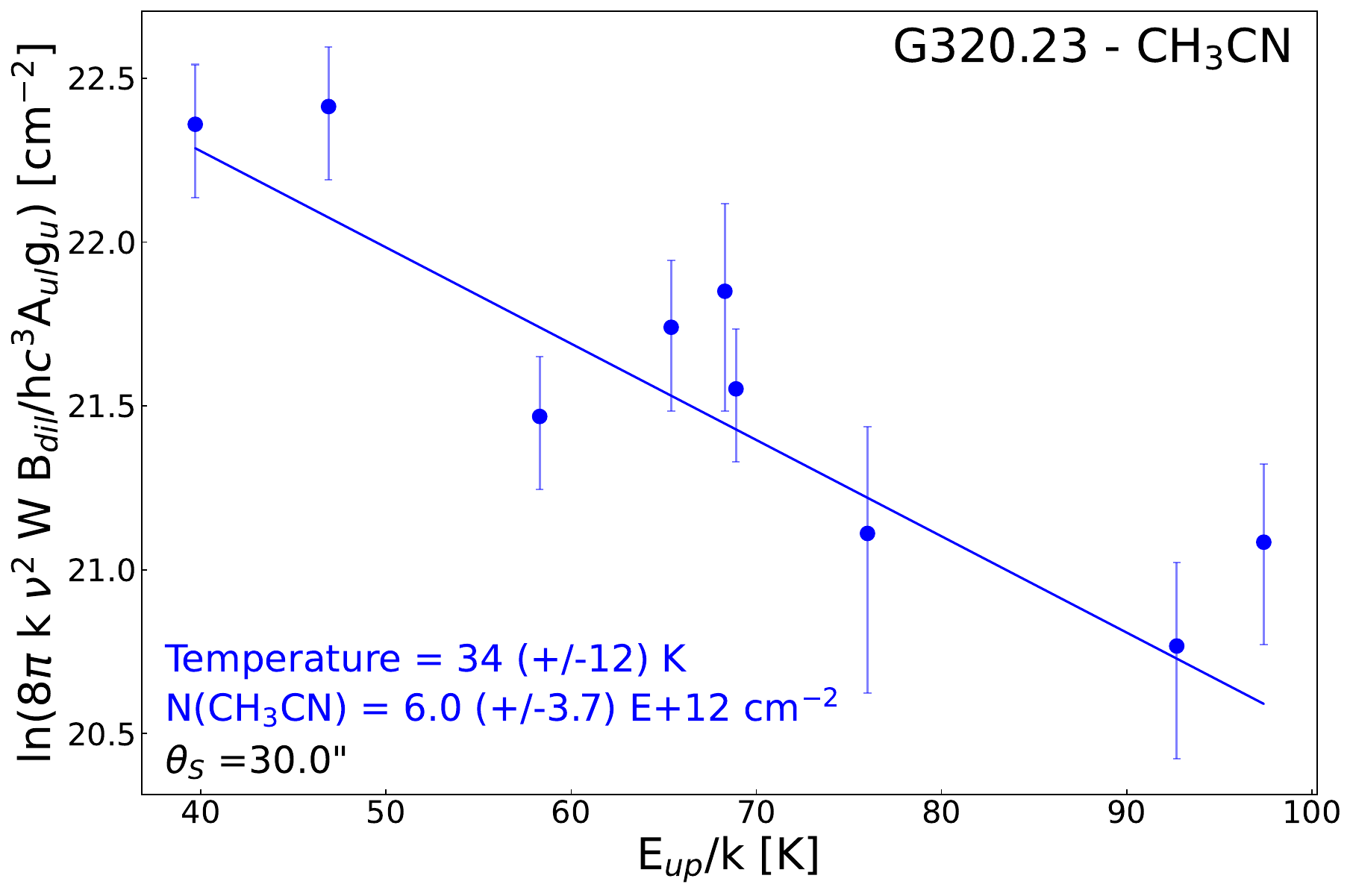}
    \includegraphics[width=0.45\linewidth]{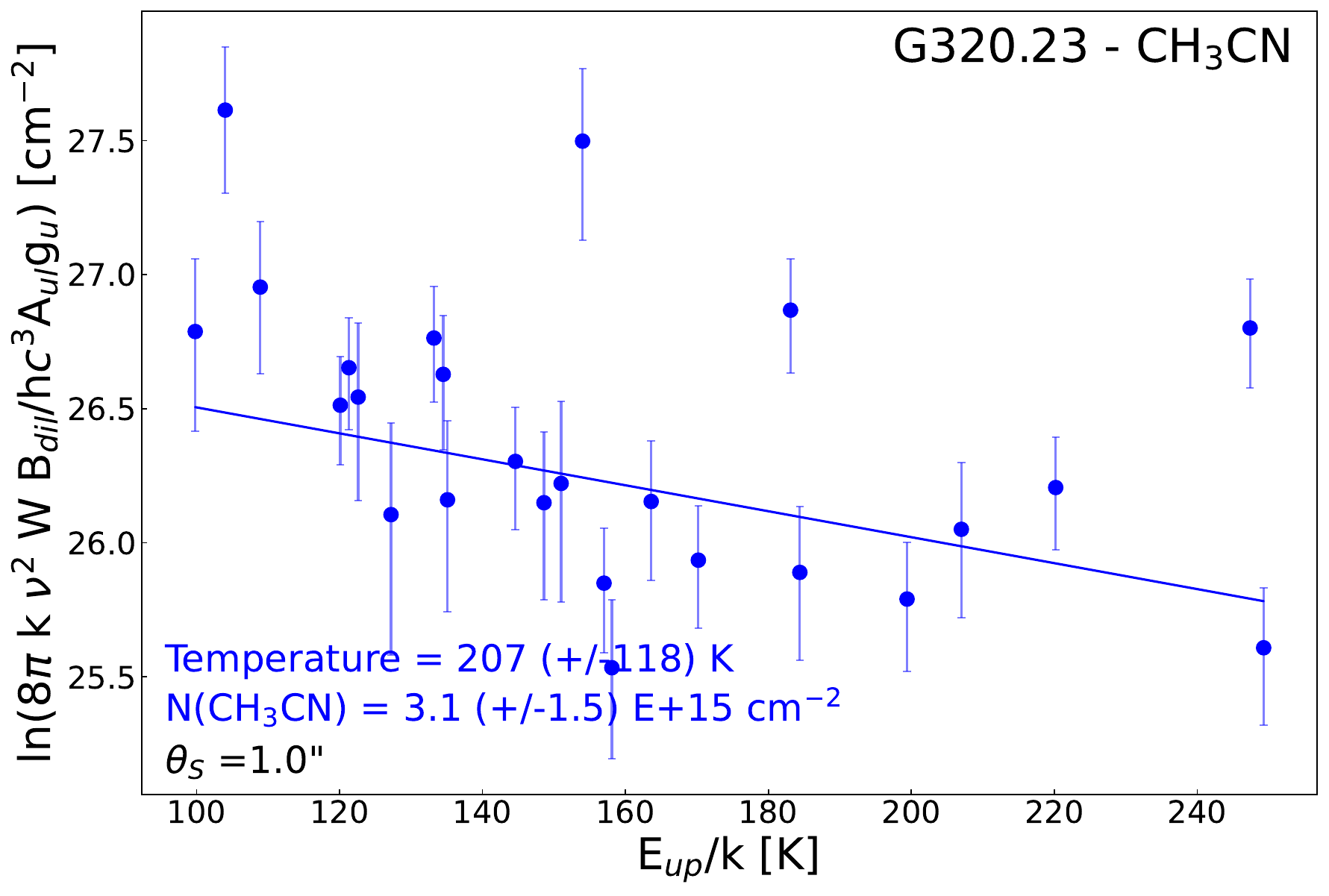}
    \caption{Rotational diagrams of the COMs in G320.23: CH$_3$OH, CH$_3$OCH$_3$, CH$_3$OCHO, CH$_3$CHO, and CH$_3$CN. Rotational temperatures and column densities are indicated with the same color code. The errorbars correspond to a 20\,\% error. To calculate the uncertainties on the rotational temperature and the column density, we used a Monte Carlo method assuming a uniform distribution of the error for each data point.}
    \label{fig:rot_diag_320p23}
\end{figure*}

\subsection{G333.46}
\begin{figure*}
    \centering
    \includegraphics[width=0.45\linewidth]{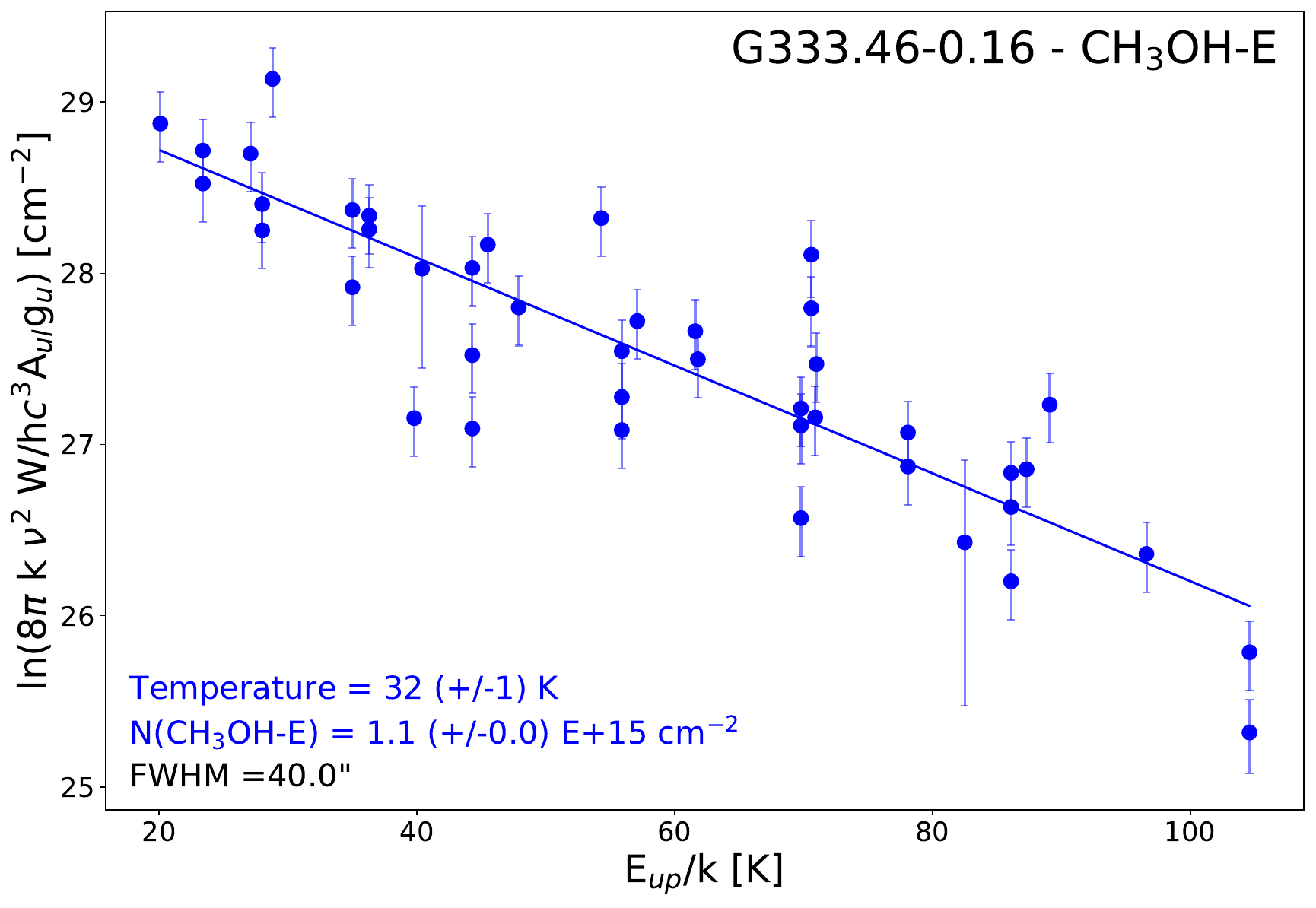}
    \includegraphics[width=0.45\linewidth]{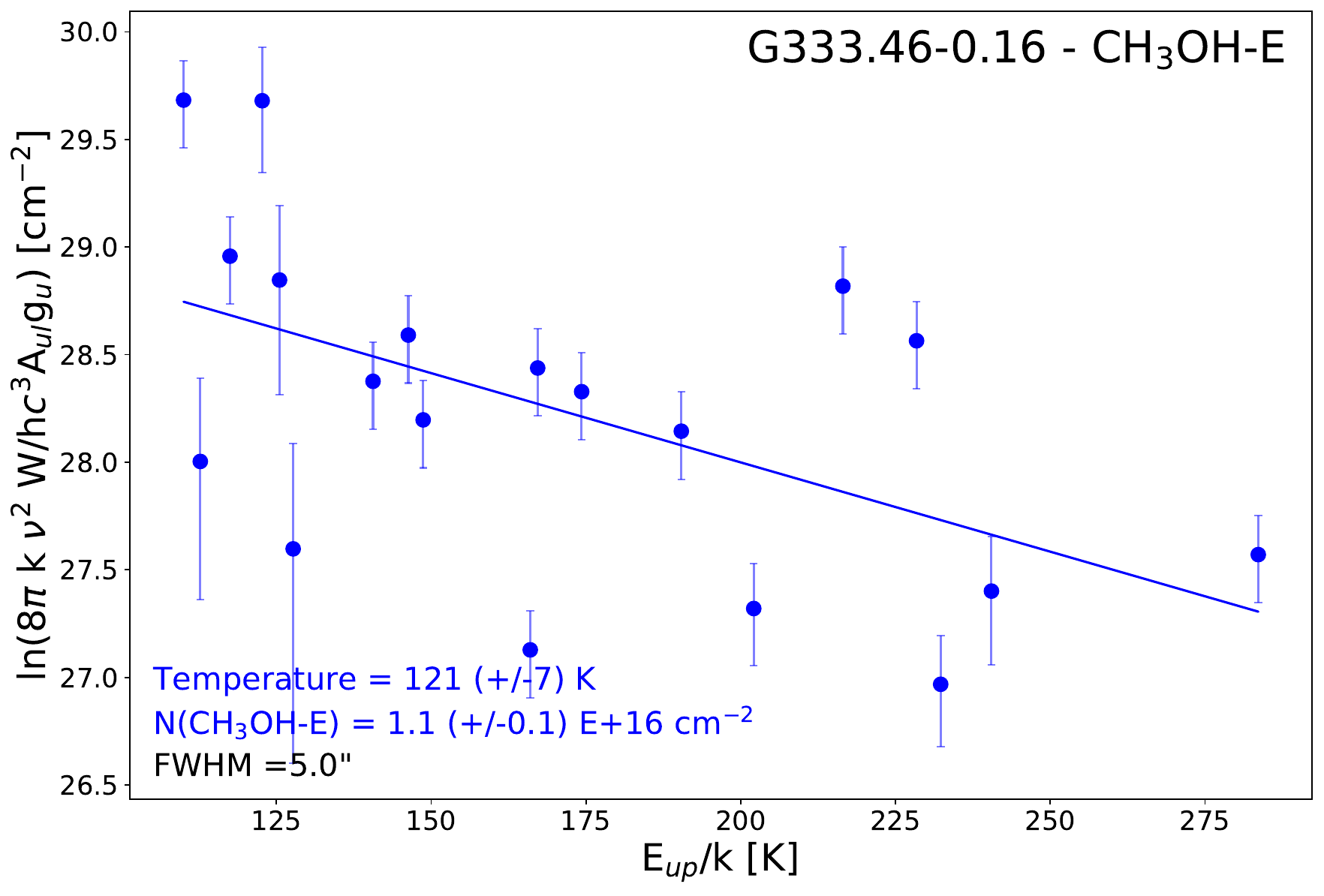}
    \includegraphics[width=0.45\linewidth]{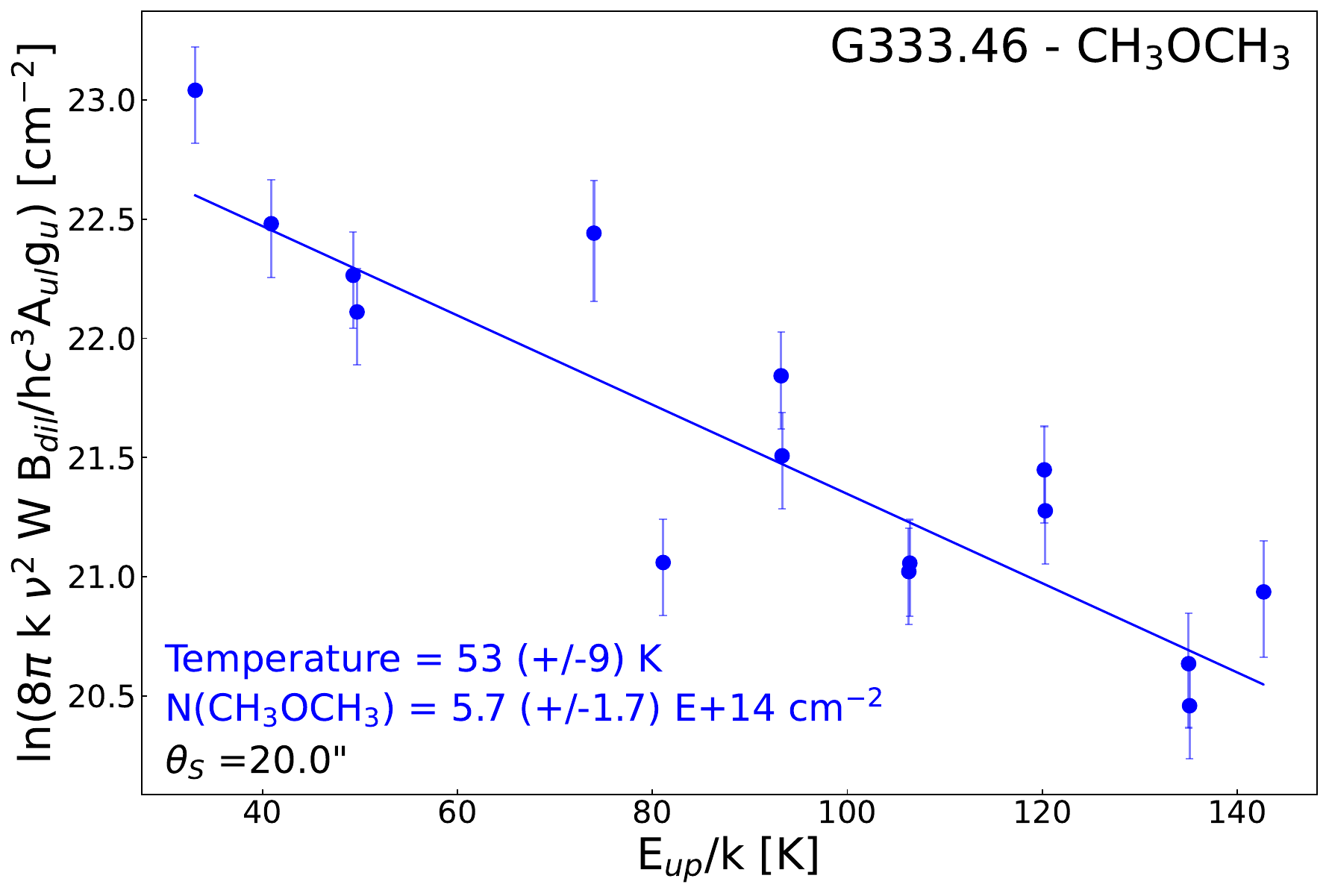}
    \includegraphics[width=0.45\linewidth]{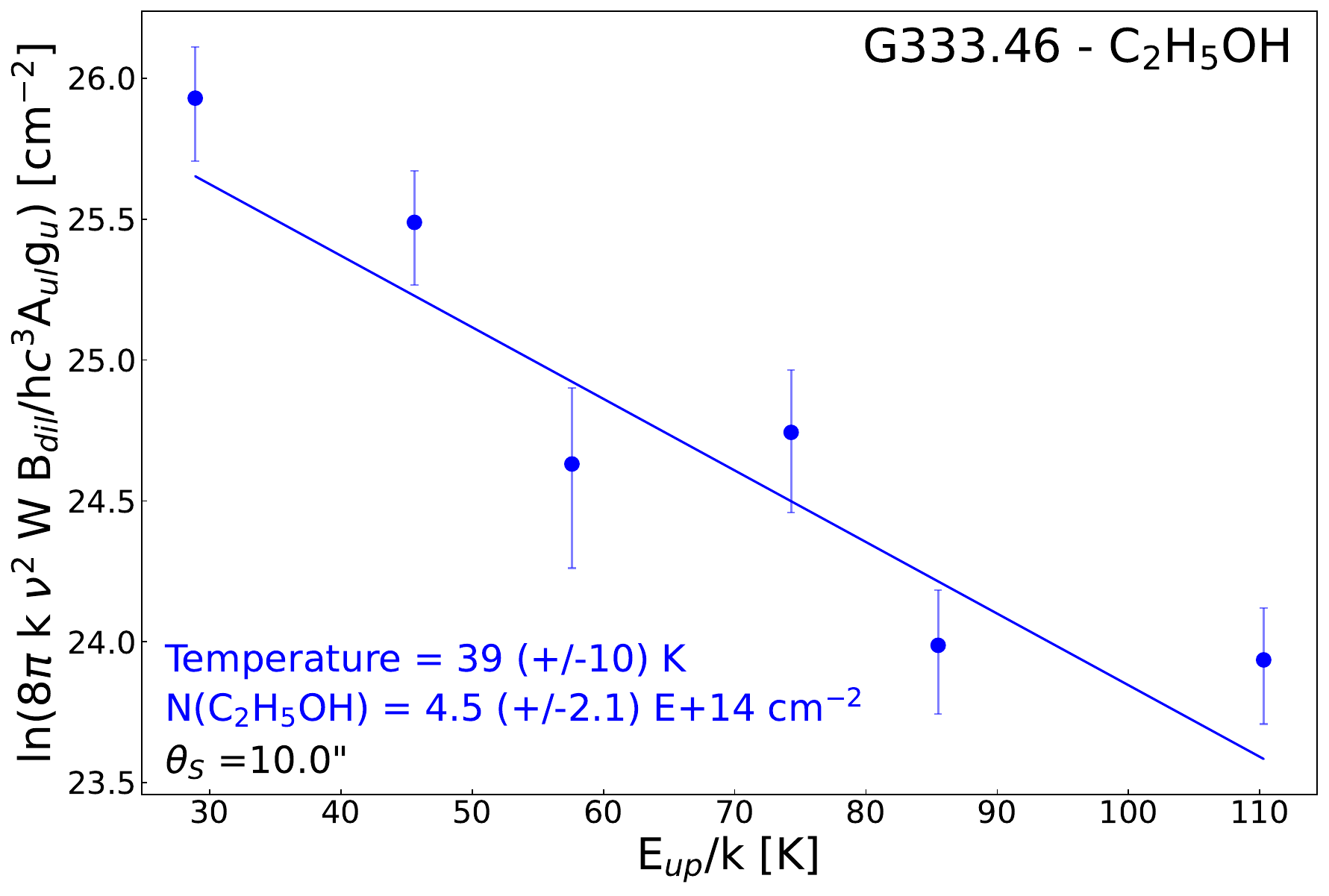}
    \includegraphics[width=0.45\linewidth]{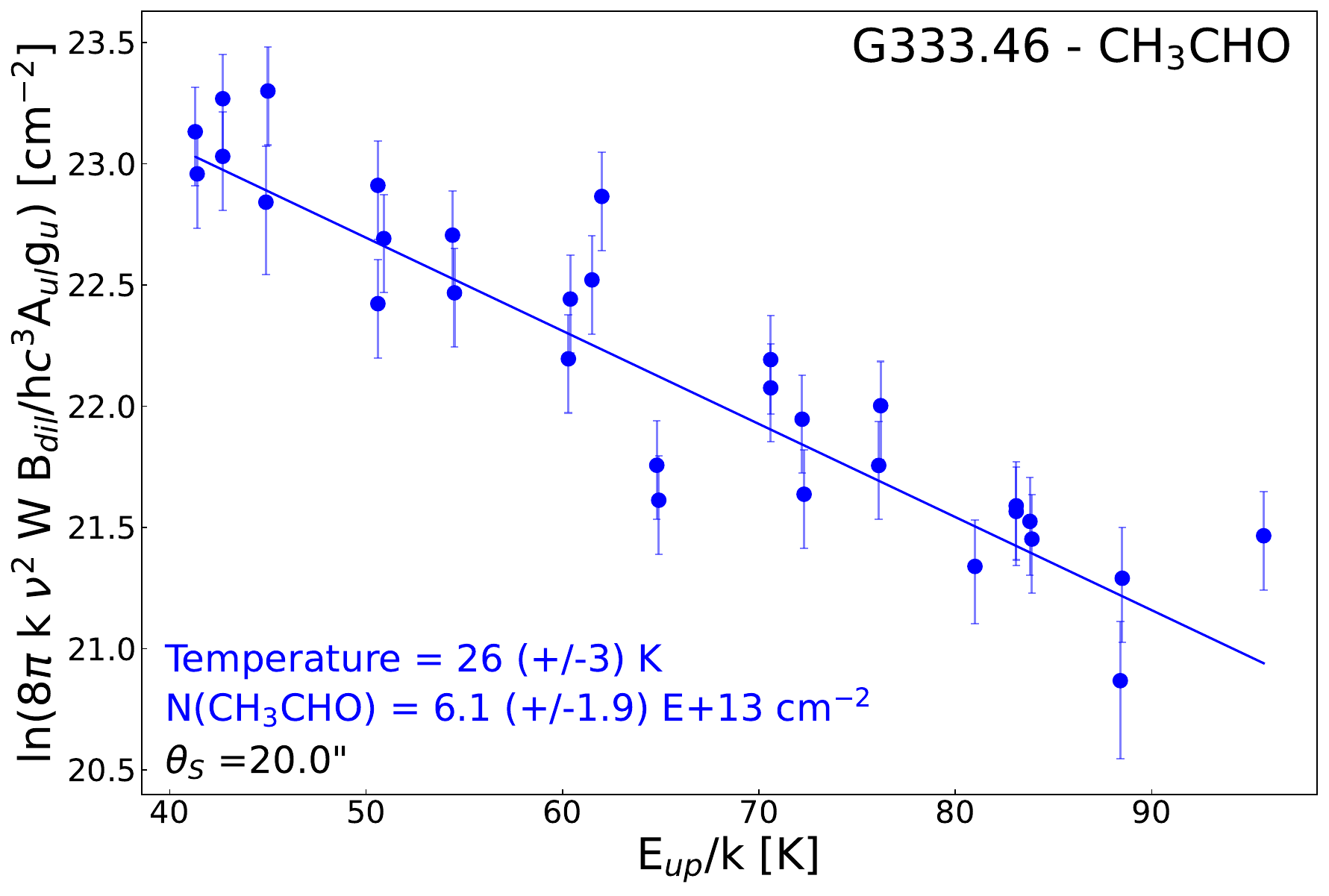}
    \includegraphics[width=0.45\linewidth]{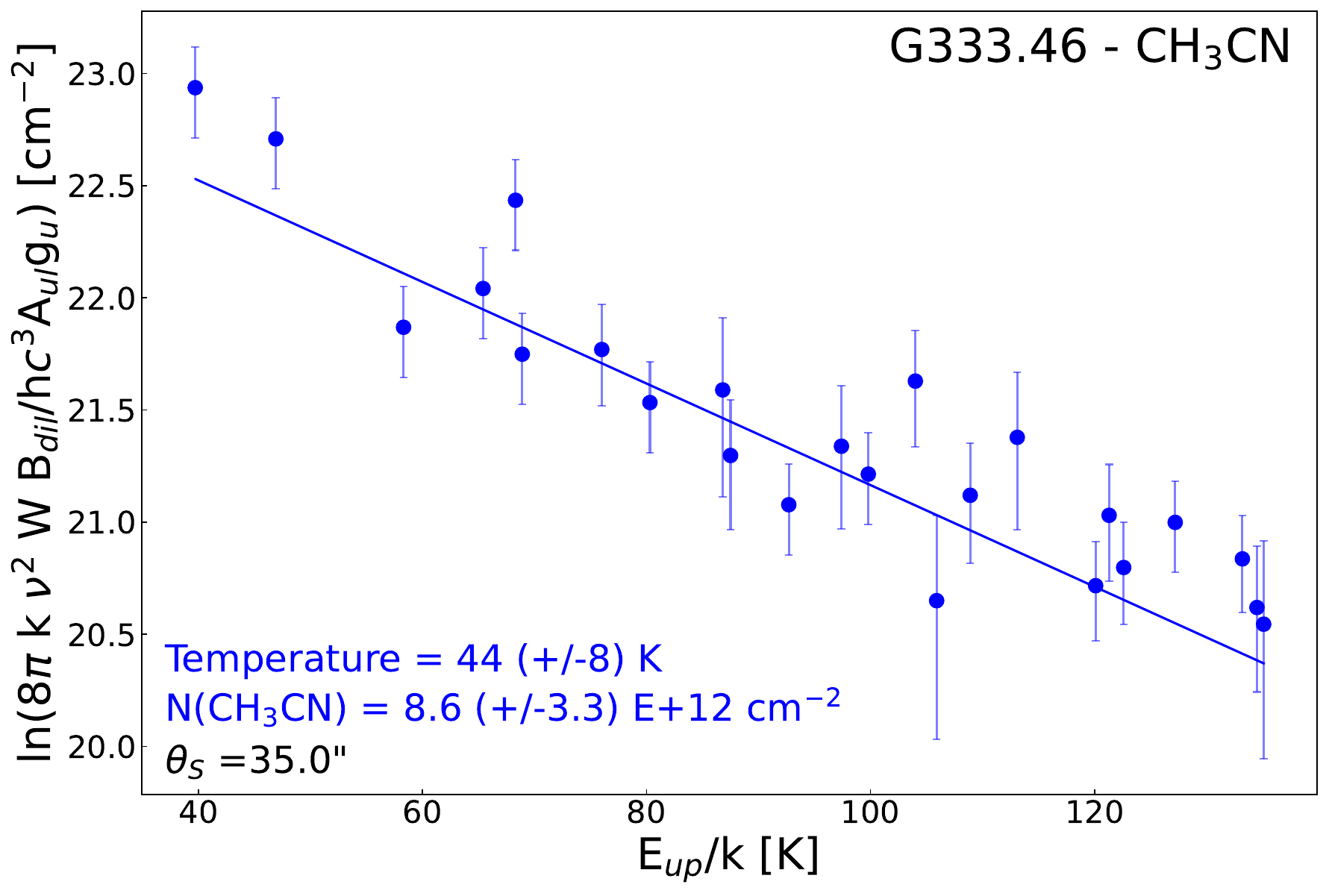}
    \includegraphics[width=0.45\linewidth]{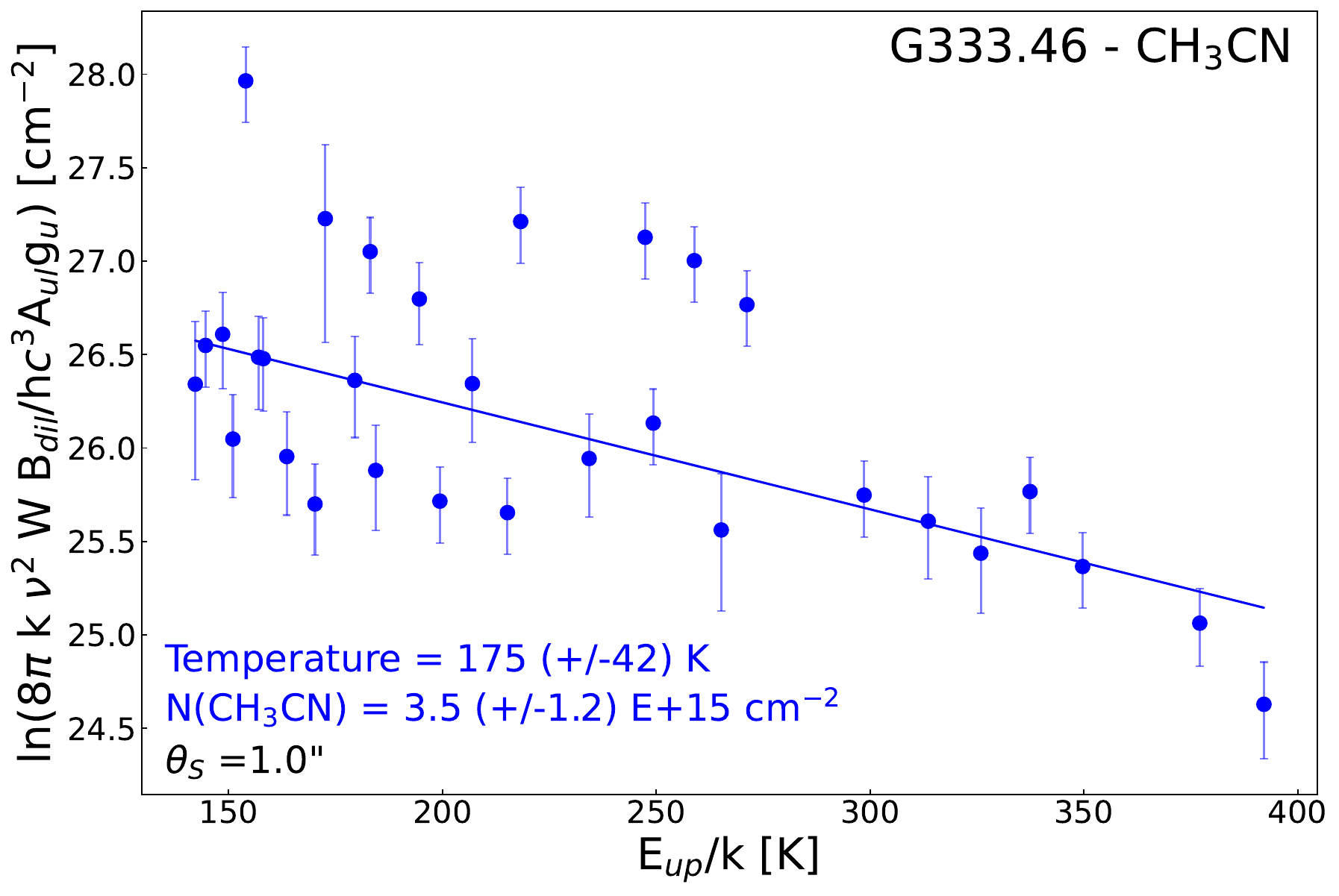}
    \caption{Rotational diagrams of the COMs in G333.46: CH$_3$OH, CH$_3$OCH$_3$, CH$_3$OCHO, CH$_3$CHO, C$_2$H$_5$OH, and CH$_3$CN. Rotational temperatures and column densities are indicated with the same color code. The errorbars correspond to a 20\,\% error. To calculate the uncertainties on the rotational temperature and the column density, we used a Monte Carlo method assuming a uniform distribution of the error for each data point.}
    \label{fig:rot_diag_333p46}
\end{figure*}

\subsection{G335.58}
\begin{figure*}
    \centering
    \includegraphics[width=0.45\linewidth]{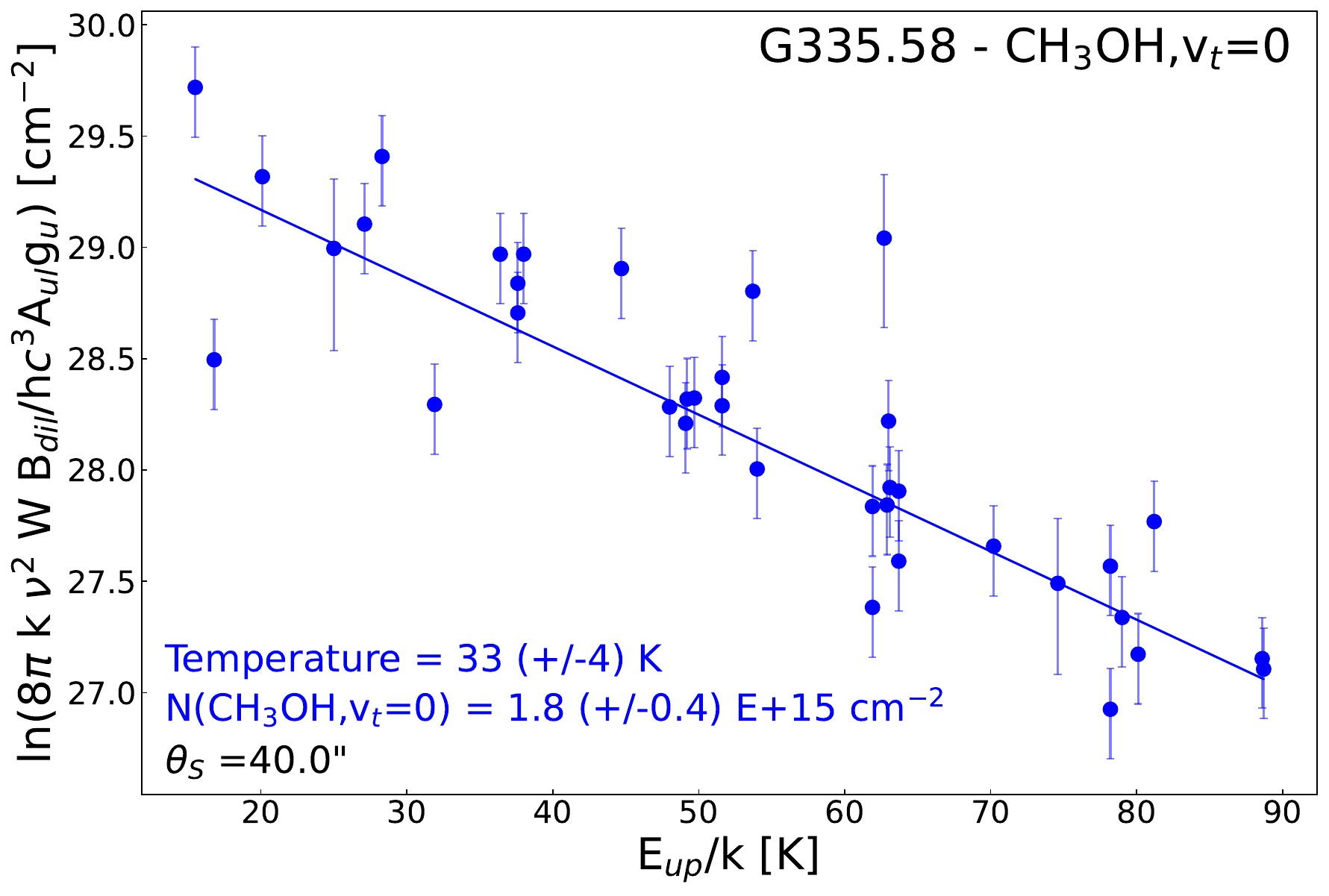}
    \includegraphics[width=0.45\linewidth]{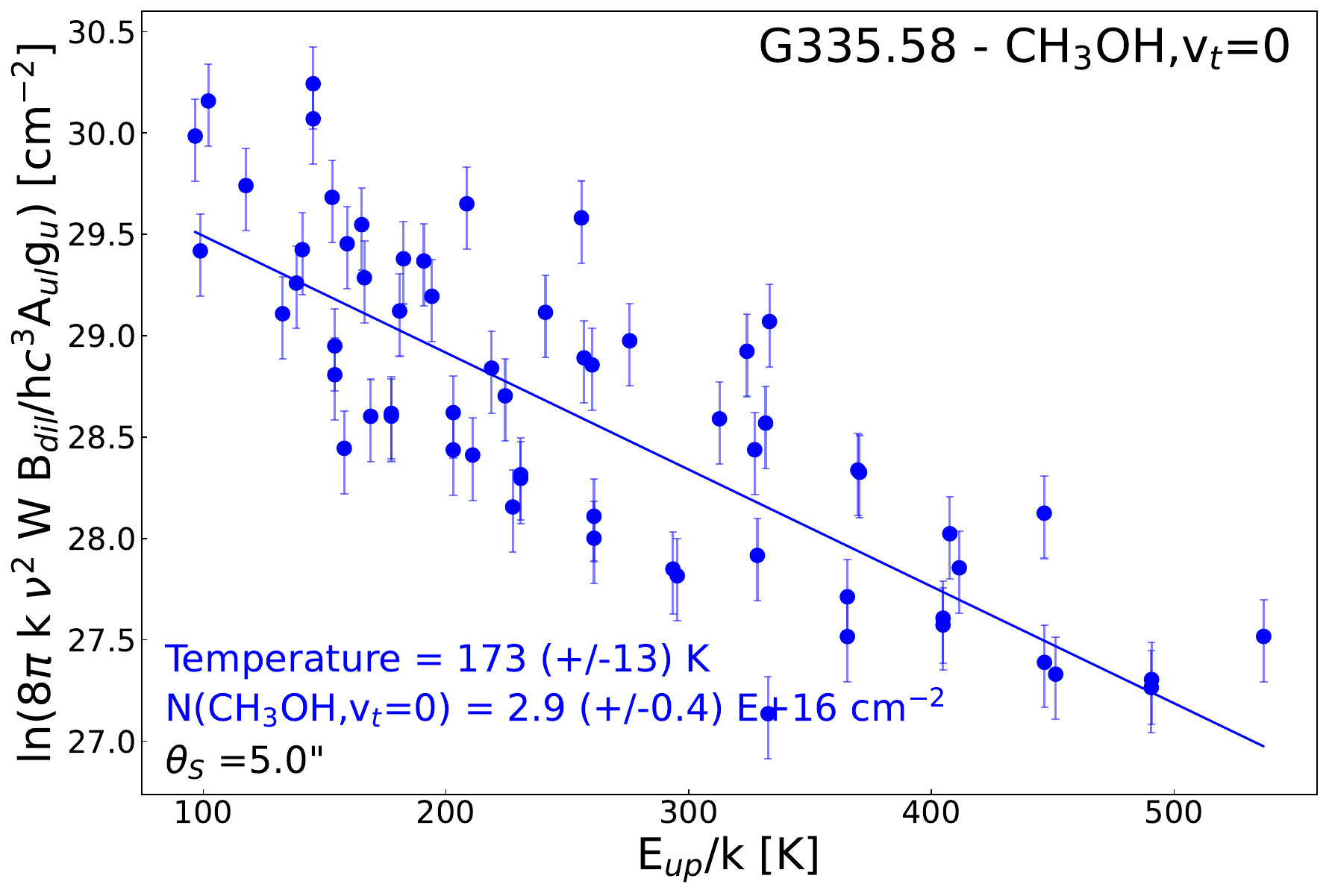}
    \includegraphics[width=0.45\linewidth]{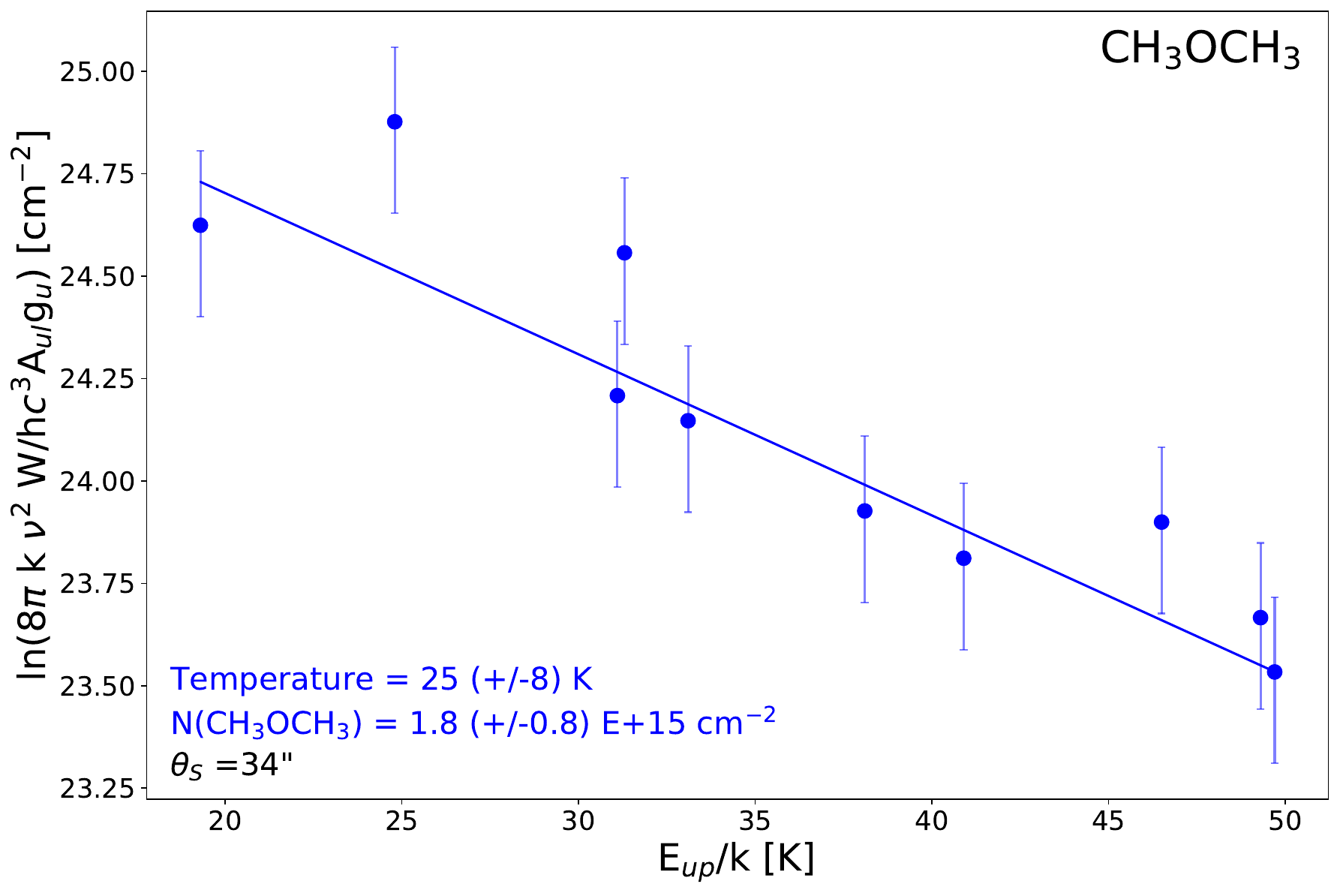}
    \includegraphics[width=0.45\linewidth]{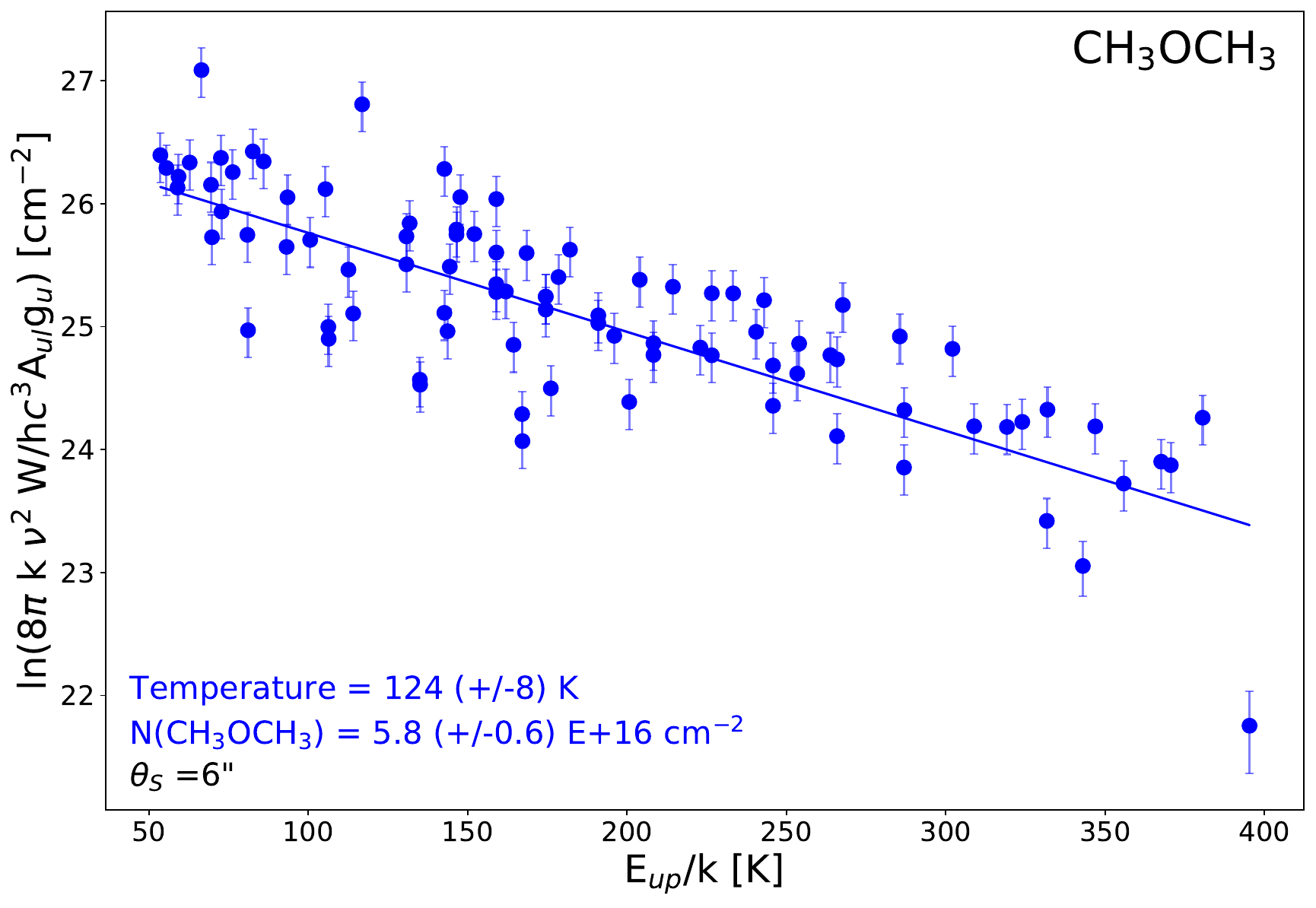}
    \includegraphics[width=0.45\linewidth]{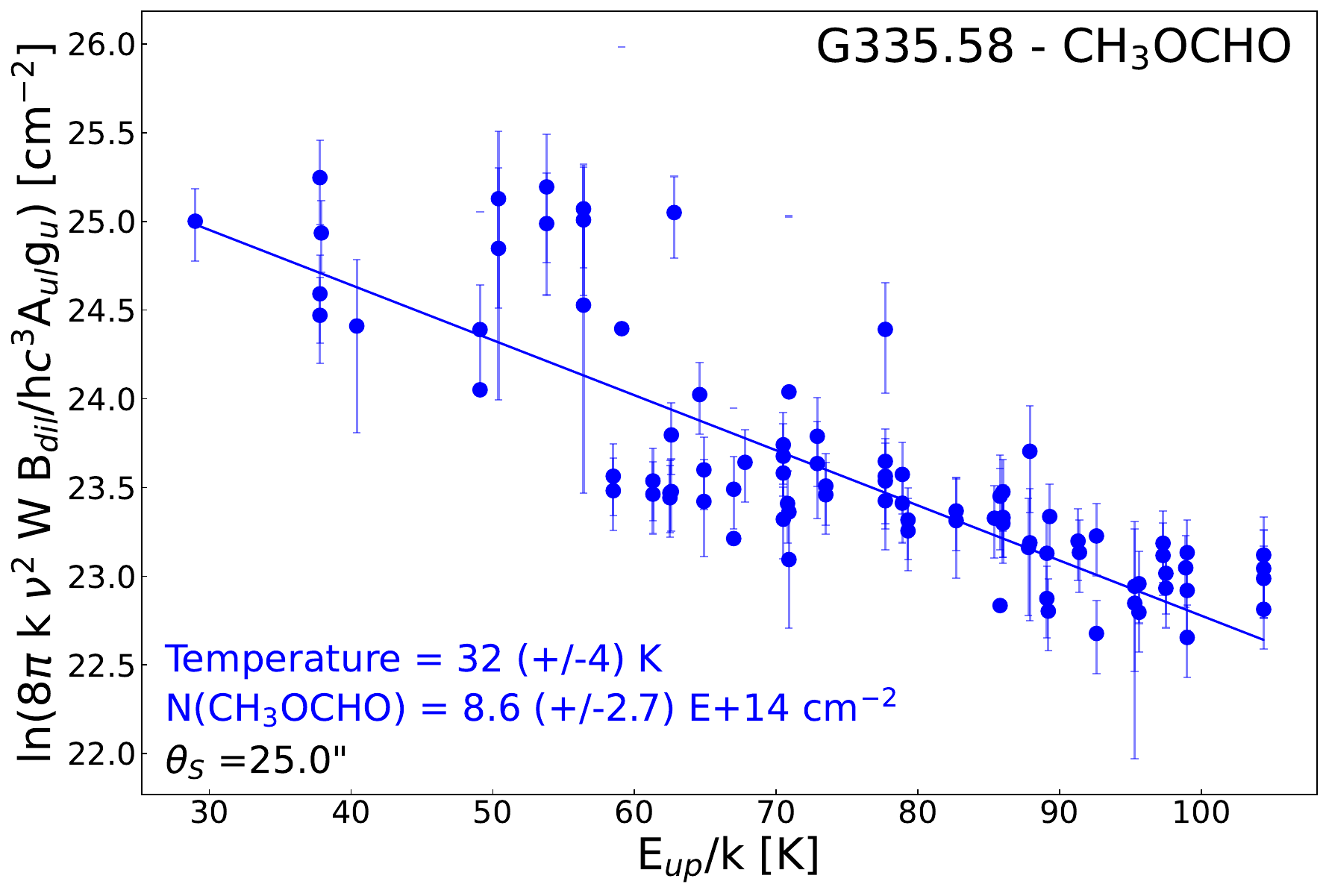}
    \includegraphics[width=0.45\linewidth]{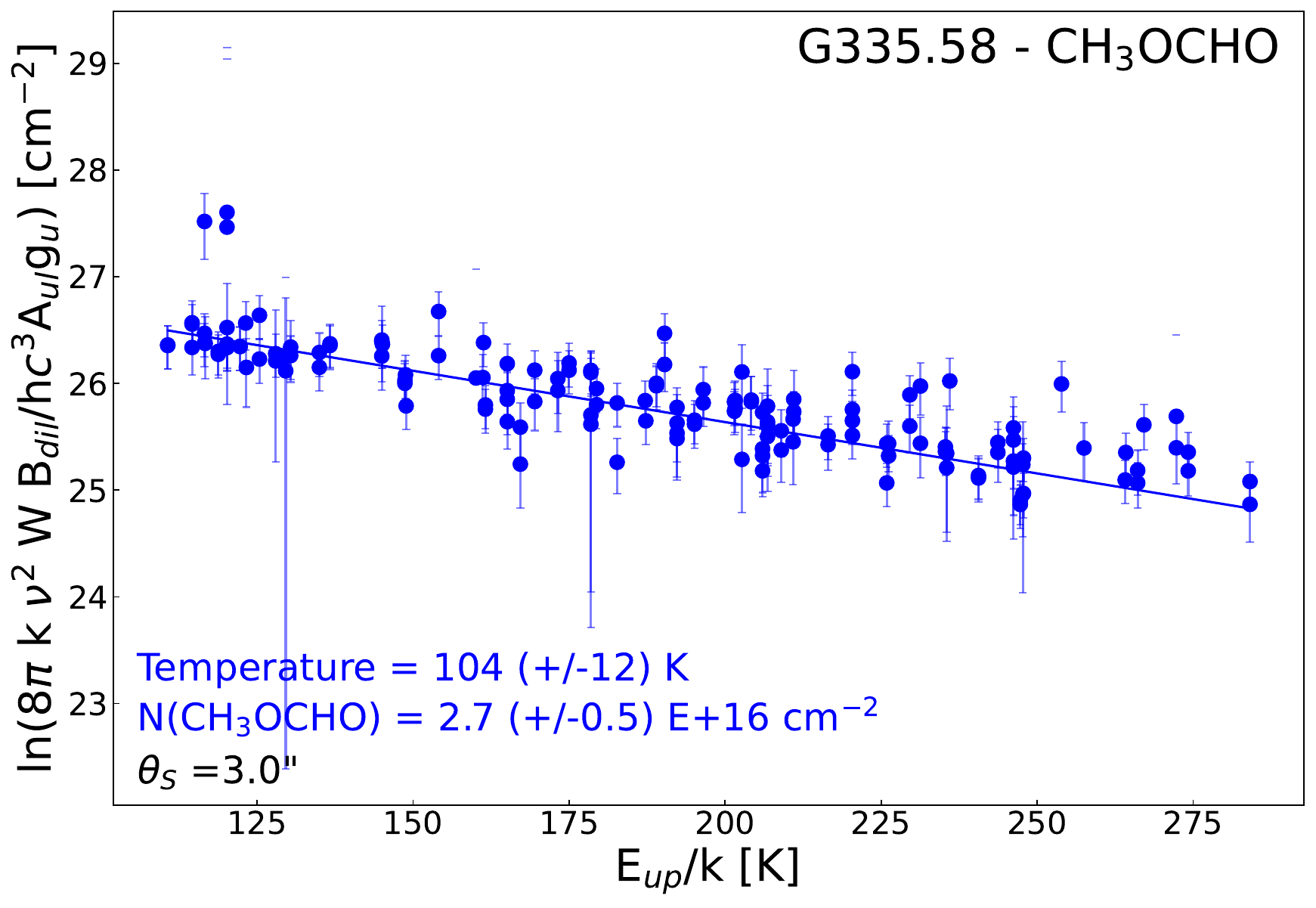}
    \includegraphics[width=0.45\linewidth]{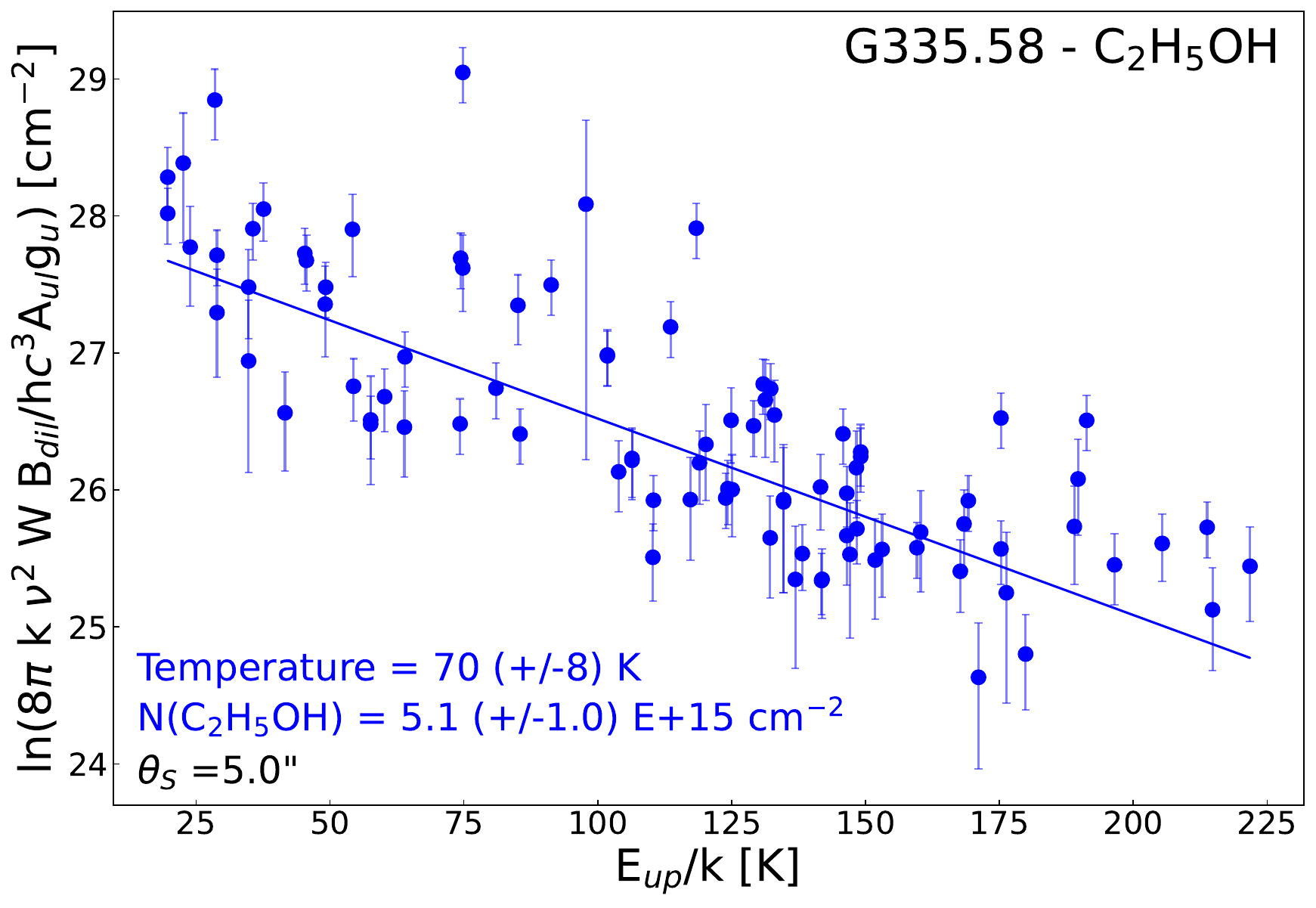}
    \includegraphics[width=0.45\linewidth]{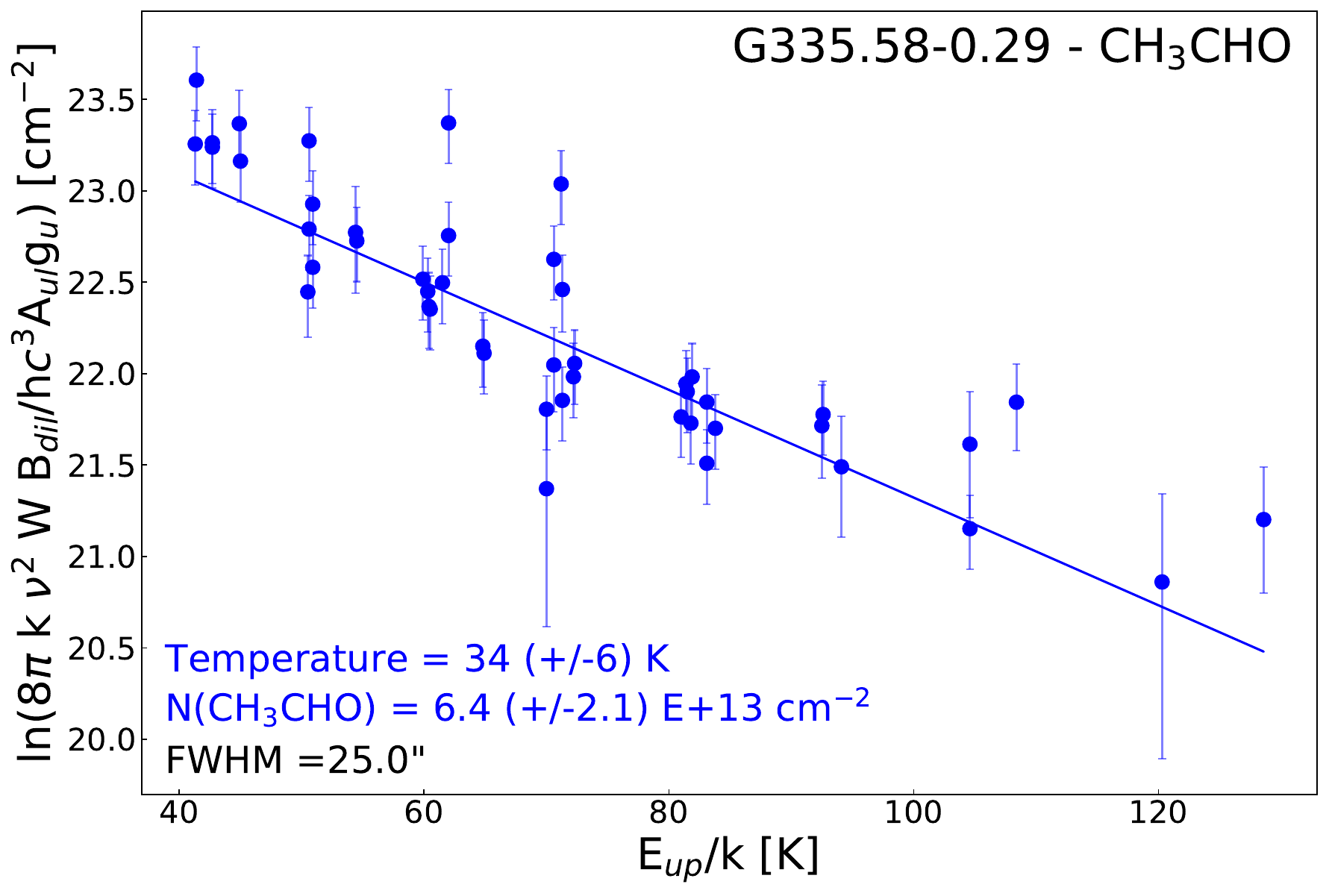}
    \caption{Rotational diagrams of the COMs in G335.58: CH$_3$OH, CH$_3$OCH$_3$, CH$_3$OCHO, C$_2$H$_5$OH, and CH$_3$CHO. Rotational temperatures and column densities are indicated with the same color code. The errorbars correspond to a 20\,\% error. To calculate the uncertainties on the rotational temperature and the column density, we used a Monte Carlo method assuming a uniform distribution of the error for each data point.}
    \label{fig:rot_diag_335p58_1}
\end{figure*}
\begin{figure*}
    \centering
    \includegraphics[width=0.45\linewidth]{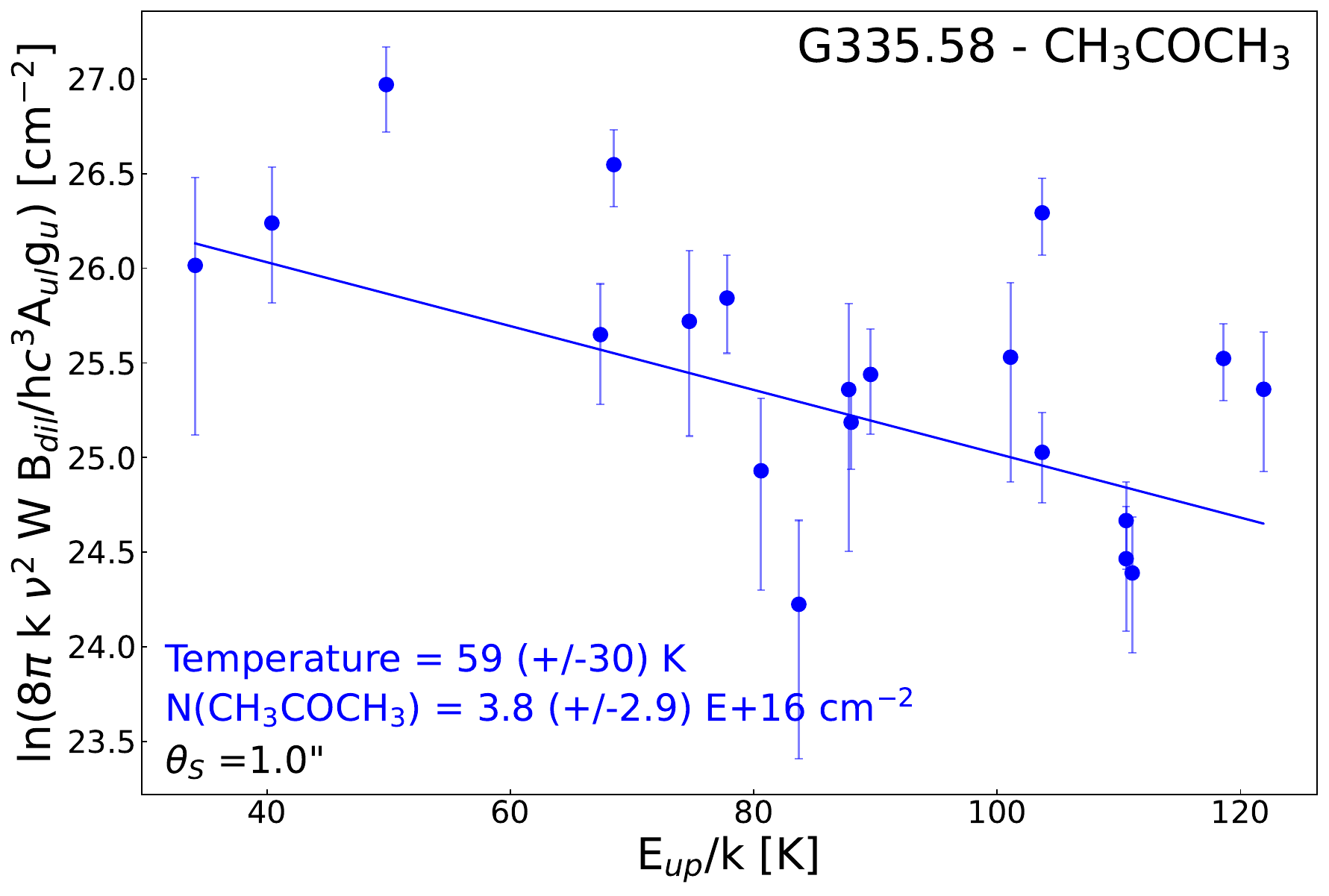}
    \includegraphics[width=0.45\linewidth]{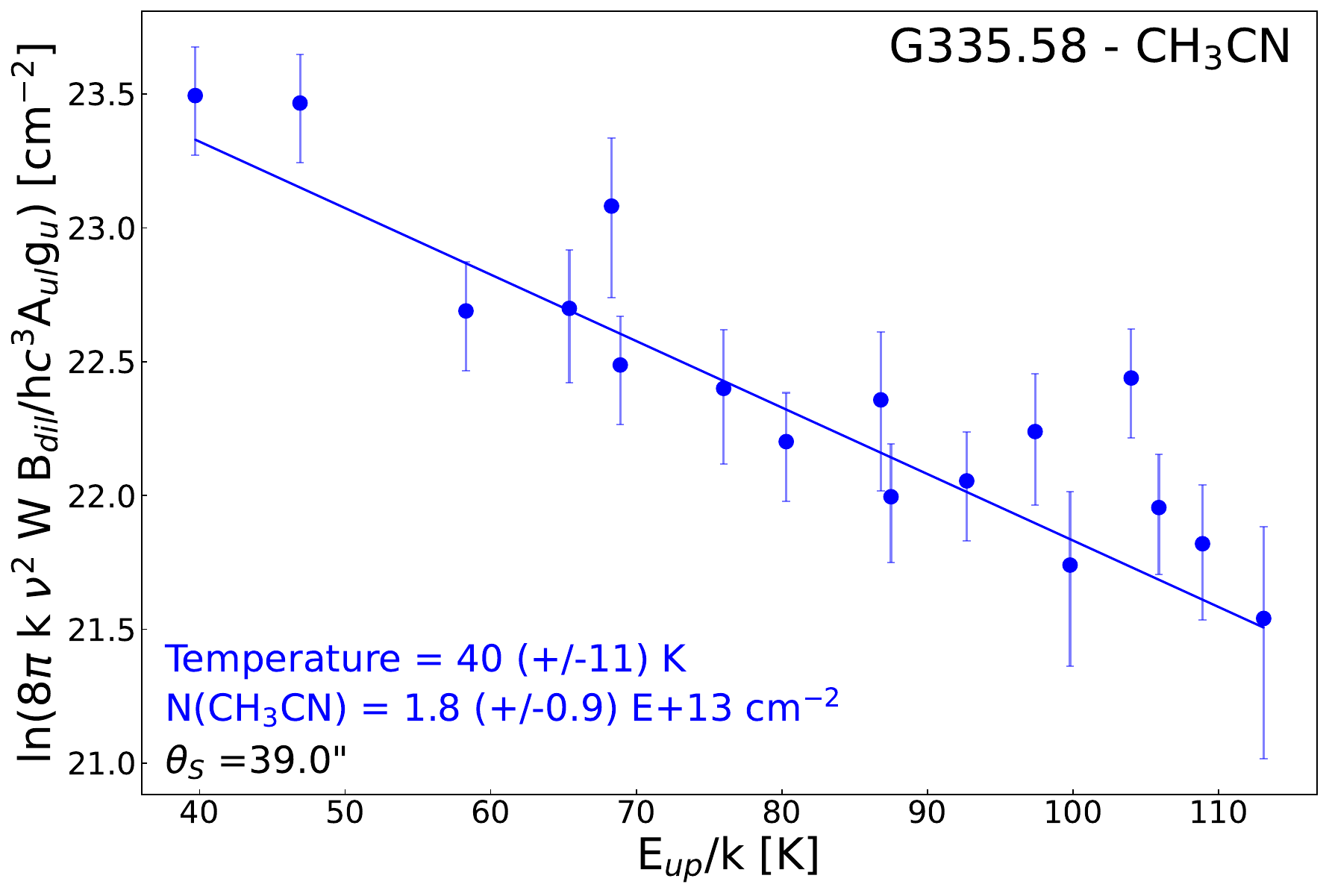}
    \includegraphics[width=0.45\linewidth]{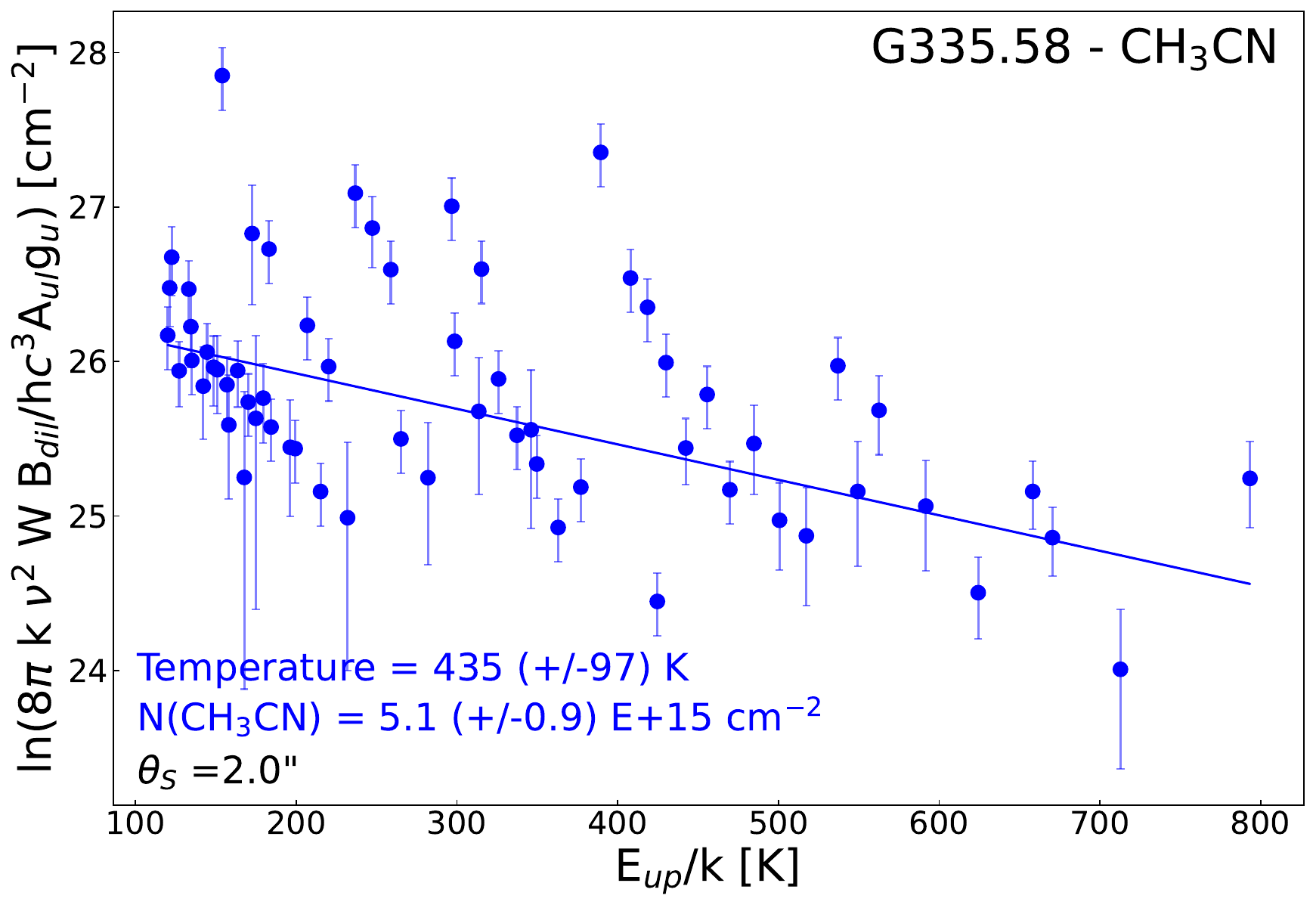}
    \includegraphics[width=0.45\linewidth]{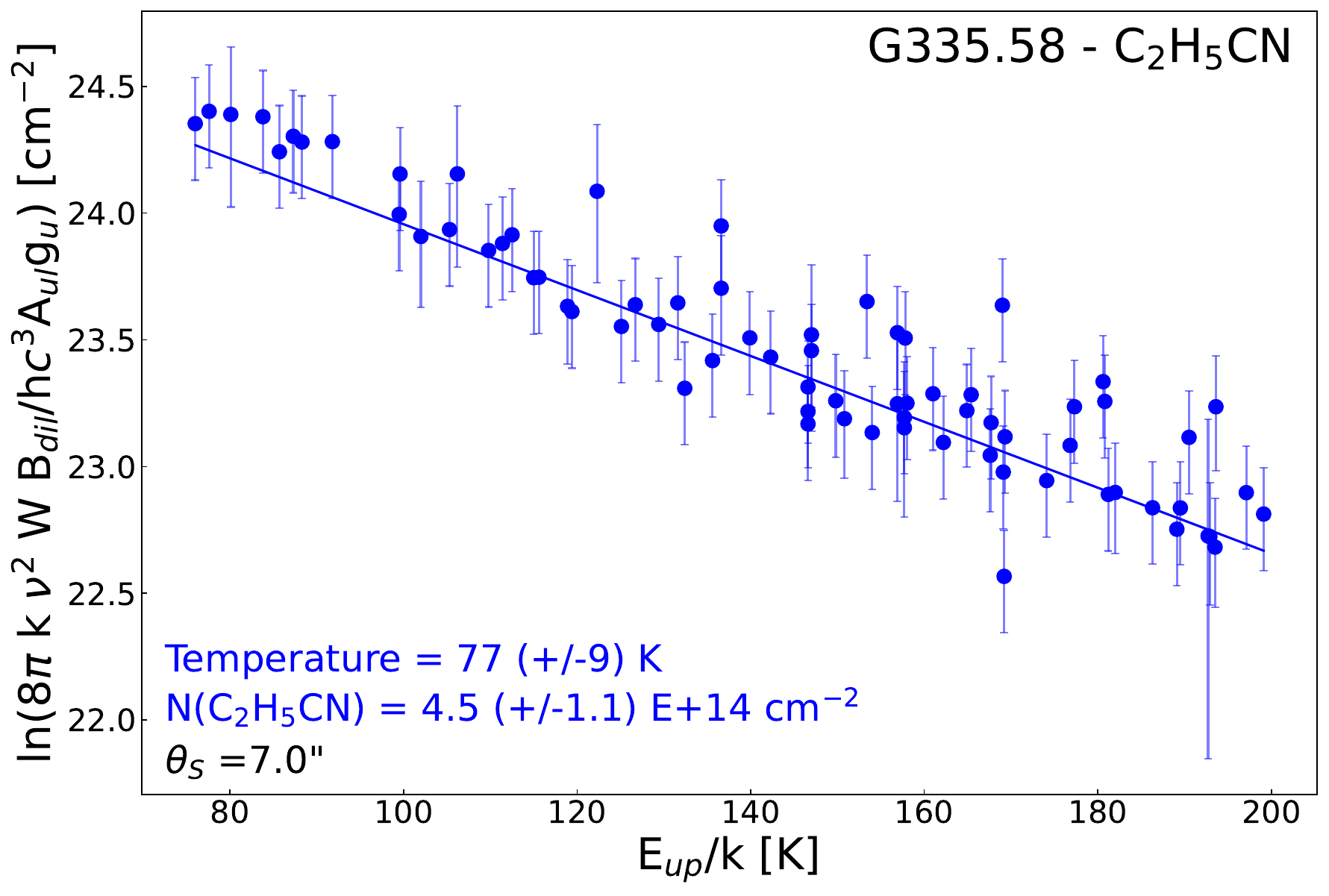}
    \includegraphics[width=0.45\linewidth]{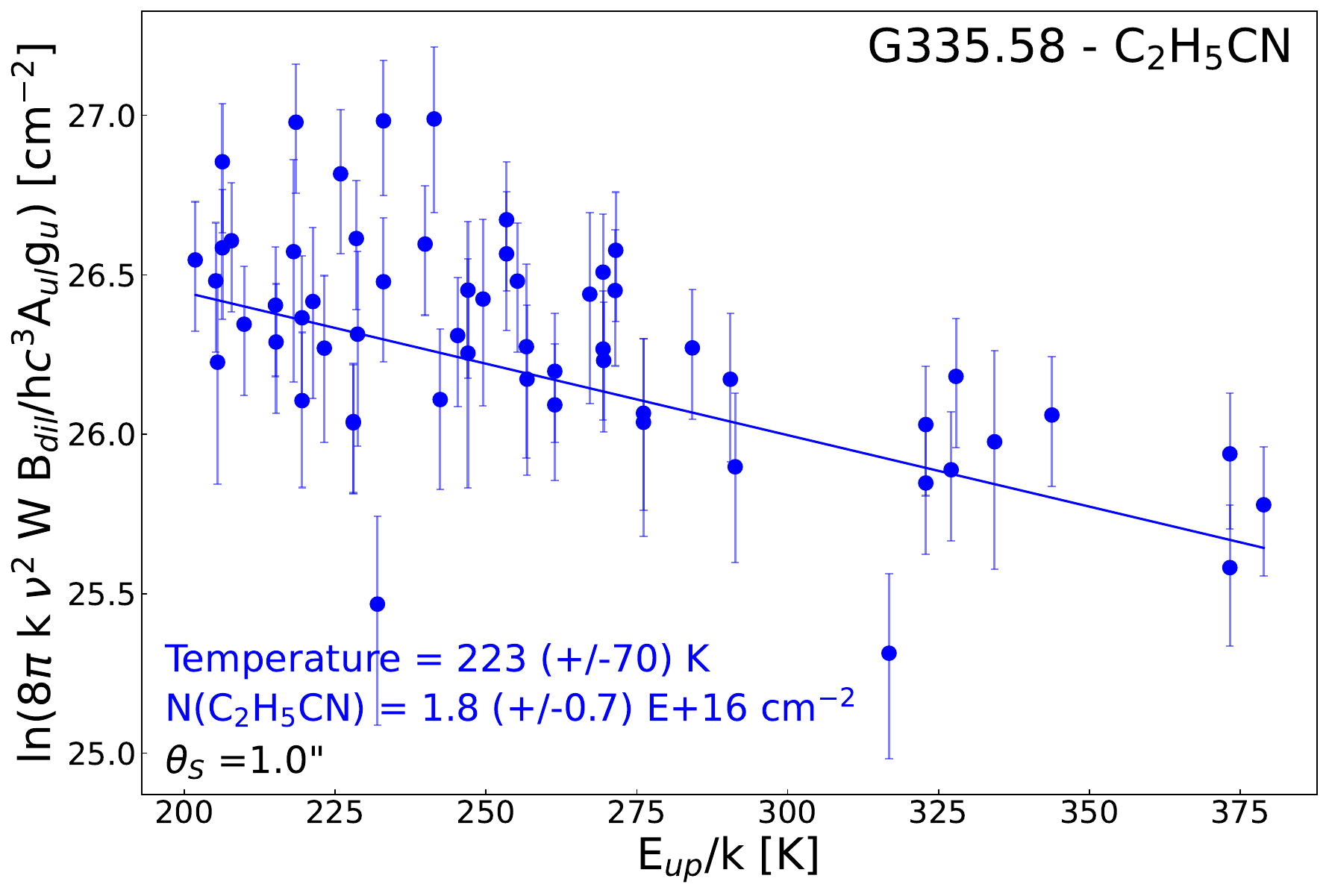}
    \includegraphics[width=0.45\linewidth]{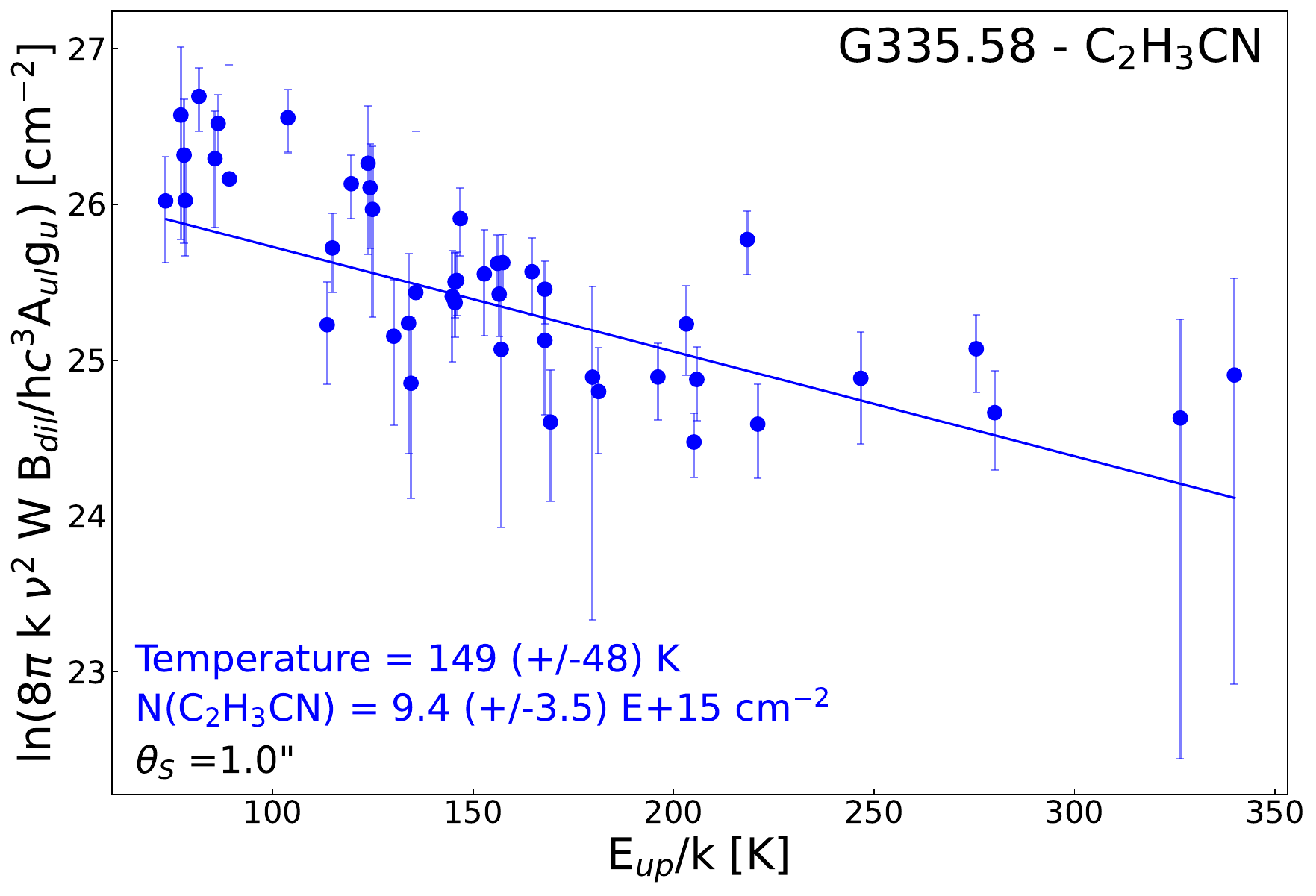}
    \caption{Rotational diagrams of the COMs in G335.58: CH$_3$COCH$_3$, CH$_3$CN, C$_2$H$_5$CN, and C$_2$H$_3$CN. Rotational temperatures and column densities are indicated with the same color code. The errorbars correspond to a 20\,\% error. To calculate the uncertainties on the rotational temperature and the column density, we used a Monte Carlo method assuming a uniform distribution of the error for each data point.}
    \label{fig:rot_diag_335p58_2}
\end{figure*}
\subsection{G335.78}
\begin{figure*}
    \centering
    \includegraphics[width=0.45\linewidth]{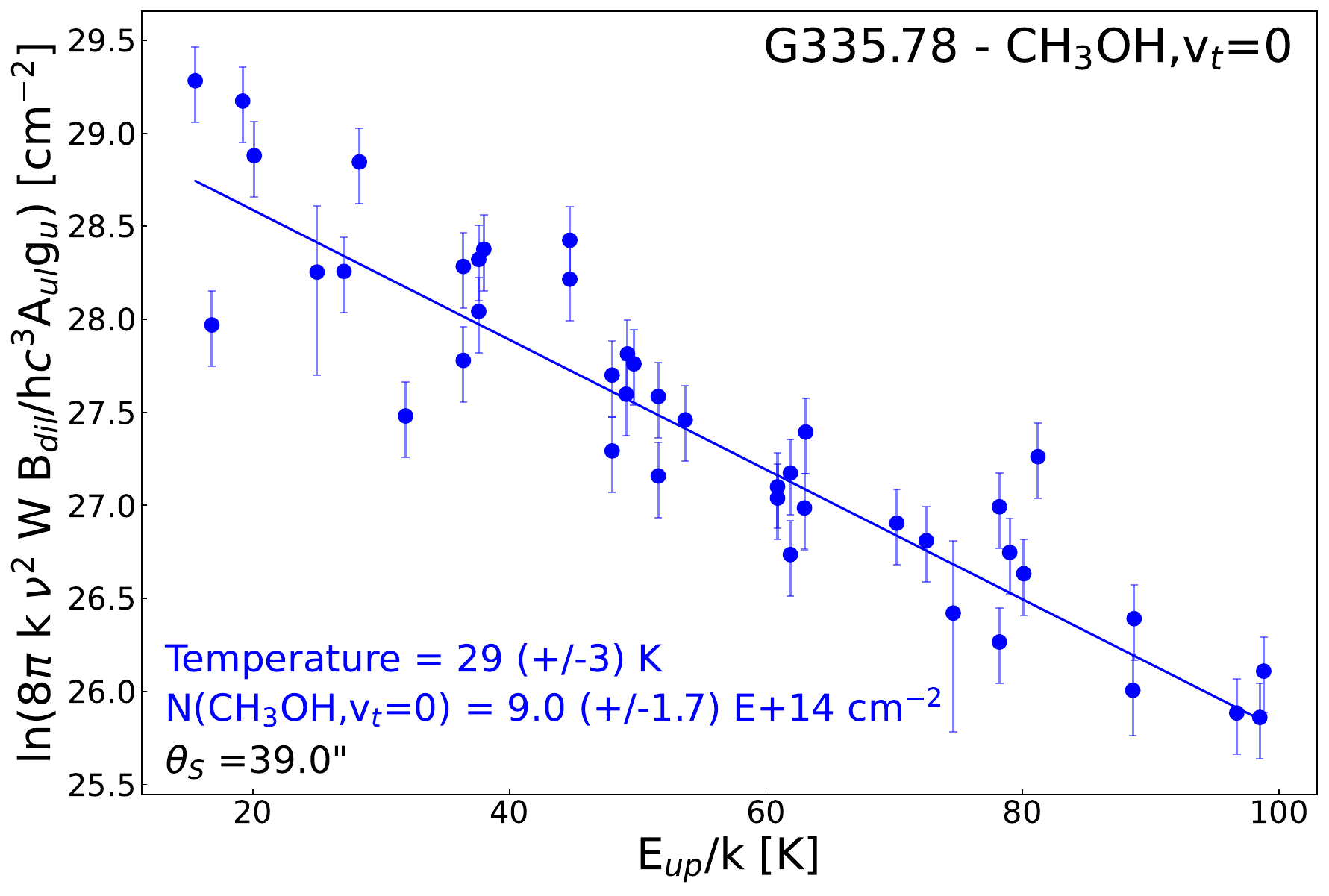}
    \includegraphics[width=0.45\linewidth]{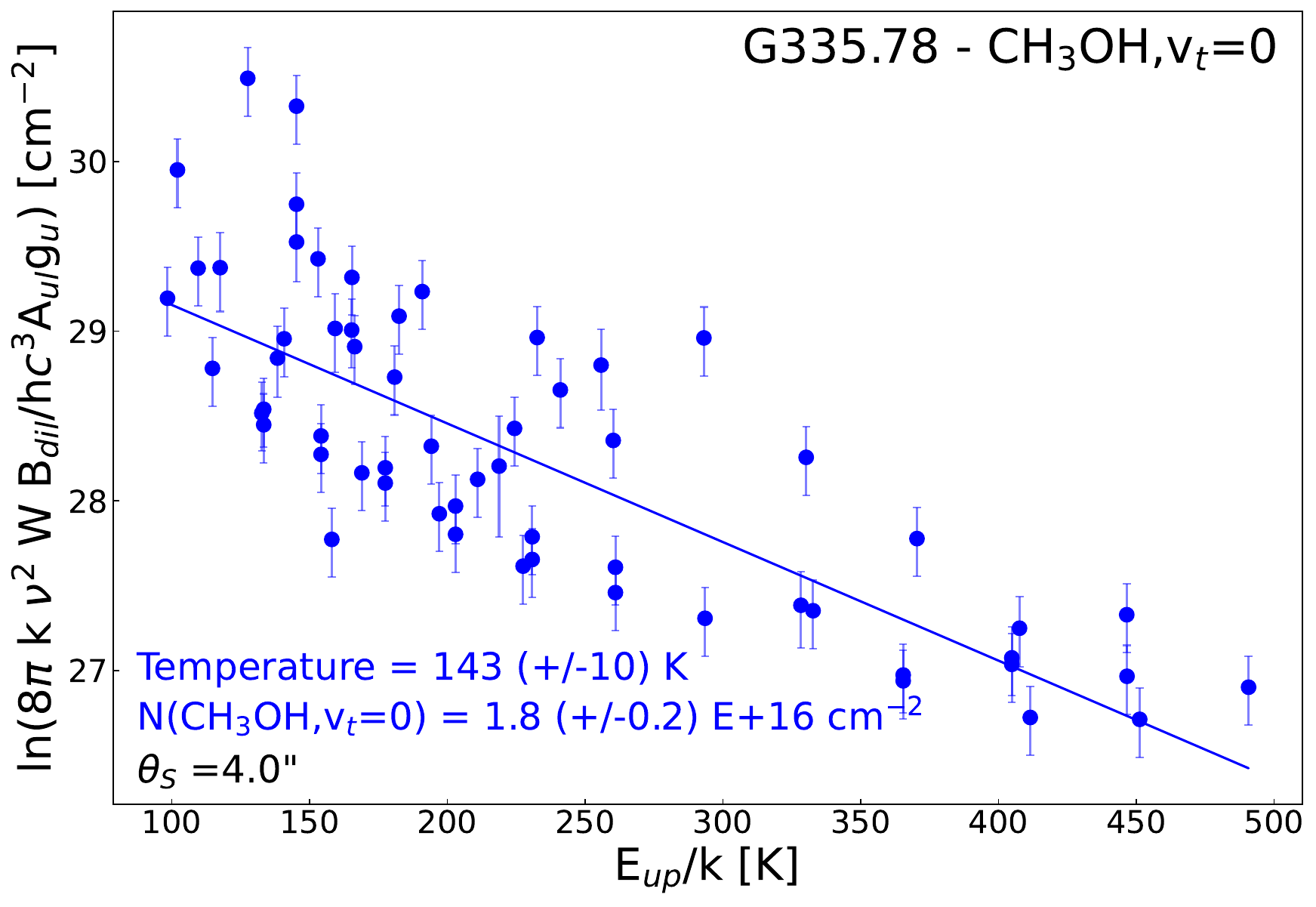}
    \includegraphics[width=0.45\linewidth]{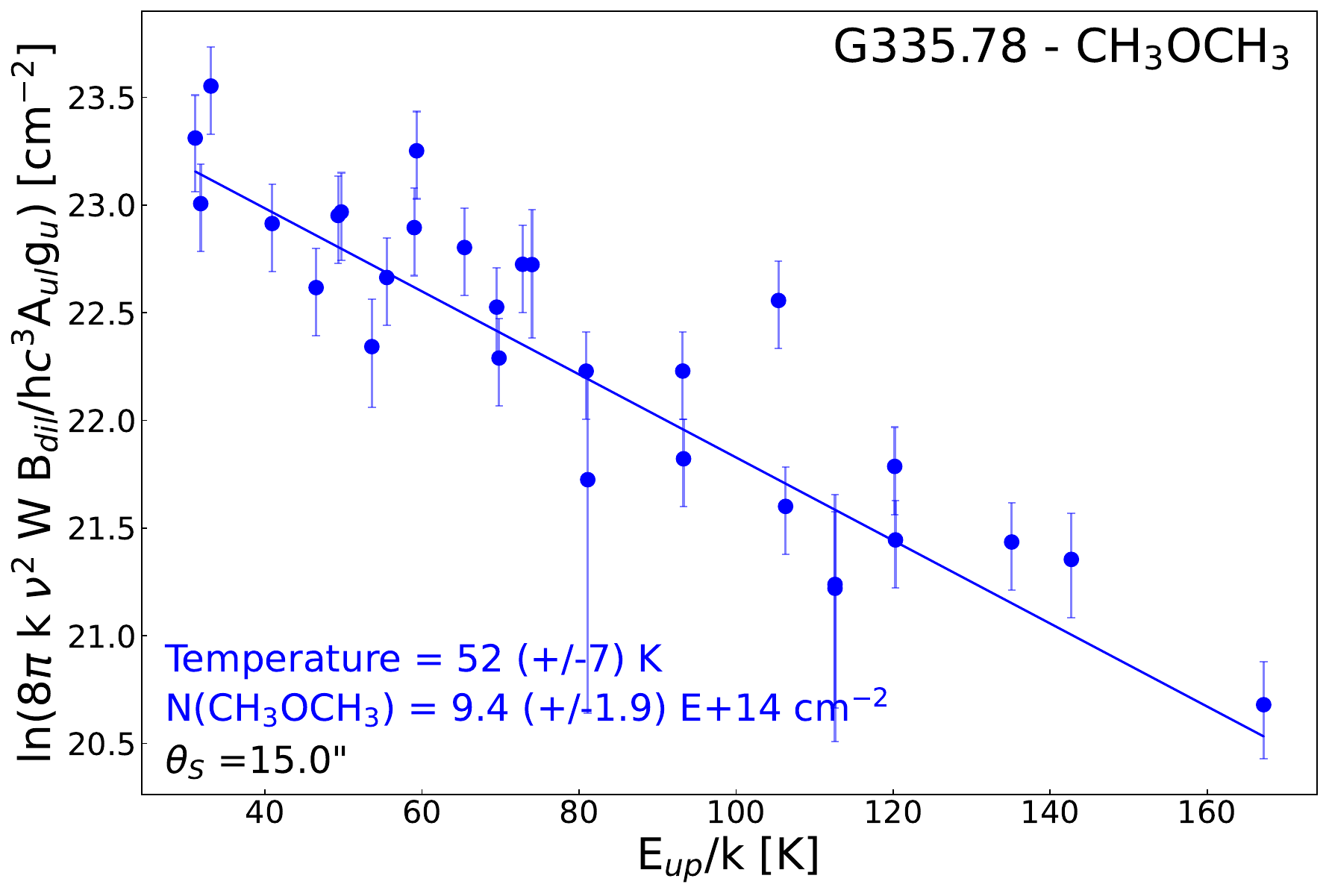}
    \includegraphics[width=0.45\linewidth]{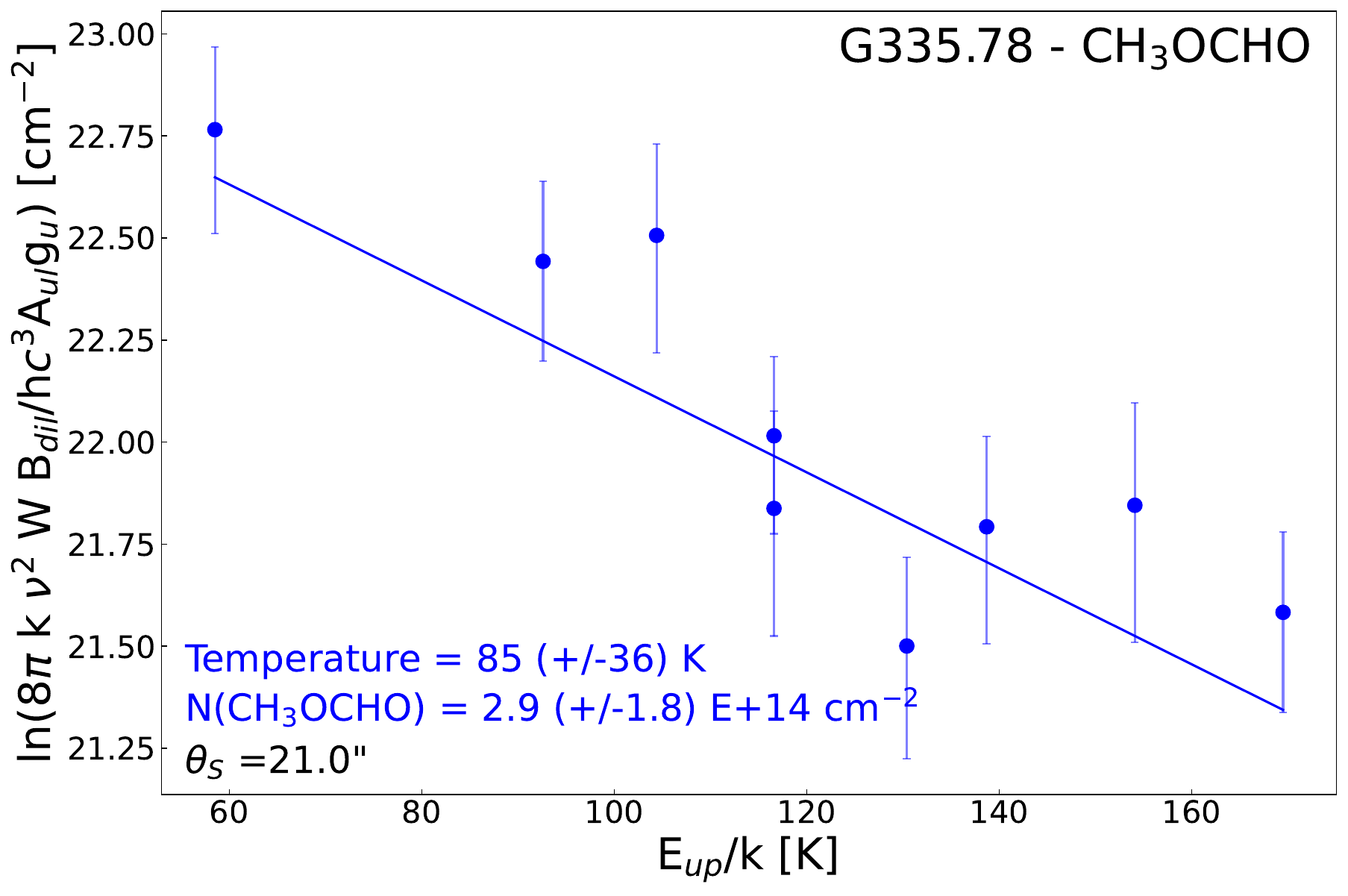}
    \includegraphics[width=0.45\linewidth]{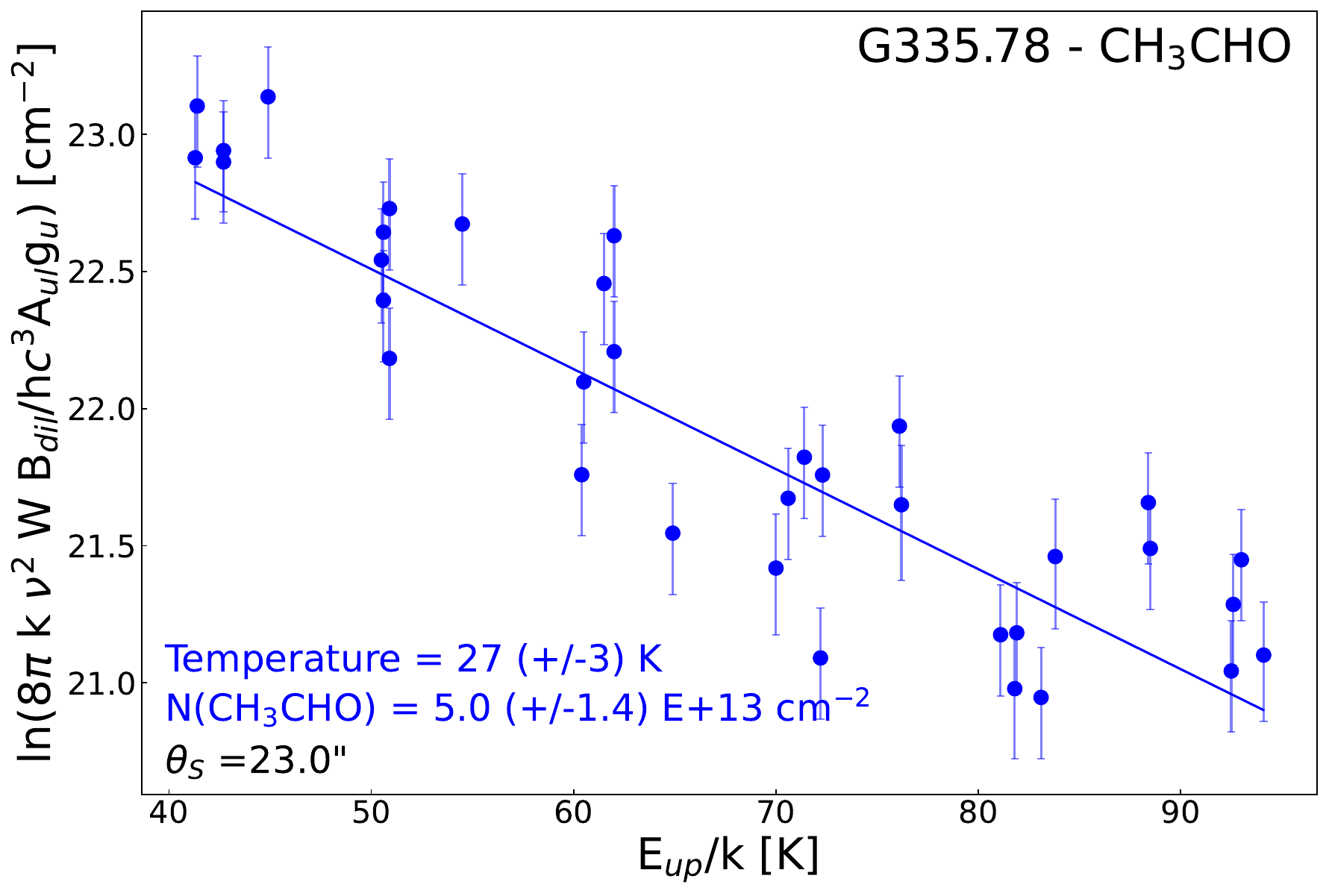}
    \includegraphics[width=0.45\linewidth]{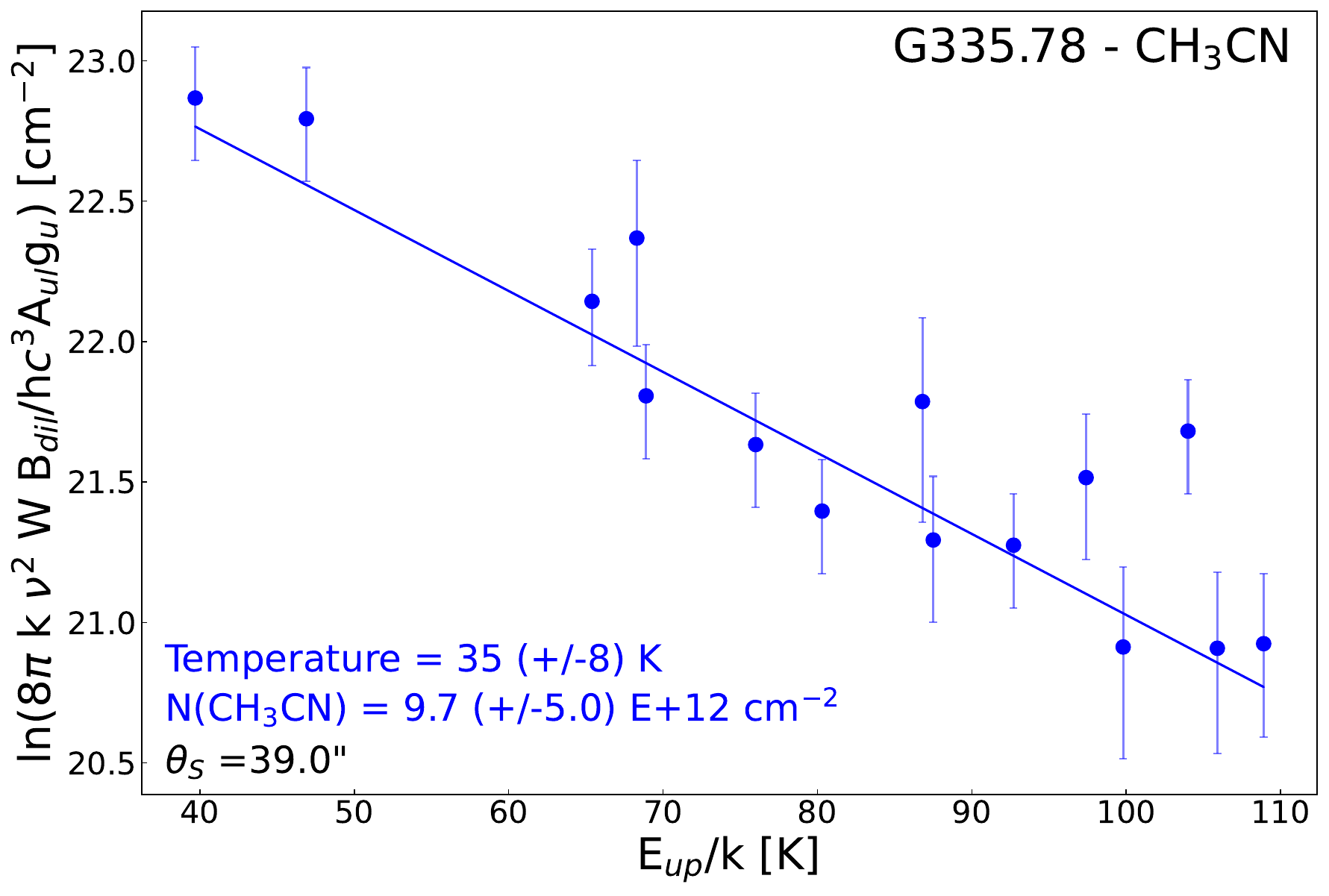}
    \includegraphics[width=0.45\linewidth]{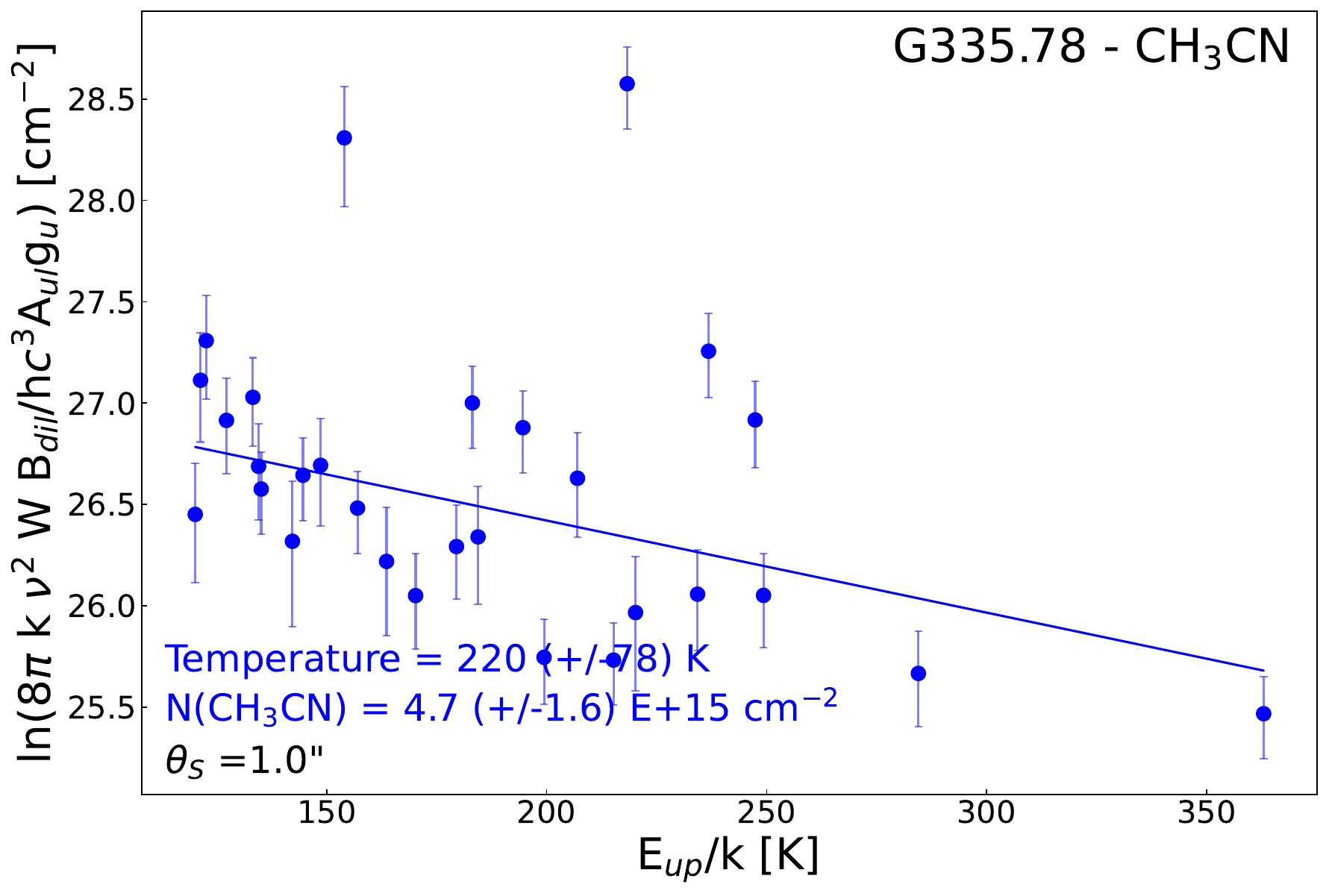}
    \caption{Rotational diagrams of the COMs in G335.78: CH$_3$OH, CH$_3$OCH$_3$, CH$_3$OCHO, CH$_3$CHO, and CH$_3$CN. Rotational temperatures and column densities are indicated with the same color code. The errorbars correspond to a 20\,\% error. To calculate the uncertainties on the rotational temperature and the column density, we used a Monte Carlo method assuming a uniform distribution of the error for each data point.}
    \label{fig:rot_diag_335p78}
\end{figure*}

\subsection{G343.75}
\begin{figure*}
    \centering
    \includegraphics[width=0.45\linewidth]{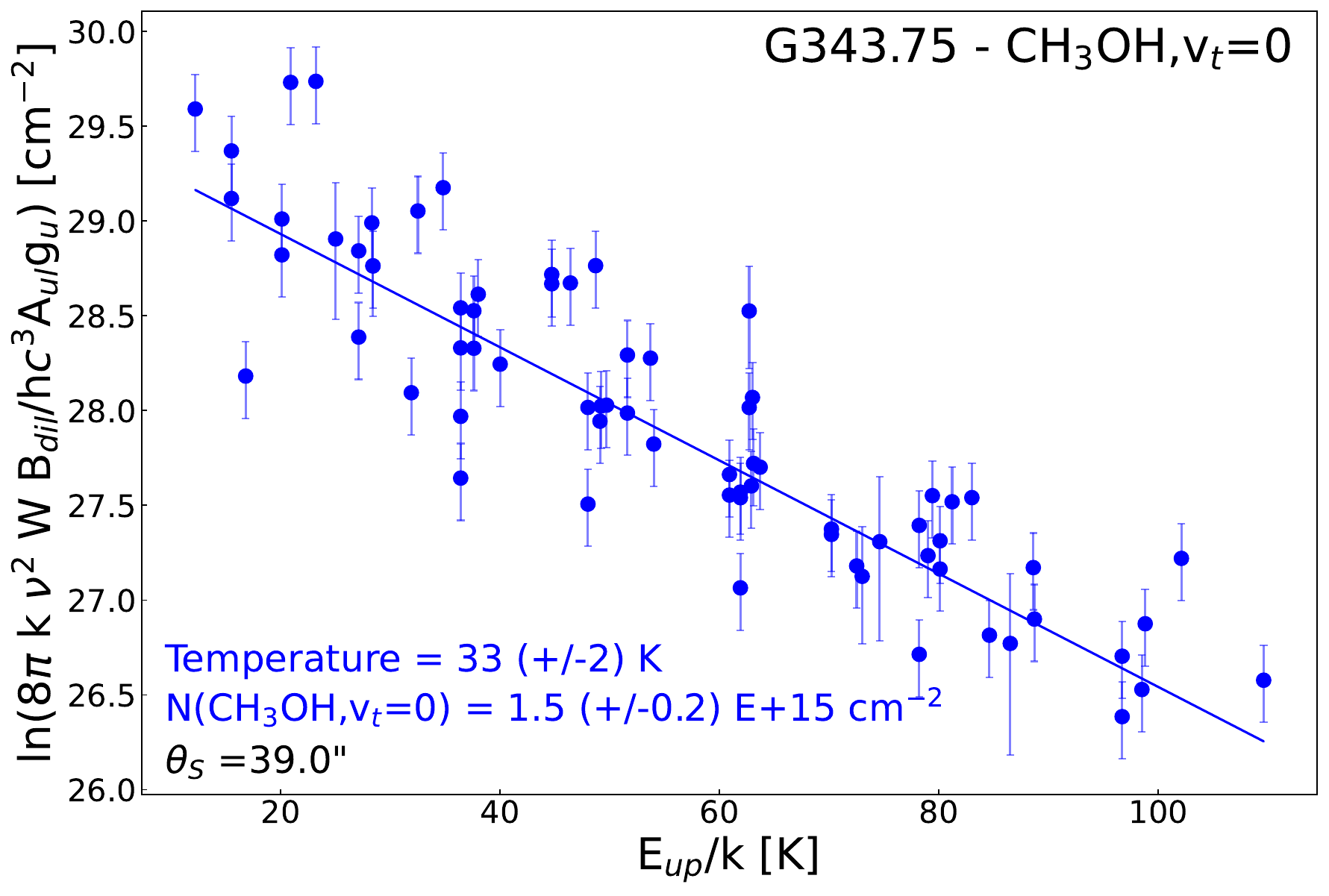}
    \includegraphics[width=0.45\linewidth]{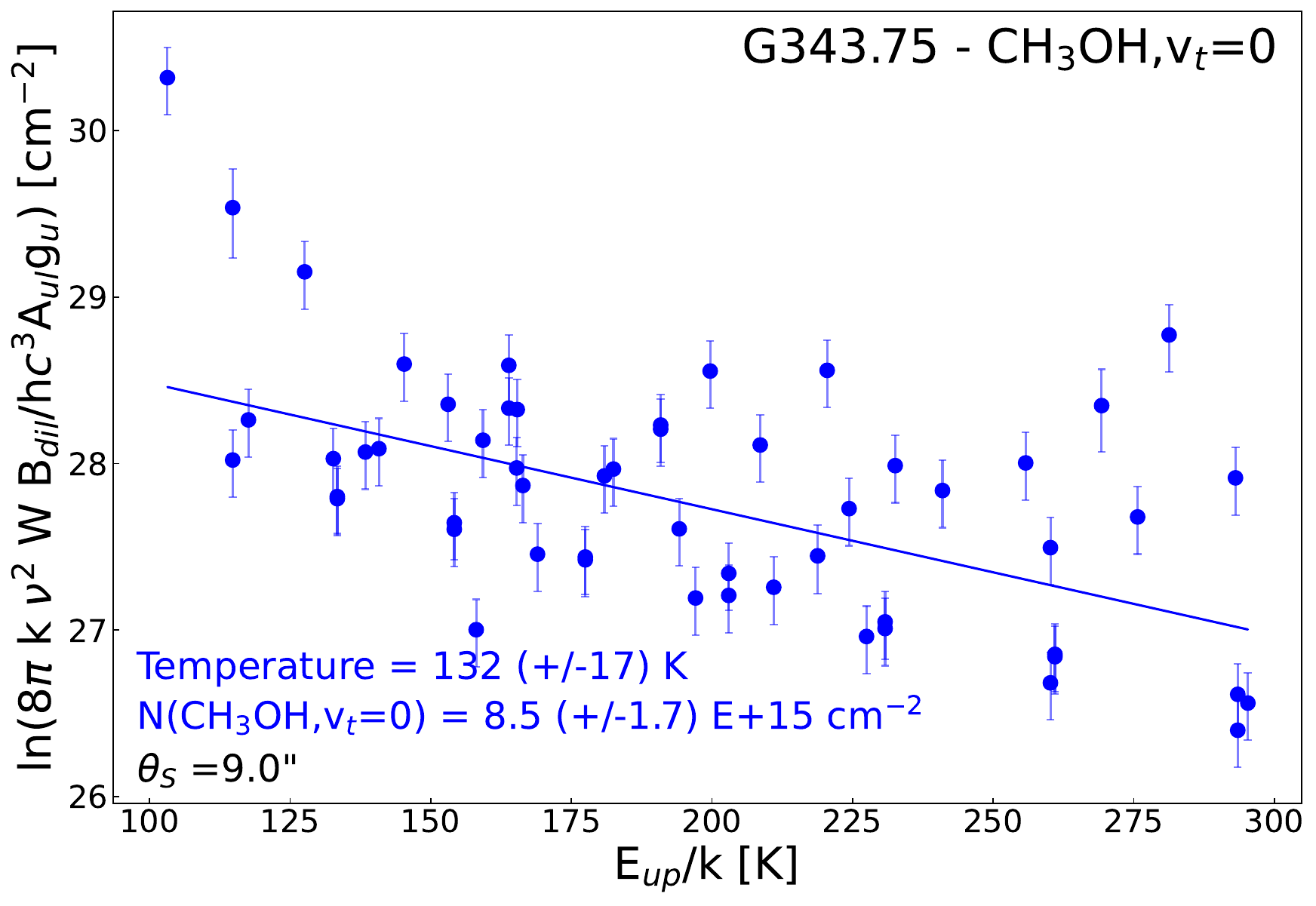}
    \includegraphics[width=0.45\linewidth]{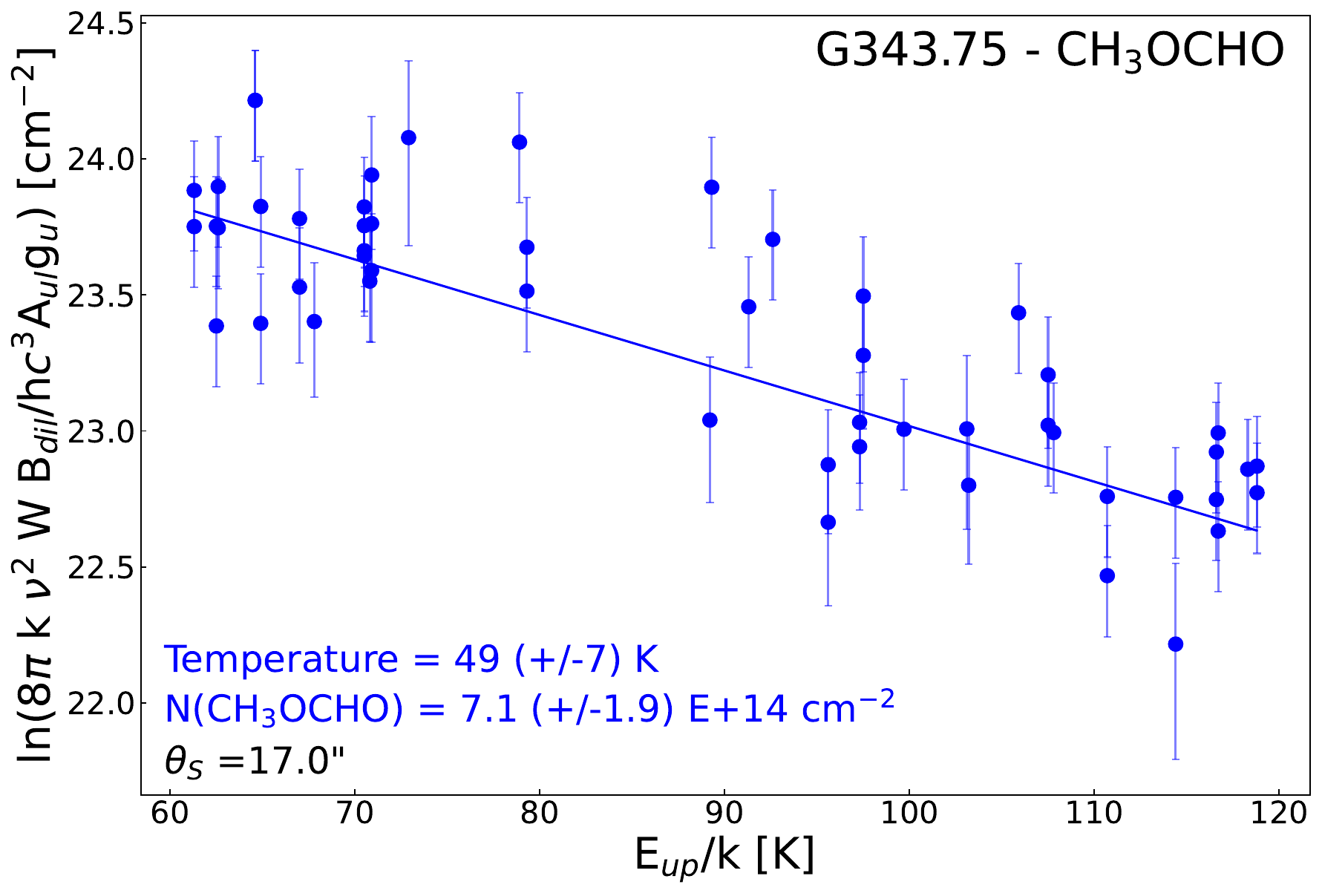}
    \includegraphics[width=0.45\linewidth]{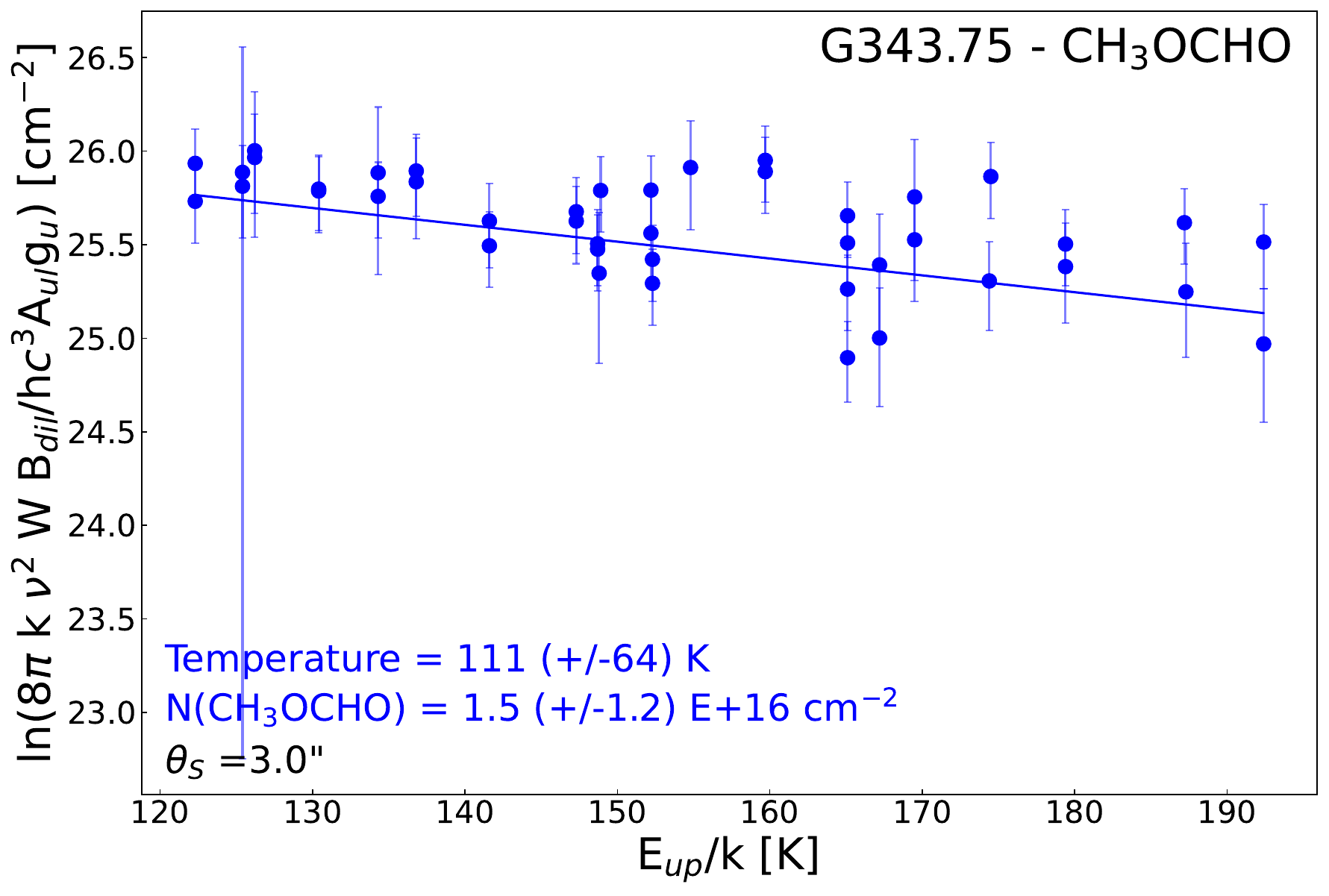}
    \includegraphics[width=0.45\linewidth]{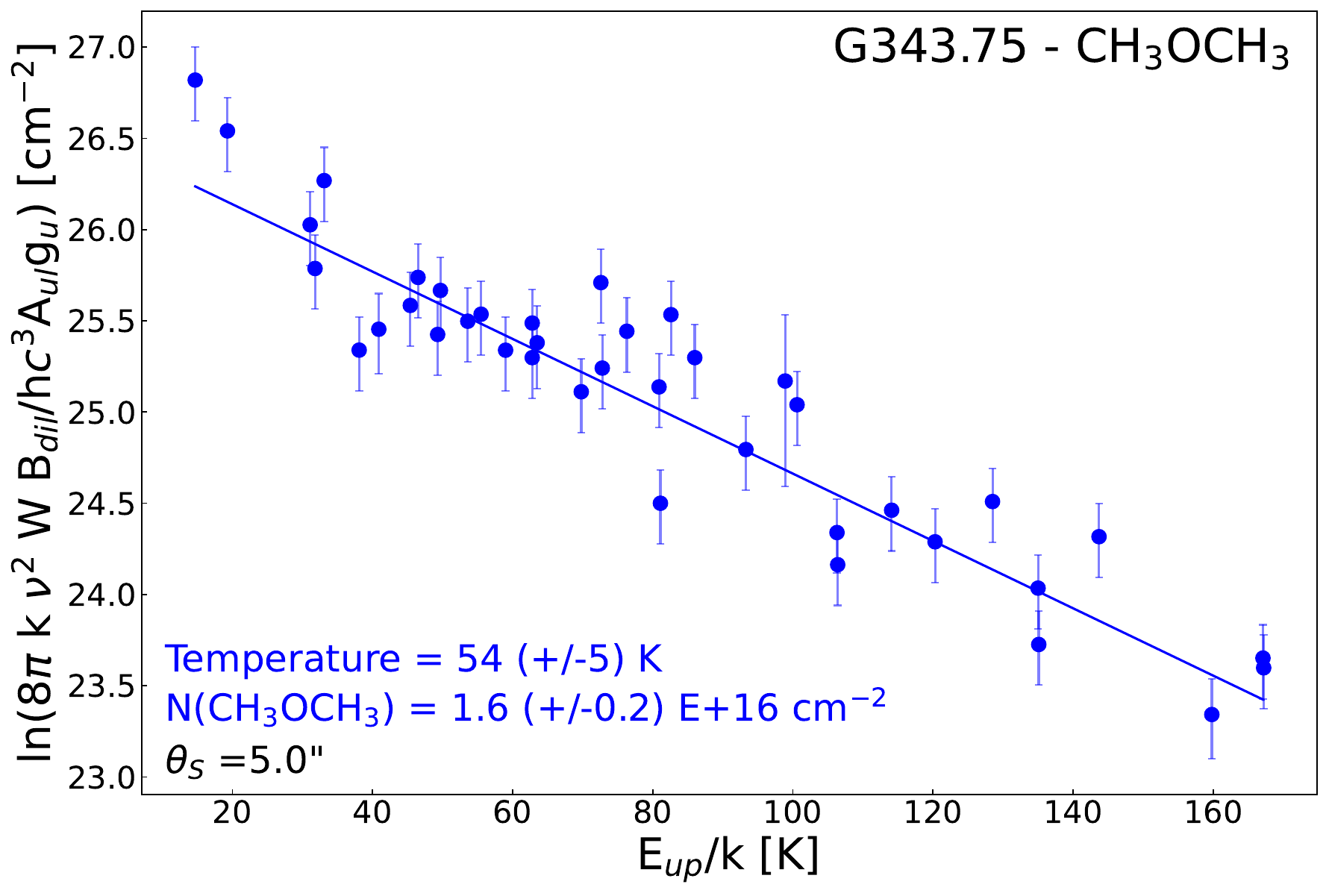}
    \includegraphics[width=0.45\linewidth]{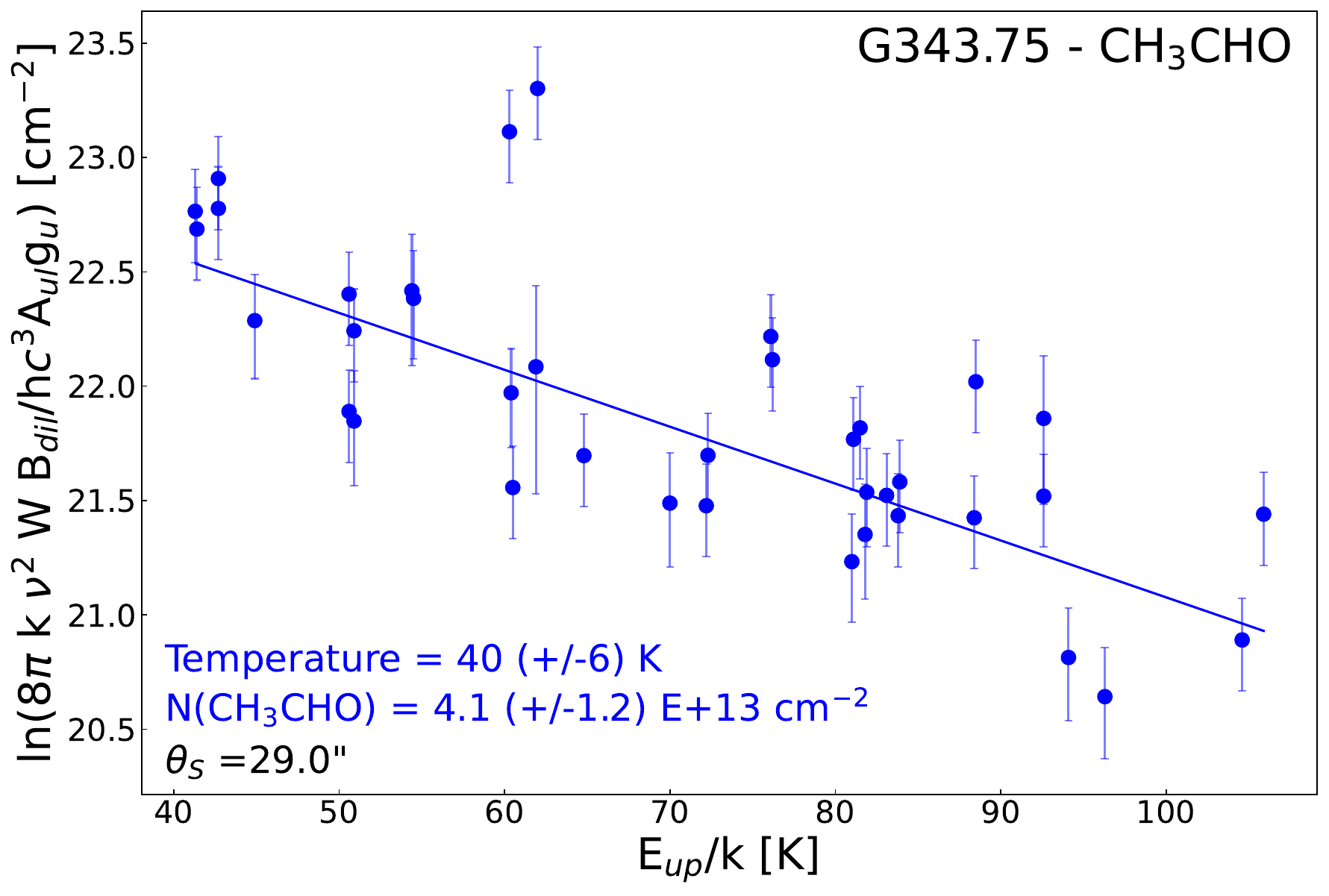}
    \caption{Rotational diagrams of the COMs in G343.75: CH$_3$OH, CH$_3$OCHO, CH$_3$OCH$_3$, and CH$_3$CHO. Rotational temperatures and column densities are indicated with the same color code. The errorbars correspond to a 20\,\% error. To calculate the uncertainties on the rotational temperature and the column density, we used a Monte Carlo method assuming a uniform distribution of the error for each data point.}
    \label{fig:rot_diag_343p75_1}
\end{figure*}
\begin{figure*}
    \centering
    \includegraphics[width=0.45\linewidth]{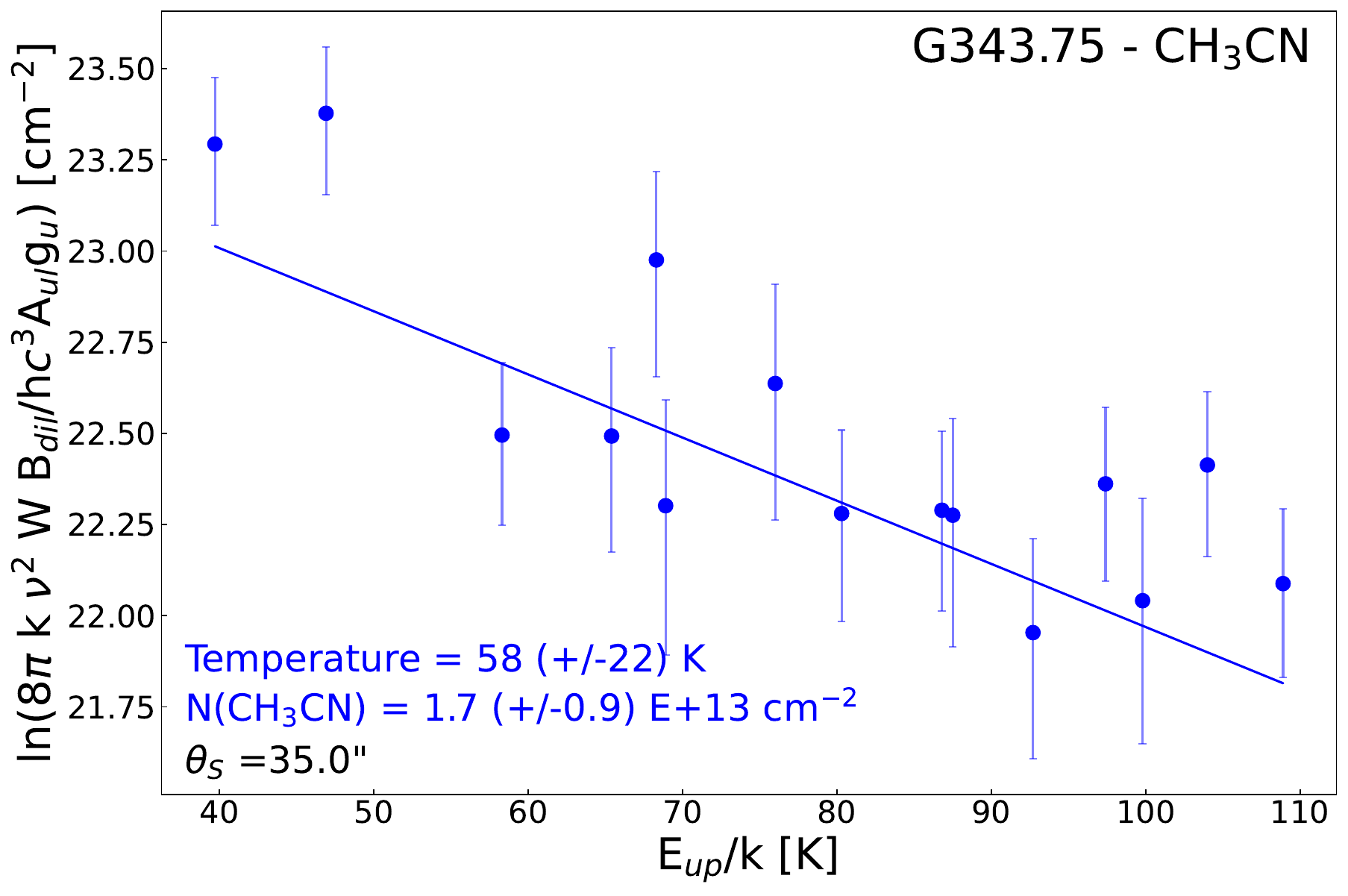}
    \includegraphics[width=0.45\linewidth]{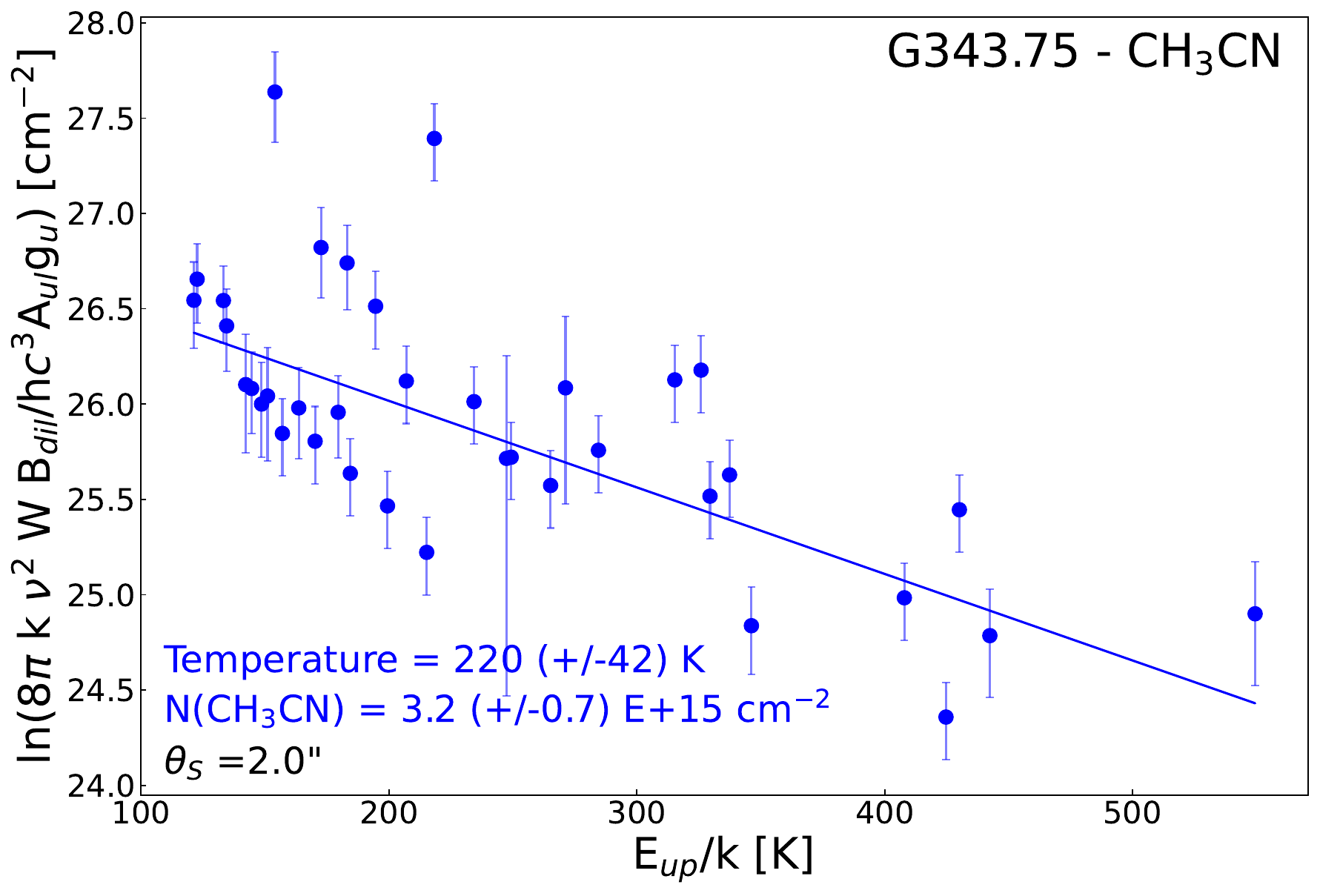}
    \includegraphics[width=0.45\linewidth]{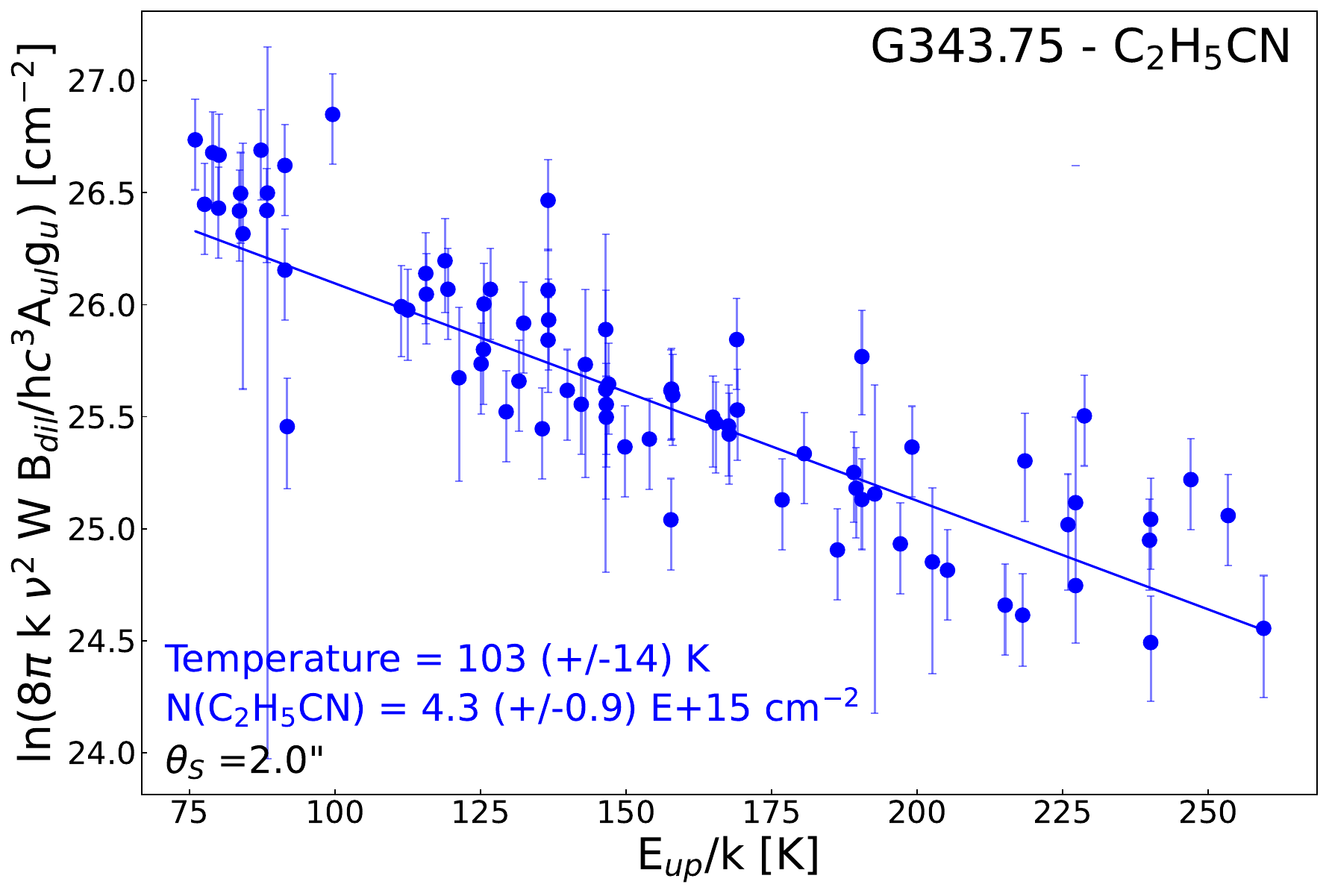}
    \caption{Rotational diagrams of the COMs in G343.75: CH$_3$CN and C$_2$H$_5$CN. Rotational temperatures and column densities are indicated with the same color code. The errorbars correspond to a 20\,\% error. To calculate the uncertainties on the rotational temperature and the column density, we used a Monte Carlo method assuming a uniform distribution of the error for each data point.}
    \label{fig:rot_diag_343p75_2}
\end{figure*}

\section{Column densities from the literature}
\label{tab:literature}
\begin{landscape}
    \begin{table}[h]
    \centering
    \small
    \setlength{\tabcolsep}{0.2pt}
    \caption{Column densities for all the sources extracted from the literature.}
    \label{tab:ref_comp} 
    \begin{tabular}{lccccccccccc}
\hline 
\hline
Source & Luminosity & {$N$(CH$_3$OH)} & {$N$(CH$_3$OCH$_3$)} & {$N$(CH$_3$OCHO)} & {$N$(C$_2$H$_5$OH)} & {$N$(CH$_3$COCH$_3$)} & {$N$(CH$_3$SH)} & {$N$(CH$_3$CN)} & {$N$(C$_2$H$_5$CN)} & {$N$(C$_2$H$_3$CN)}
& References \\
& {[L$_\odot$]} & {\unidens} &  {\unidens} & {\unidens} & {\unidens} & {\unidens} & {\unidens} & {\unidens} & {\unidens} & {\unidens} & ~ \\
\hline 
Hot cores \\
G19.61 & 1.0$\times$10$^5$ & 1.8$\times$10$^{15}$ & 4.1$\times$10$^{14}$ & 1.9$\times$10$^{14}$ & -- & -- & -- & 1.2$\times$10$^{14}$ & 5.9$\times$10$^{13}$ & 3.3$\times$10$^{13}$ & \citet{Widicus2017}\\
&  & & & & & & & & & & \citet{Wu2010}\\
G10.47 & 3.0$\times$10$^5$ & 1.1$\times$10$^{16}$ & 5.0$\times$10$^{15}$ & 2.8$\times$10$^{15}$ & 1.7$\times$10$^{15}$ & -- & -- & 5.4$\times$10$^{14}$ & 2.1$\times$10$^{14}$ & 4.5$\times$10$^{14}$ & \citet{Widicus2017}\\
&  & & & & & & & & & & \citet{Cesaroni2010}\\
G31 & 2.0$\times$10$^5$ & 3.7$\times$10$^{15}$ & 2.6$\times$10$^{15}$ & 1.4$\times$10$^{15}$ & -- & -- & -- & 1.2$\times$10$^{14}$ & 7.6$\times$10$^{13}$ & 2.2$\times$10$^{13}$ & \citet{Widicus2017}\\
&  & & & & & & & & & & \citet{Immer2019, Osorio2009}\\
SgrB2(N2) & 2.6$\times$10$^5$ & 4.0$\times$10$^{19}$ &  & 1.2$\times$10$^{18}$ & 3.1$\times$10$^{17}$ & 2.24.0$\times$10$^{18}$ & 3.4$\times$10$^{17}$ & 2.0$\times$10$^{18}$ & 6.2$\times$10$^{18}$ & 4.2$\times$10$^{17}$ & \citet{Belloche2016}\\
SgrB2(N3) & 4.5$\times$10$^4$ & 4.0$\times$10$^{18}$ & -- & 1.2$\times$10$^{18}$ & 3.1$\times$10$^{17}$ & -- & 6.0$\times$10$^{16}$ & 4.2$\times$10$^{17}$ & 2.3$\times$10$^{17}$ & 5.0$\times$10$^{16}$ & \citet{Bonfand2017, Bonfand2019}\\
&  & & & & & & & & & & \citet{Reid2019} \\
SgrB2(N4) & 3.9$\times$10$^5$ & 2.5$\times$10$^{17}$ & -- & 8.0$\times$10$^{16}$ & 3.1$\times$10$^{17}$ & -- & 1.5$\times$10$^{16}$ & 3.2$\times$10$^{16}$ & 1.2$\times$10$^{16}$ & 1.5$\times$10$^{15}$ & \citet{Bonfand2017, Bonfand2019}\\
&  & & & & & & & & & & \citet{Reid2019} \\
SgrB2(N5) & 3.8$\times$10$^5$ & 9.5$\times$10$^{17}$ & -- & 1.8$\times$10$^{17}$ & 9.9$\times$10$^{16}$ & -- & 2.5$\times$10$^{16}$ & 9.6$\times$10$^{16}$ & 7.7$\times$10$^{16}$ & 1.0$\times$10$^{16}$ & \citet{Bonfand2017, Bonfand2019}\\
&  & & & & & & & & & & \citet{Reid2019} \\
18089-1732 & 5.0$\times$10$^4$ & -- & 2.9$\times$10$^{17}$ & 2.5$\times$10$^{17}$ & -- & -- & -- & -- & 8.6$\times$10$^{15}$ & -- & \citet{Coletta2020}\\
18517+0437 & 1.0$\times$10$^4$ & -- & 1.6$\times$10$^{16}$ & 4.4$\times$10$^{16}$ & -- & -- & -- & -- & -- & -- & \citet{Coletta2020}\\
W3OH & 1.0$\times$10$^5$ & -- & 6.8$\times$10$^{16}$ & 1.0$\times$10$^{17}$ & -- & -- & -- & -- & 4.1$\times$10$^{15}$ & -- & \citet{Coletta2020}\\
W51 & 7.0e+06 & -- & 1.8$\times$10$^{18}$ & 2.7$\times$10$^{18}$ & -- & -- & -- & -- & 7.3$\times$10$^{16}$ & -- & \citet{Coletta2020}\\
ON1 & 2.0$\times$10$^4$ & -- & 8.1$\times$10$^{16}$ & 3.2$\times$10$^{16}$ & -- & -- & -- & -- & -- & -- & \citet{Coletta2020}\\
AFGL4176 & -- & 5.5$\times$10$^{18}$ & 1.3$\times$10$^{17}$ & 1.7$\times$10$^{17}$ & 7.6$\times$10$^{16}$ & 5.3$\times$10$^{16}$ & -- & 3.4$\times$10$^{16}$ & 6.4$\times$10$^{15}$ & 6.2$\times$10$^{15}$ & \citet{Bogelund2019}\\
AFGL5142-MM1 & 4.0$\times$10$^3$ & -- & 4.1$\times$10$^{15}$ & 8.1$\times$10$^{15}$ & -- & -- & -- & -- & -- & -- & \citet{Coletta2020}\\
G24p78 & 2.0$\times$10$^5$ & -- & 1.8$\times$10$^{18}$ & 1.7$\times$10$^{17}$ & -- & -- & -- & -- & -- & -- & \citet{Coletta2020}\\
G75 & 6.3$\times$10$^4$ & -- & 2.3$\times$10$^{16}$ & 1.0$\times$10$^{16}$ & -- & -- & -- & -- & -- & -- & \citet{Coletta2020}\\
G14p33 & 2.0$\times$10$^4$ & -- & 7.4$\times$10$^{16}$ & 1.9$\times$10$^{16}$ & -- & -- & -- & -- & -- & -- & \citet{Coletta2020}\\
OrionKL & 1.0$\times$10$^5$ & 2$\times$10$^{16}$ & 1$\times$10$^{16}$ & 6$\times$10$^{15}$ & 2$\times$10$^{15}$ & -- & -- & 2$\times$10$^{15}$ & 5$\times$10$^{15}$ & 5$\times$10$^{15}$ & \citet{Beuther2009}\\
101899 & 9.2$\times$10$^4$ & -- & -- & -- & -- & -- & -- & 2.2$\times$10$^{16}$ & 8.2$\times$10$^{15}$ & 1.8$\times$10$^{15}$ & \citet{Nazari2022}\\
126348 & 6.8$\times$10$^3$ & -- & -- & -- & -- & -- & -- & 4.2$\times$10$^{15}$ & 1.0$\times$10$^{15}$ & 1.7$\times$10$^{14}$ & \citet{Nazari2022}\\
615590 & 5.5$\times$10$^3$ & 1.9$\times$10$^{18}$ & -- & -- & -- & -- & -- & 1.5$\times$10$^{16}$ & 9.7$\times$10$^{14}$ & 6.2$\times$10$^{14}$ & \citet{Nazari2022}\\
693050 & 1.2$\times$10$^4$ & -- & -- & -- & -- & -- & -- & 1.9$\times$10$^{16}$ & 3.7$\times$10$^{15}$ & 6.4$\times$10$^{14}$ & \citet{Nazari2022}\\
705768 & 9.2$\times$10$^4$ & -- & -- & -- & -- & -- & -- & 5.2$\times$10$^{15}$ & 8.6$\times$10$^{14}$ & -- & \citet{Nazari2022}\\
707948 & 2.0$\times$10$^5$ & -- & -- & -- & -- & -- & -- & 1.2$\times$10$^{17}$ & 2.9$\times$10$^{16}$ & 8.0$\times$10$^{15}$ & \citet{Nazari2022}\\
717461A & 3.3$\times$10$^3$ & -- & -- & -- & -- & -- & -- & 4.5$\times$10$^{15}$ & 3.3$\times$10$^{14}$ & 2.1$\times$10$^{14}$ & \citet{Nazari2022}\\
881427C & 4.2$\times$10$^2$ & -- & -- & -- & -- & -- & -- & 9.3$\times$10$^{16}$ & 1.3$\times$10$^{16}$ & 2.2$\times$10$^{15}$ & \citet{Nazari2022}\\
G023p39 & 3.1$\times$10$^2$ & -- & -- & -- & -- & -- & -- & 5.4$\times$10$^{15}$ & 7.2$\times$10$^{14}$ & 2.0$\times$10$^{14}$ & \citet{Nazari2022}\\
G025p65 & 3.9$\times$10$^2$ & -- & -- & -- & -- & -- & -- & 1.9$\times$10$^{16}$ & 1.7$\times$10$^{15}$ & 1.9$\times$10$^{15}$ & \citet{Nazari2022}\\
G305p20 & 4.2$\times$10$^2$ & -- & -- & -- & -- & -- & -- & 1.1$\times$10$^{16}$ & 1.6$\times$10$^{15}$ & 9.6$\times$10$^{14}$ & \citet{Nazari2022}\\
G314p32 & 4.1$\times$10$^2$ & -- & -- & -- & -- & -- & -- & 3.1$\times$10$^{16}$ & 6.7$\times$10$^{15}$ & 1.3$\times$10$^{15}$ & \citet{Nazari2022}\\
G316p64 & 4.1$\times$10$^2$ & -- & -- & -- & -- & -- & -- & 3.1$\times$10$^{16}$ & 6.7$\times$10$^{15}$ & 1.3$\times$10$^{15}$ & \citet{Nazari2022}\\
G318p05 & 3.9$\times$10$^2$ & -- & -- & -- & -- & -- & -- & 8.8$\times$10$^{15}$ & 1.0$\times$10$^{15}$ & 2.2$\times$10$^{14}$ & \citet{Nazari2022}\\
G318p94 & 4.4$\times$10$^2$ & -- & -- & -- & -- & -- & -- & 5.4$\times$10$^{16}$ & 6.2$\times$10$^{15}$ & 1.2$\times$10$^{15}$ & \citet{Nazari2022}\\
G323p74 & 3.8$\times$10$^2$ & -- & -- & -- & -- & -- & -- & 1.7$\times$10$^{16}$ & 3.0$\times$10$^{15}$ & 6.2$\times$10$^{14}$ & \citet{Nazari2022}\\
G326p48 & 4.1$\times$10$^2$ & -- & -- & -- & -- & -- & -- & 2.1$\times$10$^{15}$ & 3.4$\times$10$^{14}$ & 1.5$\times$10$^{14}$ & \citet{Nazari2022}\\
G326p66 & 4.0$\times$10$^2$ & -- & -- & -- & -- & -- & -- & 5.2$\times$10$^{15}$ & 3.4$\times$10$^{14}$ & 2.0$\times$10$^{14}$ & \citet{Nazari2022}\\
G327p12 & 3.2$\times$10$^2$ & -- & -- & -- & -- & -- & -- & 2.6$\times$10$^{16}$ & 4.2$\times$10$^{15}$ & 1.3$\times$10$^{15}$ & \citet{Nazari2022}\\
G343p13 & 4.0$\times$10$^2$ & -- & -- & -- & -- & -- & -- & 3.0$\times$10$^{16}$ & 6.1$\times$10$^{15}$ & 4.5$\times$10$^{15}$ & \citet{Nazari2022}\\
G345p50 & 4.0$\times$10$^2$ & -- & -- & -- & -- & -- & -- & 1.3$\times$10$^{17}$ & 1.1$\times$10$^{16}$ & 4.5$\times$10$^{15}$ & \citet{Nazari2022}\\
\hline
\end{tabular}
\end{table}
\end{landscape}

\begin{landscape}
\begin{table}[h]
    \centering
    \small
    \setlength{\tabcolsep}{0.5pt}
    \caption{Column densities for all the sources extracted from the literature.} 
    \label{tab:ref_comp2} 
    \begin{tabular}{lccccccccccc}
Hot corinos \\
IRAS16293B & 2.7$\times$10$^1$ & 1.0$\times$10$^{19}$ & 2.4$\times$10$^{17}$ & 2.6$\times$10$^{17}$ & 2.3$\times$10$^{17}$ & -- & 5.5$\times$10$^{15}$ & 4.0$\times$10$^{16}$ & 3.6$\times$10$^{15}$ & 7.4$\times$10$^{14}$ & \citet{Jorgensen2020} and references therein \\
IRAS4A2 & 4.7 & 6.0$\times$10$^{17}$ & 6.0$\times$10$^{16}$ & 8.9$\times$10$^{16}$ & 5.3$\times$10$^{16}$ & -- & -- & 1.1$\times$10$^{16}$ & 2.3$\times$10$^{15}$ & -- & \citet{Belloche2020}\\
L483 & 1.3$\times$10$^1$ & 1.7$\times$10$^{19}$ & 8.0$\times$10$^{16}$ & 1.3$\times$10$^{17}$ & 1.0$\times$10$^{17}$ & -- & -- & -- & -- & -- & \citep{Oya2017}\\
L1448C & 1.1$\times$10$^1$ & 8.0$\times$10$^{16}$ & 1.0$\times$10$^{16}$ & 5.2$\times$10$^{15}$ & 4.4$\times$10$^{15}$ & -- & -- & 2.5$\times$10$^{15}$ & 3.9$\times$10$^{14}$ & -- & \citep{Belloche2020}\\
SerpSmm18a & 1.3$\times$10$^1$ & 2.2$\times$10$^{17}$ & 2.6$\times$10$^{16}$ & 3.0$\times$10$^{16}$ & 1.7$\times$10$^{16}$ & -- & -- & 1.3$\times$10$^{16}$ & 1.9$\times$10$^{15}$ & -- & \citep{Belloche2020}\\
SVS13-A & 4.4$\times$10$^1$ & 2.0$\times$10$^{18}$ & 2.0$\times$10$^{17}$ & 2.1$\times$10$^{17}$ & 1.5$\times$10$^{17}$ & 1.0$\times$10$^{16}$ & -- & 4.0$\times$10$^{16}$ & 4.1$\times$10$^{15}$ & -- & \citep{Belloche2020}\\
IRAS4B & 2.3 & 1.5$\times$10$^{17}$ & 2.0$\times$10$^{16}$ & 5.4$\times$10$^{16}$ & 1.5$\times$10$^{16}$ & -- & -- & 2.5$\times$10$^{15}$ & 6.9$\times$10$^{14}$ & -- & \citep{Belloche2020}\\
Per-emb-11A & -- & 9.1$\times$10$^{15}$ & 3.4$\times$10$^{15}$ & 7.3$\times$10$^{15}$ & 2.3$\times$10$^{15}$ & 2.0$\times$10$^{15}$ & -- & 1.4$\times$10$^{14}$ & 1.3$\times$10$^{14}$ & -- & \citet{Yang2021}\\
Per-emb-12B & -- & 2.0$\times$10$^{17}$ & 5.1$\times$10$^{16}$ & 9.2$\times$10$^{16}$ & 5.2$\times$10$^{16}$ & 3.9$\times$10$^{16}$ & -- & 2.9$\times$10$^{15}$ & 2.3$\times$10$^{15}$ & -- & \citet{Yang2021}\\
Per-emb-13 & -- & 5.0$\times$10$^{16}$ & 2.9$\times$10$^{16}$ & 3.7$\times$10$^{16}$ & 1.5$\times$10$^{16}$ & 9.6$\times$10$^{15}$ & -- & 7.3$\times$10$^{14}$ & 5.7$\times$10$^{14}$ & -- & \citet{Yang2021}\\
Per-emb-27 & -- & 1.1$\times$10$^{18}$ & 5.7$\times$10$^{16}$ & 8.4$\times$10$^{16}$ & 4.8$\times$10$^{16}$ & 3.0$\times$10$^{16}$ & -- & 6.5$\times$10$^{15}$ & 3.1$\times$10$^{15}$ & -- & \citet{Yang2021}\\
Per-emb-29 & -- & 9.0$\times$10$^{16}$ & 2.8$\times$10$^{16}$ & 5.9$\times$10$^{16}$ & 3.6$\times$10$^{16}$ & 2.5$\times$10$^{16}$ & -- & 1.1$\times$10$^{15}$ & 1.5$\times$10$^{15}$ & -- & \citet{Yang2021}\\
Per-emb-44 & -- & 8.1$\times$10$^{17}$ & 1.1$\times$10$^{17}$ & 1.6$\times$10$^{17}$ & 9.6$\times$10$^{16}$ & 6.3$\times$10$^{16}$ & -- & 5.9$\times$10$^{15}$ & 2.8$\times$10$^{15}$ & -- & \citet{Yang2021}\\
\hline 
\end{tabular}
\end{table}
\end{landscape} 
\end{appendix}
\end{document}